\documentclass[11pt,a4paper]{article}

\usepackage{jheppub}
\pdfoutput=1                     % when submitting pdflatex (i.e. with pdf, png or jpg images)

\usepackage{graphicx}
\graphicspath{{./Figs/}}
\usepackage{slashed}
\usepackage{mathtools}        % Fixes and extends (and also loads) amsmath.
\usepackage{bbm}              % Blackboard math variants of CM fonts. (For \mathbbm{1}.)
\usepackage{bm}               % Bolded math symbols (e.g. bolded greek alphabet).
\usepackage{amsmath, amssymb}
\usepackage{bbold}
\usepackage{subcaption} % subfigure-environment
\usepackage{tikz} % graphics macro package (load after hyperref, apparently)
\usetikzlibrary{positioning,arrows.meta,decorations.markings,decorations.pathmorphing}

\usepackage{cleveref} % easier cross-refs using \cref (load after hyperref!)
    \crefname{equation}{}{}
    \crefname{figure}{}{} % example: \crefformat{figure}{#figure~#1#3}
    \crefname{table}{}{}
    \crefname{section}{}{} % example: \crefformat{section}{#2section~#1#3}
    \crefname{appendix}{}{}
    \crefname{footnote}{}{}
    
    \crefalias{subequation}{equation}

\def\nc{\newcommand}

\nc{\nn}{\nonumber}

\nc{\deldag}{{\mathbin{\partial\mkern-10mu/}}}
\def\dag#1{{\mathbin{#1\mkern-9mu /}}}

\def\Slashnew#1{#1\kern-0.55em\raise.05ex\hbox{/}}
\def\slashnew#1{#1\kern-0.5em\raise.05ex\hbox{{$\scriptstyle /$}}}

\def\lsim{\mathrel{\raise.3ex\hbox{$<$\kern-.75em\lower1ex\hbox{$\sim$}}}}
\def\gsim{\mathrel{\raise.3ex\hbox{$>$\kern-.75em\lower1ex\hbox{$\sim$}}}}

\nc{\shalf}{\ensuremath{\textstyle \frac{1}{2}}}
\nc{\ihalf}{\ensuremath{\textstyle \frac{i}{2}}}

\def\sfrac#1#2{\ensuremath{\textstyle \frac{#1}{#2}}}

\def\emph#1{{\em #1}}

\nc{\sss}{\scriptscriptstyle}
\nc{\W}{{\sss W}}
\nc{\sparallel}{{\sss\parallel}}
\nc{\tGamma}{{\tilde \Gamma}}

\newcommand{\sing}{\mathrm{sing}} % singular
\newcommand{\herm}{\mathrm{H}}    % Hermitian part
\newcommand{\evec}[1]{{\bm{#1}}}  % Euclidean vector

\def\sR{{\scriptscriptstyle\rm R}}
\def\sL{{\scriptscriptstyle\rm L}}
\def\H{{\rm H}}
\def\R{{\rm R}}
\def\L{{\rm L}}
\def\I{{\rm I}}

\def\FD{{\scriptscriptstyle {\rm FD}}}
\def\LO{{\scriptscriptstyle {\rm LO}}}
\def\TOT{{\scriptscriptstyle {\rm TOT}}}
\def\PA{{\scriptscriptstyle ||}}
\def\l{{\scriptscriptstyle <}}
\def\g{{\scriptscriptstyle >}}
\def\sl{{s\scriptscriptstyle <}}
\def\sg{{s\scriptscriptstyle >}}

\def\CP{{\scriptscriptstyle {\rm CP}}}
\nc{\CPslash}{{{\scriptscriptstyle {\rm CP}}\mkern-18mu / \mkern 10mu}}

\nc{\HF}{{\sss \rm FH}}

\DeclarePairedDelimiter{\expval}{\langle}{\rangle} % expectation value
\DeclarePairedDelimiter{\acomm}{\{}{\}} % anticommutator

\def\op{{p_0}}
\def\bp{{|\evec{p}|}}
\def\opl{\omega_\pm^{\rm \scriptscriptstyle pl}}
\def\m2p{{|m|^{2\prime}}}
\def\x2p{{|x|^{2\prime}}}
\def\SigA{{\Sigma_{\mathcal A}}}

\title{CP-violating transport theory for Electroweak Baryogenesis with thermal corrections}

\author[a,b]{Kimmo Kainulainen}
\affiliation[a]{Department of Physics, PL 35 (YFL), 40014 University of Jyv\"askyl\"a, Finland}
\affiliation[b]{Helsinki Institute of Physics, PL 64, 00014 University of Helsinki, Finland}
\emailAdd{kimmo.kainulainen@jyu.fi}

\abstract{We derive CP-violating transport equations for fermions for electroweak baryogenesis from the CTP-formalism including thermal corrections at the one-loop level. We consider both the VEV-insertion approximation (VIA) and the semiclassical (SC) formalism. We show that the VIA-method is based on an {\em assumption} that leads to an ill-defined source term containing a pinch singularity, whose regularisation by thermal effects leads to ambiguities including spurious ultraviolet and infrared divergences.
We then carefully review the derivation of the semiclassical formalism and extend it to include thermal corrections. We present the semiclassical Boltzmann equations for thermal WKB-quasiparticles with source terms up to the second order in gradients that contain both dispersive and finite width corrections. We also show that the SC-method reproduces the current divergence equations and that a correct implementation of the Fick's law captures the semiclassical source term even with conserved total current $\partial_\mu j^\mu = 0$. Our results show that the VIA-source term is not just ambiguous, but that it does not exist. Finally, we show that the collisional source terms reported earlier in the semiclassical literature are also spurious, and vanishes in a consistent calculation.}


\arxivnumber{}

\begin{document}
\maketitle

%%%%%%%%%%%%%%%%%%%%%%%%%%%%%%%%%%%%%%%%%%%%%%%%%%%%%%%%%%%%%%%%%%%%%%%%%%%%%%%%%%%%%%%%%%%%%%
%%%%%%%%%%%%%%%%%%%%%%%%%%%%%%%%%%%%%%%%%%%%%%%%%%%%%%%%%%%%%%%%%%%%%%%%%%%%%%%%%%%%%%%%%%%%%%
%
\section{Introduction}
%
%%%%%%%%%%%%%%%%%%%%%%%%%%%%%%%%%%%%%%%%%%%%%%%%%%%%%%%%%%%%%%%%%%%%%%%%%%%%%%%%%%%%%%%%%%%%%%
%%%%%%%%%%%%%%%%%%%%%%%%%%%%%%%%%%%%%%%%%%%%%%%%%%%%%%%%%%%%%%%%%%%%%%%%%%%%%%%%%%%%%%%%%%%%%%

The idea of creating baryons in electroweak phase transition has been around for more than thirty years~\cite{Kuzmin:1985mm,Arnold:1987mh,Bochkarev:1990fx,Cohen:1990py,Cohen:1990it,Turok:1990zg}. While the mechanism does not work in the standard model (SM)~\cite{Kajantie:1996mn}, it can be viable in beyond standard model contexts~\cite{Cline:2017qpe}. There is also a growing interest on models with very strong transitions that could give an observable gravitational wave signal~\cite{Espinosa:2011ax,Kainulainen:2019kyp}. 
An essential part of the electroweak baryogenesis (EWBG) program is a computation of the CP-violating perturbations, which induce the creation of the baryon asymmetry via the electroweak anomaly.
Since the early days, the field has been divided into two distinct approaches: the ``VEV-insertion'' approximation~\cite{Huet:1994jb,Huet:1995mm,Huet:1995sh,Riotto:1995hh,Riotto:1997vy,Carena:1996wj,Carena:1997gx,Lee:2004we,Postma:2019scv,Ramsey-Musolf:2017tgh,Chiang:2016vgf,Blum:2010by,Postma:2021zux} (VIA) and the ``semiclassical force mechanism'' (SC)~\cite{Joyce:1994zt,Cline:2000nw,Kainulainen:2001cn,Kainulainen:2002th,Prokopec:2003pj,Prokopec:2004ic,Fromme:2006wx,Cline:2020jre}, which lead to very different results for the baryon asymmetry~\cite{Cline:2020jre,Basler:2021kgq,Cline:2021dkf}. 

The VIA-method made its first appearance in~\cite{Huet:1994jb,Huet:1995mm,Huet:1995sh}. It was implemented in the context of the closed-time-path (CTP) formalism in~\cite{Riotto:1995hh,Riotto:1997vy} and this derivation was repeated many times since~\cite{Carena:1996wj,Carena:1997gx,Lee:2004we,Chung:2009cb,Chung:2009qs,Postma:2019scv,Postma:2021zux}. All VIA variants are based on computing CP-violating sources for current divergences $\partial_\mu j_a^\mu$, and turning these into diffusion equations using phenomenological Fick's law $j^\mu = (n;D\nabla n)$~\cite{Cohen:1994ss}. In semiclassical method~\cite{Joyce:1994zt} one derives Boltzmann equations for  particle distribution functions, which contain a CP-violating semiclassical force. The SC method has been derived using both the WKB approximation~\cite{Cline:1997vk,Cline:2000nw,Cline:2001rk,Kainulainen:2002th} and the CTP-formalism~\cite{Kainulainen:2001cn,Kainulainen:2002th,Prokopec:2003pj,Prokopec:2004ic}, working in a controlled expansion in gradients. The SC-Boltzmann equation can be integrated into a set of moment equations~\cite{Kainulainen:2001cn,Fromme:2006wx,Cline:2020jre} or fluid equations~\cite{Moore:1995si,Laurent:2020gpg,Friedlander:2020tnq,Dorsch:2021ubz} and eventually to a diffusion equation~\cite{Cline:2000nw,Cline:2020jre}, with a source term induced by the SC-force~\cite{Cline:2000nw,Kainulainen:2001cn,Cline:2020jre}.

The CP-violating sources predicted by the two approaches are parametrically different and give very different results for baryon asymmetry~\cite{Cline:2020jre,Basler:2021kgq,Cline:2021dkf}. Here we will carefully review the derivation of both methods from the CTP-formalism. We first show that the VIA-method is based on an incorrect treatment of the singular mass operator, which gives rise to an ambiguous source term containing a pinch singularity. The standard regularisation of the singularity by thermal corrections is shown to lead to a number of problems including spurious ultraviolet and infrared divergences, which have been either addressed by {\em ad hoc} arguments or just ignored in the VIA literature. 
In the second part of the paper we extend the SC-formalism to include thermal corrections. The ensuing thermal WKB-quasiparticle dispersion relations display a rich structure including WKB-particle and WKB-hole excitations. We also incorporate the coherence and collisional damping corrections, elevating the semiclassical formalism to contain all elements invoked to regularise the VIA-method, but now implemented in a controlled expansion both in gradients and in coupling constants. Our results contain no VIA-type sources, suggesting that these are but artefacts of a deficient scheme.

We also show that Fick's law has been used too naively in the VEV-insertion literature, where it  is applied to the full vector and axial currents. In reality Fick's law should be applied only to the {\em diffusive} part of the total current, which contains also advective and drag-force parts. We show that when proper a division of the current is made, the vector current conservation equation $\partial_\mu j^\mu = 0$ is fully consistent with the first moment of the SC-equations of motion and gives a non-vanishing source for the diffusion equation. 

Finally, we show that {\em collisional sources} predicted in the context of the SC-formalism in ref.~\cite{Prokopec:2004ic} are also spurious. A troubling aspect of these sources always was that they did not vanish in thermal equilibrium. We show that this result is due to inconsistent ans\"aze for thermal background solutions. Our results then show that one can compute collision terms for the SC-Boltzmann equations using the standard field theory methods ignoring all gradient corrections. The main result of this paper, far more important than the criticism of the earlier work, is the semiclassical equation network~\cref{eq:bolzmann_equationdewKB,eq:CPodd_source,eq:colltermB,eq:full_thermal_source} for  CP-violating perturbations in thermal WKB-quasiparticles, with source and collision terms that include all thermal corrections up to one-loop and gradient corrections up to second order. 

This paper is organised as follows. In section~\cref{sec:CTP} we review the basic CTP formalism, paying special attention to the role of singular operators. In section~\cref{sec:VEV_insertion} we review the VIA-method in the toy model considered in~\cite{Postma:2019scv}, discussing carefully the pinch singularity and the resulting ambiguities in the VIA-source. A reader familiar with the CTP-formalism and not interested in this critique may skip directly to section~\cref{sec:SC}, where we review the derivation of  the SC-method and derive the Boltzmann equation for the WKB-quasiparticles. In section~\cref{VEV:correct} we show how  one consistently implements the Fick's law in the current divergence equations. 
In section~\cref{sec:SC_thermal} we extend the semiclassical treatment to include thermal corrections: in section~\cref{sec:SC_disp} we show how the dispersive corrections are implemented and in sections~\cref{sec:SC_collisional_damping}-\cref{sec:SC_coherence_damping} we include the coherent and collisional damping by a thermal operator. In section~\cref{sec:SC_collisional} we evaluate more general collision integrals, including the example discussed in~\cite{Prokopec:2004ic}. Finally, section~\cref{sec:conclusion} contains our conclusions.

%%%%%%%%%%%%%%%%%%%%%%%%%%%%%%%%%%%%%%%%%%%%%%%%%%%%%%%%%%%%%%%%%%%%%%%%%%%%%%%%%%%%%%%%%%%%%%
%%%%%%%%%%%%%%%%%%%%%%%%%%%%%%%%%%%%%%%%%%%%%%%%%%%%%%%%%%%%%%%%%%%%%%%%%%%%%%%%%%%%%%%%%%%%%%
%
\section{CTP-formalism}
\label{sec:CTP}
%
%%%%%%%%%%%%%%%%%%%%%%%%%%%%%%%%%%%%%%%%%%%%%%%%%%%%%%%%%%%%%%%%%%%%%%%%%%%%%%%%%%%%%%%%%%%%%%
%%%%%%%%%%%%%%%%%%%%%%%%%%%%%%%%%%%%%%%%%%%%%%%%%%%%%%%%%%%%%%%%%%%%%%%%%%%%%%%%%%%%%%%%%%%%%%

Both the semiclassical and the VEV-insertion mechanism have been derived starting from the Closed Time Path formalism for the out-of-equilibrium quantum field theory. The main quantity of interest in the CTP formalism is the contour-time ordered 2-point function
\begin{equation}
   i S_{\alpha\beta}(u,v) \equiv
   \expval[\big]{{\mathcal{T}}_{\mathcal{C}}\bigl(\psi_{\alpha}(u)\bar\psi_{\beta}(v)\bigr)}
   \text{,} \label{eq:def-fermion-ctp-propagator}
\end{equation}
defined on some suitable time contour ${\mathcal{C}}$. The expectation value in equation \cref{eq:def-fermion-ctp-propagator} is defined as a trace over states weighted by the non-equilibrium density operator. In what follows, we shall usually suppress the spin indices, which simply follow the space-time coordinate of the field.
The 2-point function obeys the Schwinger--Dyson equation
\begin{equation}
   \int_{\mathcal{C}} {\rm d}^4w S_0^{-1}(u,w) S(w,v)
   = \delta_{\mathcal{C}}^{(4)}(u-v) + \int_{\mathcal{C}} {\rm d}^4w \Sigma(u,w) S(w,v)
   \text{,} 
\label{eq:schwinger-dyson}
\end{equation}
where $S_0^{-1}$ is the free inverse fermion propagator, $S$ is the full fermion propagator
\cref{eq:def-fermion-ctp-propagator} and $\Sigma$ is the fermion self-energy. It is crucial to observe that $\Sigma$ can be divided to singular and nonlocal parts:
\begin{equation}
   \Sigma(u,v) \equiv \delta_{\mathcal{C}}^{(4)}(u-v)\Sigma_\sing(u) + \widetilde\Sigma(u,v)
   \text{.}
\end{equation}
A space-time dependent mass term $m(u)$ is a particular example of a singular operator. In this paper we shall be interested in a complex mass operator
\begin{equation}
 m(u) = m_\R(u) + i m_\I(u)\gamma^5.
\end{equation}
The essential difference between the SC- and the VEV-insertion approaches is in how they treat the singular self-energy contributions. In the SC approach, the mass term is fully re-summed to all orders by absorbing it into the free propagator:
\begin{equation}
   S_0^{-1}(u,v) \equiv \bigl[i\slashed\partial_u - m(u)\bigr]
   \delta_{\mathcal{C}}^{(4)}(u - v)
   \text{.} \label{eq:inverse-free-propagator-contour}
\end{equation}
It is indeed obvious that the singular self-energy terms can be moved freely between the right hand side (including them in $\Sigma$) and the left hand side (including them in $S_0^{-1}$) of the equation~\cref{eq:schwinger-dyson}. However, in the VEV-insertion mechanism mass is not treated as a singular operator, but via a particular ansaz for the nonlocal part of the self-energy.

%%%%%%%%%%%%%%%%%%%%%%%%%%%%%%%%%%%%%%%%%%%%%%%%%%%%%%%%%%%%%%%%%%%%%%%%%%%%%%%%%%%%%%%%%%%%%%
%
\subsection{Kadanoff-Baym equations}
\label{sec:kb-equations}
%
%%%%%%%%%%%%%%%%%%%%%%%%%%%%%%%%%%%%%%%%%%%%%%%%%%%%%%%%%%%%%%%%%%%%%%%%%%%%%%%%%%%%%%%%%%%%%%

The contour-time correlation function $S$ can be parametrised in terms of four real-time 2-point function. We choose them to be the statistical Wightman functions
\begin{equation}
   \begin{split}
      i S^\l(u,v) &\equiv \expval{ \bar\psi(v) \psi(u) }
      \text{,} \\
      i S^\g(u,v) &\equiv \expval{ \psi(u) \bar \psi(v) }
      \text{,}
   \end{split}
   \label{eq:Wightman}
\end{equation}
and the retarded and advanced pole functions $i S^r(u,v) = 2\theta(u_0 - v_0){\mathcal{A}}(u,v)$ and $i S^a(u,v) = - 2\theta(v_0 - u_0){\mathcal{A}}(u,v)$, where ${\mathcal{A}}(u,v) \equiv \frac{1}{2} \expval{ \acomm{\psi(u), \bar\psi(v)} }$ is the spectral function. Using the definition of the spectral function one can also show that $S^{r} - S^{a} = S^\g + S^\l = -2i{\mathcal{A}}$. One can also decompose the pole functions as $S^{r,a} = S_\herm \mp i {\mathcal A}$, where $S_\herm$ and $\mathcal{A}$ obey the spectral relation: $S_\herm(u,v) = -i\, {\rm sign}(u^0 - v^0){\mathcal{A}}(u,v)$.

The contour self-energy function $\Sigma$ can be analogously parametrised in terms of real-time self-energies $\Sigma^\l$, $\Sigma^\g$, $\Sigma^r$ and $\Sigma^a$. Again the retarded and advanced self-energy functions can be decomposed as
\begin{equation}
  \Sigma^{r,a}(u,v) = \Sigma_\herm(u,v)  \mp i \Sigma_{\mathcal{A}}(u,v).
\end{equation}
Note that the singular part of the self-energy is non-absorptive and belongs to $\Sigma_\herm$:
\begin{equation}
  \Sigma_\herm(u,v) = \delta^{(4)}(u-v) \Sigma_{\sing}(u) + \tilde \Sigma_\herm(u,v),
\end{equation}
where $\tilde \Sigma_\herm(u,v)$ is the nonlocal part of the self-energy.
Different self-energy functions are related analogously to the correlation functions: $\Sigma^r - \Sigma^a = \Sigma^\g + \Sigma^\l = -2i\Sigma_{\mathcal{A}}$.

Written in terms of real-time propagators and self-energies, the contour
Scwinger-Dyson equation \cref{eq:schwinger-dyson} breaks into four Kadanoff--Baym (KB) equations
in real time:
\begin{align}
   \bigl((S_0^{-1} - \Sigma^p) * S^p\bigr)(u,v) &= \delta^{(4)}(u - v)
   \text{,} \label{eq:pole} \\
   \bigl((S_0^{-1} - \Sigma^r) * S^s\bigr)(u,v) &= (\Sigma^s * S^a)(u,v)
   \text{,} 
\label{eq:kb}
\end{align}
where $p=r,a$ and $s=<,>$. We continued to suppress the spin and flavour coordinates and defined a shorthand notation for the convolution
\begin{equation}
   (F * G)(u,v) \equiv
   \int_{t_{\rm in}}^\infty {\rm d}{w_0} \int {\rm d}^3{\evec{w}} F(u,w) G(w,v)
   \text{,} 
\label{eq:convolution-real-time-def}
\end{equation}
where ${t_{\rm in}}$ is the initial time of the closed time path. In equations \cref{eq:pole} and \cref{eq:kb} the free inverse propagator $S_0^{-1}$ is understood as a real-time operator with the usual delta function instead of the contour delta function in equation \cref{eq:inverse-free-propagator-contour}.

%%%%%%%%%%%%%%%%%%%%%%%%%%%%%%%%%%%%%%%%%%%%%%%%%%%%%%%%%%%%%%%%%%%%%%%%%%%%%%%%%%%%%%%%%%%%%%
%
\subsection{Current divergences}
\label{sec:currents}
%
%%%%%%%%%%%%%%%%%%%%%%%%%%%%%%%%%%%%%%%%%%%%%%%%%%%%%%%%%%%%%%%%%%%%%%%%%%%%%%%%%%%%%%%%%%%%%%

Equations~\cref{eq:kb} can be expressed in several equivalent forms. For example, employing the time-ordered and the anti time-ordered functions: $S_t = S^r - S^\l = S^a + S^\g$ and $S_{\bar t} = -S^a - S^\l = -S^r + S^\g$ (and similarly for the self-energies) one can write the equation for $S^\l$ as follows
\begin{equation}
   i\deldag_x S^\l(x,y) = (\Sigma_{t} * S^\l - \Sigma^\l * S_{\bar t})(x,y). 
\label{eq:Sless}
\end{equation}
Here $i\deldag$ is the free massless inverse propagator and the singular mass term was treated as an interaction. Technically, it is hidden in the time-ordered self-energy function 
$\Sigma_t$:
\begin{equation}
\Sigma_t(u,v) = \delta^{(4)}(u-v)m(u) + \theta(u_0-v_0) \Sigma^\g(u,v) - \theta(v_0-u_0) \Sigma^\l(u,v)\big).
\label{eq:sigmat}
\end{equation}
Here we assume that $m(u)$ was the only singular operator. Other singular terms could arise from other classical fields and from tadpole diagrams.

Equation~\cref{eq:Sless} serves as the starting point for the VEV-insertion formalism in the CTP approach, where it is used to derive approximations for current divergence equations. However, we can also use it to derive the following {\em exact} results for the vector and the axial current divergences:
\begin{align}
  \phantom{m}
  \partial_\mu j^\mu(x) &= \lim_{y\rightarrow x}
                            {\rm Tr}[ (m(x) - m(y)) S^< (x,y) ]
\nn \\     
      &+ 2{\rm Re} \int {\rm d}^3\evec{w}\int_{t_{\rm in}}^{x_0} {\rm d}w_0 
             {\rm Tr}\big[ \Sigma^>(x,w)S^<(w,x) - \Sigma^<(x,w)S^>(w,x)\big].
\label{eq:vec-current} \\     
  \phantom{m}
      \partial_\mu j_5^\mu(x)  &= -\lim_{y\rightarrow x}
                                {\rm Tr}[ \gamma^5(m(x) + m(y)) S^< (x,y) ]
\nn \\      &+ 2{\rm Re} \int {\rm d}^3\evec{w}\int_{t_{\rm in}}^{x_0} {\rm d}w_0 
           {\rm Tr}\big[ \gamma^5 \big( \Sigma^>(x,w)S^<(w,x) - \Sigma^<(x,w)S^>(w,x)\big)\big].
\label{eq:ax-current}
\end{align}
In practice one can set $t_{\rm in}\rightarrow -\infty$. The currents were defined as follows:
\begin{equation}
j_{\cal O}(x) = \langle \bar \psi (x) {\cal O} 
\psi(x)\rangle = \int \frac{{\rm d}^4k}{(2\pi)^4}{\rm Tr}\big[{\cal O}iS^\l(k,x)\big],
\label{eq:currents_def}
\end{equation}
where ${\cal O} = \gamma^\mu$ for vector current and ${\cal O}_5 = \gamma^5$ for the axial current. First lines in expressions~\cref{eq:vec-current,eq:ax-current} completely account for the space-time dependent singular mass operator, while subsequent memory integrals account for interactions with other particles. Operators $\Sigma^{<,>}$ are nonlocal and vanish when interaction strengths are taken to zero. 

Equations~\cref{eq:vec-current,eq:ax-current} are highly truncated expressions, which can a priori only be used to compute the current divergences when the solutions $S^{p,s}$ are known. Memory integrals in particular contain implicit information hidden in finite upper integration limits in~\cref{eq:vec-current,eq:ax-current}. One can appreciate this by writing an alternative, equivalent form, for example for the vector current divergence equation:
\begin{align}
 \partial_\mu j^\mu(x) \; &= \;\lim_{y\rightarrow x}
                            {\rm Tr}[ (m(x) - m(y)) S^< (x,y) ]
\nn \\     
&+  2{\rm Re}\int {\rm d}^4w
             {\rm Tr}\big[  \tilde \Sigma_\H(x,w)S^<(w,x) +  \Sigma^<(x,w)S_\H(w,x) \big].
\nn \\     
&+ {\rm Re}\int {\rm d}^4w
             {\rm Tr}\big[ \Sigma^>(x,w)S^<(w,x) - \Sigma^<(x,w)S^>(w,x)\big].
\label{eq:vec-current2}
\end{align}
The last term in~\cref{eq:vec-current2} is similar to the memory integral in~\cref{eq:vec-current}, except for being divided by two and with the upper time-integration limit extended to infinity. These changes are compensated by new terms containing the Hermitean self-energy function $\Sigma_\H$ and the Hermitean pole function $S_\H$.

The terms in the first lines of the current divergence equations~\cref{eq:vec-current,eq:ax-current,eq:vec-current2} give the full results from a complete resummation of all mass insertions. They must be evaluated carefully because correlation functions are divergent in the local limit. It would be easy to show that the mass-correction to the vector current divergence in~\cref{eq:vec-current} and~\cref{eq:vec-current2} vanishes, but computing the mass-term in the axial current divergence requires additional information beyond current divergence equations and we postpone these calculations to section~\cref{VEV:correct}. For now just stress again that these terms fully encompass all contributions from mass insertions. This is in stark contrast with the VIA-literature, where the mass is introduced to current equations through a particular non-local memory integral.

%%%%%%%%%%%%%%%%%%%%%%%%%%%%%%%%%%%%%%%%%%%%%%%%%%%%%%%%%%%%%%%%%%%%%%%%%%%%%%%%%%%%%%%%%%%%%%
%%%%%%%%%%%%%%%%%%%%%%%%%%%%%%%%%%%%%%%%%%%%%%%%%%%%%%%%%%%%%%%%%%%%%%%%%%%%%%%%%%%%%%%%%%%%%%
%
\section{VEV-insertion method}
\label{sec:VEV_insertion}
%
%%%%%%%%%%%%%%%%%%%%%%%%%%%%%%%%%%%%%%%%%%%%%%%%%%%%%%%%%%%%%%%%%%%%%%%%%%%%%%%%%%%%%%%%%%%%%%
%%%%%%%%%%%%%%%%%%%%%%%%%%%%%%%%%%%%%%%%%%%%%%%%%%%%%%%%%%%%%%%%%%%%%%%%%%%%%%%%%%%%%%%%%%%%%%

%=============================================================================================
%
\begin{figure}[t!]
\begin{center}
\includegraphics[scale=0.5]{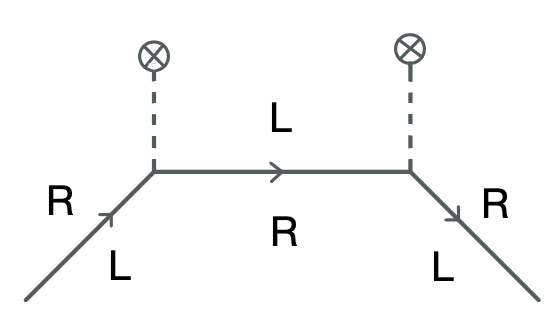}
\end{center}
\vskip-0.5cm
\caption{The self-energy function with two mass insertions on a fermion line.}
\label{fig:self-energy}
\end{figure}
%
%=============================================================================================

We will study the VIA-formalism in the CTP-approach. All existing literature~\cite{Riotto:1995hh,Riotto:1997vy,Carena:1996wj,Lee:2004we,Ramsey-Musolf:2017tgh,Chiang:2016vgf,Blum:2010by,Postma:2019scv} agrees with the current divergence equations~\cref{eq:vec-current,eq:ax-current}, but none of it includes the correction from a singular spatially varying mass derived in previous section. Instead, they add a new {\em non-local} self-energy function induced by {\em two}\footnote{\label{fn:first}The self-energy~\cref{eq:sigma_m} is called the lowest order (LO) correction, while the true LO correction of course is the single mass insertion diagram. This term is presumably neglected for convenience, because it would invoke new chirality mixing components that would spoil the closure of the current divergence equations. However, recovering the local singular mass limit would require summing over {\em all} mass-insertion diagrams.}
mass insertions, corresponding to the diagram shown in figure~\cref{fig:self-energy}: 
\begin{equation}
i\Sigma_{\sL,\sR}^{\l,\g}(x,w) \equiv [-im(x)] i S_{\sR,\sL}^{\l,\g}(x,w) [-im(w)].
\label{eq:sigma_m}
\end{equation}
where $m = m_R + i\gamma^5 m_I$. This term creates new sources to the divergence equations, which are very different from the exact results found in the previous section. While this {\em ansaz} looks obviously deficient, the method is widely used~\cite{Riotto:1995hh,Riotto:1997vy,Carena:1996wj,Carena:1997gx,Lee:2004we,Chung:2009cb,Chung:2009qs,Postma:2019scv,Postma:2021zux} and we will follow the derivation in detail. In the notation of~\cite{Postma:2019scv} the right chiral current divergence equation now becomes
\begin{equation}
  \partial_\mu j_\R^\mu(x) \equiv S_\R^\CP(x) + S_\R^\CPslash(x),
\label{eq:VEV-equationsA}
\end{equation}
were the source terms induced by the operator~\cref{eq:sigma_m} are
\begin{align}
  S_\R^\CP(x)  &= \phantom{-} 2 \int {\rm d}^4w \Theta_{xw} {\rm Re}(m_xm_w^*)
             {\rm Tr}\big[ iS_{\L,xw}^\g iS^\l_{\R,wx} - iS_{\L,xw}^\l iS^\g_{\R,wx} \big],
\label{eq:Source-equationsV1}\\ 
  S_\R^\CPslash(x)        &= -2 \int {\rm d}^4w \Theta_{xw} {\rm Im}(m_xm_w^*)
             {\rm Im}\big[ iS_{\L,xw}^\g iS^\l_{\R,wx} - iS_{\L,xw}^\l iS^\g_{\R,wx} \big], 
\label{eq:Source-equationsA1}
\end{align}
with $\Theta_{xw}\equiv\theta(x^0-w^0)$ and $m_x\equiv m_\R(x)+im_\I(x)$. Left current sources are just the negative of these $S^a_\L = - S^a_\R$, so that $\partial_\mu j_\L^\mu(x) =-\partial_\mu j_\R^\mu(x)$. Vector current is thus conserved and both sources arise from the axial current~\cref{eq:ax-current}. One evaluates these integals by moving to Wigner space (explicit Wigner transform is given by~\cref{eq:wignertot} below) and using massless thermal propagators for $S^{\l,\g}_{\sR,\sL}$:
\begin{equation}
iS^{\l,\g}_{\sL,\sR}(k) = 2\pi \dag{k} \, {\rm sgn}(u\cdot k) f_{\sL,\sR}^{\l,\g}(u\cdot k+\mu_{\sL,\sR}) \delta (k^2)P_{\R,\L}.
\label{eq:eqsless}
\end{equation}
Here $u^\mu$ is the plasma 4-velocity, $P_{\R,\L}=\frac{1}{2}(1\pm \gamma^5)$ and 
\begin{equation}
f^\l_{\rm th}(x) = f(x) \quad {\rm and} \quad f^\g_{\rm th}(x) = 1-f(x),\quad {\rm with} 
\quad f(x) = \frac{1}{e^{x/T}+1}.
\end{equation}
After moving to a new integration variable $r=x-w$ one finds:
\begin{align}
   S_\R^\CP(x) &=
   4 \int {\rm d}^4r \theta(r_0) \int_{k_1,k_2}
   a^+_{x,r}\; k_1 \cdot k_2 \; c_\R(k_1,k_2) \cos((k_1-k_2)\cdot r)
\label{eq:Source-equationsV2}\\ 
   S_\R^\CPslash(x) &=
   4 \int {\rm d}^4r \theta(r_0) \int_{k_1,k_2}
    b^-_{x,r}\; k_1 \cdot k_2 \; c_\R(k_1,k_2) \sin((k_1-k_2)\cdot r),
\label{eq:Source-equationsA2}
\end{align}
where we defined $\int_k \equiv \int \frac{{\rm d}^4k}{(2\pi)^4}$, and
\begin{equation}
c_\R(k_1,k_2) \equiv 4\pi^2{\rm sgn}(u\cdot k_1){\rm sgn}(u\cdot k_2)
\big(f(u\cdot k_1 + \mu_{\R})-f(u\cdot k_2 + \mu_{\L})\big)\delta(k_1^2)\delta(k_2^2).
\label{eq:ik1k2}
\end{equation}
Finally to leading order in gradients
\begin{align}
a^+_{x,r} \equiv m_{\R x}m_{\I x+r} + m_{\I x}m_{\R x+r} &\approx |m_x^2| + \cdots
\nn\\
b^-_{x,r} \equiv m_{\R x}m_{\I x+r} - m_{\I x}m_{\R x+r} &\approx
|m_x^2|\partial_\mu \theta \,r^\mu + \cdots
\label{eq:abpm}
\end{align}
where $\cdots$ refer to higher order gradient corrections. We assume that $m_{\R,\I}(x)$ are real functions, which are time-independent in the wall frame: $m_{\R,\I}(x) =  m_{\R,\I}(z_w) = m_{\R,\I}(\gamma_w(z_{\rm pl} - v_w t_{\rm pl}))$, where $v_w$ is the velocity of the phase transition front and $\gamma_w=1/\sqrt{1-v_w^2}$. 

%%%%%%%%%%%%%%%%%%%%%%%%%%%%%%%%%%%%%%%%%%%%%%%%%%%%%%%%%%%%%%%%%%%%%%%%%%%%%%%%%%%%%%%%%%%%%%
%
\subsection{Pinch singularity}
\label{sec:pinch}
%
%%%%%%%%%%%%%%%%%%%%%%%%%%%%%%%%%%%%%%%%%%%%%%%%%%%%%%%%%%%%%%%%%%%%%%%%%%%%%%%%%%%%%%%%%%%%%%

Equations~(\ref{eq:Source-equationsV2}-\ref{eq:abpm}) agree with~\cite{Postma:2019scv}, which is the latest VIA-calculation in this model. We will  continue the calculation differently from~\cite{Postma:2019scv} however, using the fact that the integrands in both equations~\cref{eq:Source-equationsV2} and~\cref{eq:Source-equationsA2} are symmetric under $r\rightarrow -r$, so that the integration range in $r_0$ can be continued to positive infinity\footnote{This is actually more consistent to begin with. When calculating the self-energy~\cref{eq:sigma_m} and the ensuing memory integrals one is using thermal equilibrium propagators, which means that terms involving the self-energy $\Sigma_\H$ and the pole function $S_\H$ are implicitly absorbed to the definition of thermal quasiparticles and should be dropped in~\cref{eq:vec-current2}. This reduces the memory integral in~\cref{eq:vec-current} to the last line of~\cref{eq:vec-current2}, which is just what we are using here based on symmetry.}:
\begin{align}
   S_\R^\CP(x) &= 2 |m_x^2| \;{\rm Re}
    \int_{k_1,k_2}
     k_1\cdot k_2\; c_\R(k_1,k_2) \int {\rm d}^4r  \,{\mathrm e}^{-i(k_1-k_2)\cdot r}
\label{eq:Source-equationsV3}\\ 
   S_\R^\CPslash(x) &= -2|m_x^2|\partial_\mu \theta \;{\rm Im}
   \int_{k_1,k_2}
     k_1\cdot k_2\; c_\R(k_1,k_2) \int {\rm d}^4r\,    
      r^\mu \,{\mathrm e}^{-i(k_1-k_2)\cdot r}.
\label{eq:Source-equationsA3}
\end{align}
These expressions are Lorenz-covariant and can be computed either in the plasma- or in the wall frame with identical results. One can set $r^\mu \rightarrow \frac{i}{2}(\partial_{k_1^\mu}-\partial_{k_2^\mu})$, after which performing the $r$-integral gives in both cases a delta-function $\delta^4(k_1-k_2)$. With no chemical potentials $S_\R^\CP$ would vanish because of the antisymmetry of the integrand in $k_1 \rightarrow k_2$. Working to first order in chemical potentials in $S_\R^\CP$ and to the lowest order in $S_\R^\CPslash$, one  finds
\begin{equation}
   S_\R^\CP = |m|^2\beta (\mu_\R-\mu_\L) \times I_\gamma  
\qquad {\rm and} \qquad
   S_\R^\CPslash = -v_w\gamma_w |m|^2\theta' \times I_\gamma,
\label{eq:Source-equations4}
\end{equation}
where $\beta = 1/T$ and
\begin{equation}
I_\gamma = 8\pi^2\int \frac{{\rm d}^4k}{(2\pi)^4} k^2\;[{\rm sgn}(k_0)\delta_\gamma(k^2)]^2 f'(k_0).
\label{eq:Igamma}
\end{equation}
Note that in contrast with the standard VIA-literature, both CP-even and CP-odd sources are proportional to the {\em same} integral factor. 

The CP-even term $S_\R^\CP$ is not really a source, but rather a collision term that tends to bring right and left chiralities to equilibrium, but the CP-odd term $S_\R^\CPslash$ appears to have the expected form $\sim v_w\gamma_w |m|^2\theta'$. However, both terms are ill defined because of the overlapping delta functions in $I_\gamma$. Such {\em pinch singularities} often appear when a calculation does not contain all relevant terms to the order one is working. Indeed, the devastating appearance of pinch singularities in the early  formulations of finite temperature field theory was instrumental to the development of the CTP formalism. Here the singularity arises from an attempt to approximate the singular {\em forward scattering} term by a nonlocal collision integral, which is but one in the infinite series of relevant terms. Technically it arises because the mass insertions carry no momenta. As emphasised earlier, the problem would disappear if one summed over all mass insertion diagrams including those with odd number of insertions. But this is not the way chosen in the VIA-literature. Instead, the singularity is hidden by a different order of integrations and regulated by a finite width and thermal masses.

%%%%%%%%%%%%%%%%%%%%%%%%%%%%%%%%%%%%%%%%%%%%%%%%%%%%%%%%%%%%%%%%%%%%%%%%%%%%%%%%%%%%%%%%%%%%%%
%
\paragraph{Regularisation by damping}
\label{sec:simple}
%
%%%%%%%%%%%%%%%%%%%%%%%%%%%%%%%%%%%%%%%%%%%%%%%%%%%%%%%%%%%%%%%%%%%%%%%%%%%%%%%%%%%%%%%%%%%%%%

The integral~\cref{eq:Igamma} is clearly ambiguous\footnote{\label{fn:regularisation}One may think of~\cref{eq:Igamma} is a delta-function integral over a test function $g \sim k^2\delta(k^2)$, so that the result of the integral is $g(0) \sim 0\cdot\infty$, which is arbitrary. The value of this integral thus entirely depends on the regularisation. For example one could have used ${\rm sgn}(k_0)^2 = 1$ leaving out the sign-factors in~\cref{eq:regulated_delta}. Also, one could continue the $k^2$-factor in the nominator by adding a contribution $k^2 \rightarrow k^2 + \alpha \gamma^2$. This quantity, with arbitrary $\alpha$, would have the same (vanishing) $\gamma\rightarrow 0$ limit as our choice for the regulated $I_\gamma$. However, for a finite $\gamma$ this continuation could be used to give any desired value for $I_\gamma$ by varying $\alpha$.}. We anticipated this by giving the delta-function an index $\gamma$, which refers to a regulated quantity. We will eventually follow the VIA literature and attempt to interpret $\gamma$ as a finite thermal width. To this end we choose the following particular regularisation choice:
\begin{align}
{\rm sgn}(k_0) \delta_\gamma(k^2) &\rightarrow \frac{1}{2\omega_a} 
\sum_\pm \pm\delta_\gamma(k_0\mp\omega_a) 
\nonumber\\
&\equiv \frac{1}{2\pi\omega_a} \sum_\pm \frac{\pm\gamma_a}{(k_0\mp\omega_a)^2+\gamma_a^2}
\equiv \frac{1}{2\pi\omega_a} \sum_\pm g_{{\bm k}a}^{\pm}(k_0).
\label{eq:regulated_delta}
\end{align}
with $\smash{\omega_a^2 = m_a^2(T)+\bm{k}^2}$ and $m^2_a$ is the thermal mass. We allow for different thermal masses and widths for the left- and right chiral states with $a=L,R$, {\em redefining}:
\begin{equation}
I_\gamma \rightarrow I^{\rm P}_\gamma \equiv \sum_{\pm\pm'} \int \frac{{\rm d}^4k}{(2\pi)^4} \frac{2}{\omega_L\omega_R}
(k_0^2-{\bm k}^2) f'(k_0) g_{{\bm k} L}^\pm(k_0)g_{{\bm k} R}^{\pm'}(k_0),
\label{eq:Igamma_special}
\end{equation}
For left and right chiral quarks one finds $m_\L^2 = (\frac{1}{6}g_s^2+\frac{3}{32}g^2 + \frac{1}{16}y_t^2)T^2$, $m_\R^2 = (\frac{1}{6}g_s^2 + \frac{1}{8}y_t^2)T^2$ and $\gamma_{\L,\R} = \gamma \approx 0.152 g_s^2T$~\cite{Weldon:1982bn,Braaten:1992gd}. The integral~\cref{eq:Igamma_special} is 
easily evaluated numerically and the $k_0$-integral can also be performed analytically using contour integration, being careful to include all residues, including the ones associated with the special points of the function $f'(k_0)$ along the imaginary axis. To be slightly more general we control the contribution from the residues on the imaginary axes by a parameter $r$, making a further redefinition
\begin{align}
I^{\rm P}_{\gamma,r} \equiv \sum_{\pm\pm'} \int_\evec{k} 
     & \frac{1}{\omega_L\omega_R} \Big\{ 
      {\rm Re}\Big[ \pm A^{\pm'}_{{\bm k} R}(\pm\omega_L-i\gamma) 
                               \pm' A^{\pm}_{{\bm k} L}(\pm'\omega_R+i\gamma) \Big]
\nonumber \\      
    & + \;2r \; {\rm Im}\sum_{n=0}^\infty \frac{\partial}{\partial k_0}\Big[(k_0^2-{\bm k}^2)
    g_{\bm{k}\gamma L}^\pm(k_0)g_{\bm{k}\gamma R}^{\pm'}(k_0))\Big]_{k_0=i\omega_n} \Big\},
\label{eq:Igamma_special_b}
\end{align}
where $\int_\evec{k} = \int\frac{{\rm d}^3\evec{k}}{(2\pi)^3}$ and $A_{{\bm k} a}^s(k_0) \equiv  f'(k_0)(k_0^2-\bm{k}^2)g_{{\bm k} a}^{s}(k_0)$ and $\omega_n \equiv (2n+1)\pi T$. To get the last term we used $f'(i\omega_n + \delta z) = 1/\delta z^2 + 1/12 + {\mathcal O}(\delta z^2)$. Obviously $I_\gamma^{\rm P} = I_{\gamma,1}^{\rm P}$. One can define entirely new functions by deforming the integration contour such that it avoids some of the poles in the imaginary axis. Setting $r=0$ would remove the second line in~\cref{eq:Igamma_special_b} entirely, giving rise to a new regulated quantity $I^{\rm P}_{\gamma,0}$, which keeps only the quasiparticle pole contributions to the original integral.

%%%%%%%%%%%%%%%%%%%%%%%%%%%%%%%%%%%%%%%%%%%%%%%%%%%%%%%%%%%%%%%%%%%%%%%%%%%%%%%%%%%%%%%%%%%%%%
%
\subsection{VIA-literature regulators}
\label{sec:earlier}
%
%%%%%%%%%%%%%%%%%%%%%%%%%%%%%%%%%%%%%%%%%%%%%%%%%%%%%%%%%%%%%%%%%%%%%%%%%%%%%%%%%%%%%%%%%%%%%%

The usual computation of $S_\R^\CP$ and $S_\R^\CPslash$ in the VIA-literature does not use the symmetry of the $r_0$-integral, but performs the $k_{0i}$-integrations  before the $r_0$-integration in equations~\cref{eq:Source-equationsV2,eq:Source-equationsA2}. This hides the pinch singularity and apparently leads to different results. In particular, all VIA-treatments find different coefficients for the CP-even and the CP-odd terms. We shall now see that these differences are due to additional, hidden assumptions associated with the choice of integration contour. 

%%%%%%%%%%%%%%%%%%%%%%%%%%%%%%%%%%%%%%%%%%%%%%%%%%%%%%%%%%%%%%%%%%%%%%%%%%%%%%%%%%%%%%%%%%%%%%
\paragraph{CP-even integral}
%%%%%%%%%%%%%%%%%%%%%%%%%%%%%%%%%%%%%%%%%%%%%%%%%%%%%%%%%%%%%%%%%%%%%%%%%%%%%%%%%%%%%%%%%%%%%%

Let us consider the CP-even integral~\cref{eq:Source-equationsV2} first. One starts from~\cref{eq:Source-equationsA2} and~\cref{eq:ik1k2} using the regulated delta functions~\cref{eq:regulated_delta}. It is easy to write the CP-even source in the form similar to~\cref{eq:Source-equations4}, but with the integral $I^{\rm P}_\gamma$ replaced by:
\begin{align}
I^\CP_\gamma = \frac{1}{2\pi^2}{\rm Re} &\sum_{\pm,\pm'} \int_{\evec{k}}\frac{1}{\omega_\L\omega_\R} 
      \int {\rm d}k_{10}{\rm d}k_{20} \int_0^{\infty} {\rm d}r_0 \, {\mathrm e}^{-i(k_{10}-k_{20})r_0}  \times \nn \\& \times 
\big(k_{10}k_{20} - k^2\big) g_{{\bm k} L}(k_{10}) g_{{\bm k} R}(k_{20})
\big[f'(k_{10})+f'(k_{20})\big].
\label{eq:cpevenvevlit}
\end{align}
The exponential factors $e^{\pm i k_{i0} r_0}$ dictate which poles contribute: because $r_0>0$ this implies that in each $k_{10}$-integral picks only the residues in the lower part and each $k_{20}$-integral only in the upper part of the complex plane. Performing first the $k_{0}$ integrals corresponding to the variable not present in the distribution function, followed by the $r_0$-integral and finally expressing the remaining $k_{0}$ integral as sum over residues in the appropriate half-plane, one can write $I^\CP_\gamma$ as follows:
\begin{align}
I^\CP_\gamma = {\rm Im} \sum_{\pm,\pm'} \int_{\evec{k}} \frac{1}{2\omega_\L\omega_\R} 
\Big[ & \sum_{\rm res, lower}
   f'(k_{0})g_{{\bm k} L}^\pm(k_{0}) B^{(1)\pm'}_{{\bm k} R}(k_{0})
\nonumber \\
 +   & \sum_{\rm res, upper}
   f'(k_{0})g_{{\bm k} R}^{\pm'}(k_{0}) B^{(1)\pm}_{{\bm k} L}(k_{0}) \;\;\Big],
\label{eq:cpevenvevlit2}
\end{align}
where functions $g_{{\bm k} a}(k_0)$ were defined in~\cref{eq:regulated_delta} and furthermore
\begin{equation}
B^{(\ell)\pm}_{{\bm k} a}(k_0) \equiv \frac{(\omega_a \pm is_a\gamma_a )k_0\mp {\bm k}^2}{(k_0 \mp \omega_a - is_a\gamma_a )^\ell},
\end{equation}
with $s_R = 1$ and $s_L = -1$. The residues contributing to~\cref{eq:cpevenvevlit2} arise from the poles of the $g^\pm_{{\bm k}a}(k_0)$ and from the special points of the $f'(k_0)$-function. Functions $B^{(\ell )\pm}_{{\bm k}L}$ ($B^{(\ell )\pm}_{{\bm k}R}$ ) do not have poles in the upper (lower) plane by construction. After a little algebra the full result can be written as $I^\CP_\gamma = I^\CP_{\gamma,1}$ where we again define the generalised integral: 
%
%
%==============================================================================================
%
\begin{figure}[t!]
\begin{center}
\includegraphics[scale=0.4]{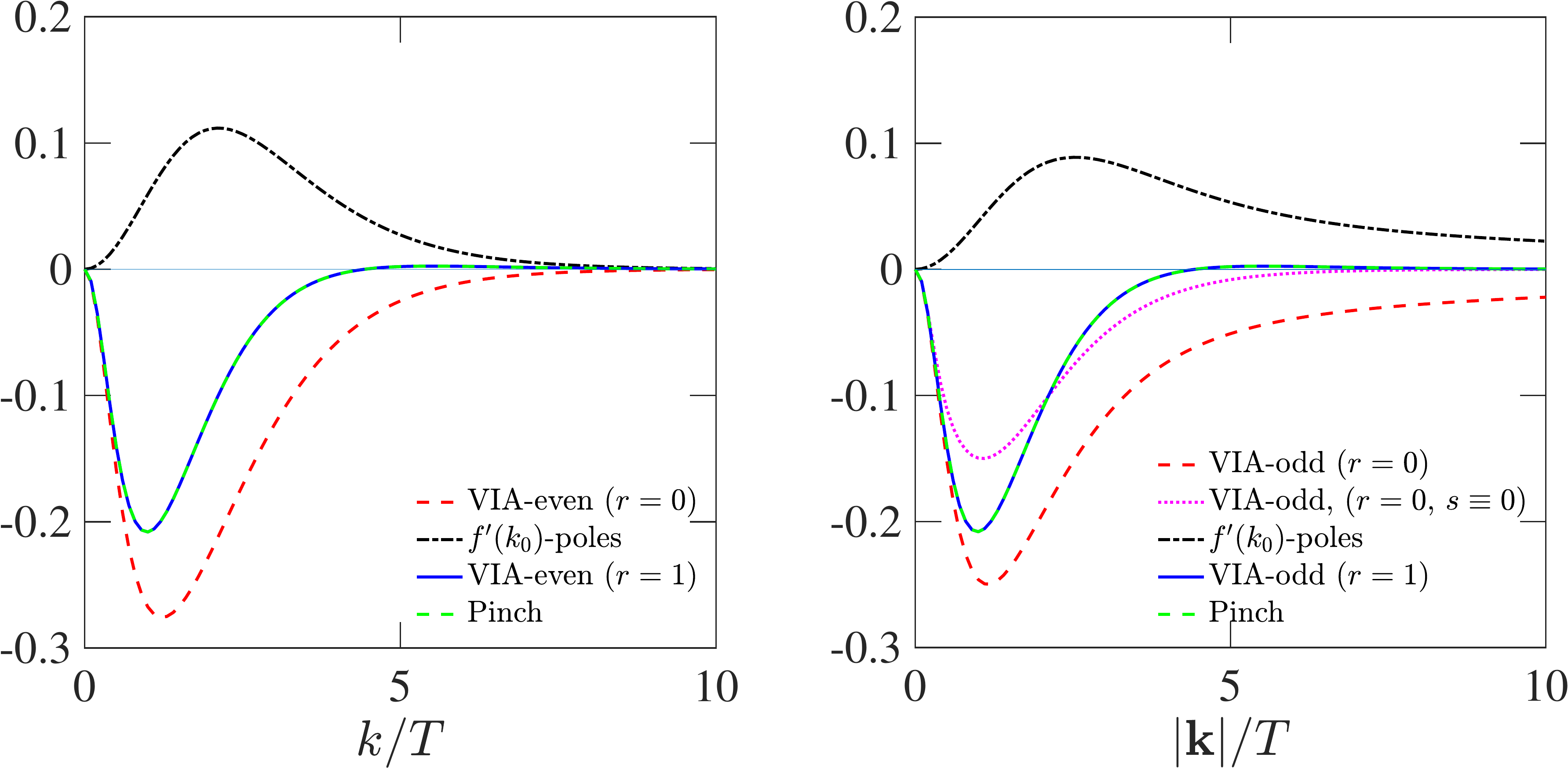}
\end{center}
\vskip-0.5truecm
\caption{(Left) shown are the distributions $i_\gamma^A(|{\bm k}|)$ corresponding to the various contributions to the CP-even function $I^\CP_\gamma$. The complete VIA-result (blue) including all residues and the numerically integrated pinch distribution from~\cref{eq:Igamma_special} (green dashed) fall on top of each others. (Right) same comparison corresponding to the contributions to the CP-odd function $I^\CPslash_\gamma$. Again the complete VIA-result (blue) agrees with the pinch distribution.}
\label{fig:vev-insertion-sources}
\end{figure}
%
%==============================================================================================
%
\begin{align}
I^\CP_{\gamma,r} \equiv  \int_{\bm k} &\frac{1}{\omega_\L\omega_\R} \Big\{
{\rm Im}\Big[ \frac{f'(E_\L^*) + f'(E_\R )}{E_\L^*-E_\R}{\rm tr}^\LO_1
           + \frac{f'(E_\L)+f'(E_\R )}{E_\L+E_\R}{\rm tr}^\LO_2 \Big] 
\nonumber \\ 
           & -r {\rm Re}\sum_{n=0}^\infty \sum_{\pm\pm'} \Big[
           \frac{\partial}{\partial k_0}\Big( g^\pm_{{\bm k}L} B^{(1)\pm'}_{{\bm k}R} \Big)_{k_0=-i\omega_n} 
         + \frac{\partial}{\partial k_0}\Big( g^{\pm'}_{{\bm k}R} B^{(1)\pm}_{{\bm k}L} \Big)_{k_0=i\omega_n} \Big]\Big\},
\label{eq:sourcepostma_even_3}
\end{align}
where $E_{\L,\R} \equiv \omega_{\L,\R} + i\gamma$ and ${\rm tr}^\LO_1 = E_L^*E_R-\evec{k}^2$ and ${\rm tr}^\LO_2 = -E_LE_R-\evec{k}^2$. The first line comes from the 
quasiparticle poles (those of the $g^\pm_{{\bm k}a}$-function) rewritten in the usual VIA-notation and it agrees with the full result of~\cite{Postma:2019scv}. The remaining poles on the second line of~\cref{eq:sourcepostma_even_3} were always ignored in the VIA-literature, which then has implicitly assumed that $I^\CP_\gamma \equiv I^\CP_{\gamma,0}$. It should not be a surprise that when these terms are included, integrals~\cref{eq:Igamma_special} and~\cref{eq:sourcepostma_even_3} are equivalent: $I^\CP_{\gamma,§1} = I_{\gamma,1}^{\rm P}$. We show this in the left panel of figure~\cref{fig:vev-insertion-sources}, at the level of differential distributions $i^A_\gamma(|{\bm k}|)$ defined by
\begin{equation}
I_\gamma^A \equiv \frac{1}{2\pi^2}\int_0^\infty {\rm d}|{\bm k}| i^A_\gamma(|{\bm k}|)
\end{equation}
for the various contributions to the even source function. The red dashed line in fig.~\cref{fig:vev-insertion-sources} is the usual VIA-result that includes only the first line in~\cref{eq:sourcepostma_even_3} and black dash-dotted lines is the contribution from the residues in the imaginary axis, and finally the blue line is the sum of the two. The full result result exactly coincides with the green dashed line that corresponds to the integrand of the regulated pinch-term defined in~\cref{eq:Igamma_special}.

%%%%%%%%%%%%%%%%%%%%%%%%%%%%%%%%%%%%%%%%%%%%%%%%%%%%%%%%%%%%%%%%%%%%%%%%%%%%%%%%%%%%%%%%%%%%%%
\paragraph{CP-odd integral}
%%%%%%%%%%%%%%%%%%%%%%%%%%%%%%%%%%%%%%%%%%%%%%%%%%%%%%%%%%%%%%%%%%%%%%%%%%%%%%%%%%%%%%%%%%%%%%

Similarly to the CP-even case, the CP-odd term~\cref{eq:Source-equationsA2} can be written in the simple form of equation~\cref{eq:Source-equations4}, in terms of a regulated integral $I^\CPslash_{\gamma,r}$. The calculation is entirely analogous to the CP-even case and we only quote the final result:
\begin{align} \phantom{m}
I^\CPslash_{\gamma,r} = 2 \int_{\evec{k}} &\frac{1}{\omega_\L\omega_\R} \Big\{
{\rm Im}\Big[\frac{f(E_\L^*)-f(E_\R )}{(E_\L^*-E_\R)^2}{\rm tr}^\LO_1
           - \frac{f(E_\L)+f(E_\R )+s}{(E_\L+E_\R)^2}{\rm tr}^\LO_2 \Big]
\nonumber\\
           & + r {\rm Re} \sum_{n=0}^\infty \sum_{\pm\pm'}
           \Big( g^\pm_{{\bm k}L}(-k_0) B^{(2)\pm'}_{{\bm k}R}(-k_0) 
               + g^{\pm'}_{{\bm k}R}(k_0) B^{(2)\pm}_{{\bm k}L}(k_0) \Big)_{k_0=i\omega_n} \Big\},
\label{eq:sourcepostma_odd}
\end{align}
where $s=-1$ for fermions. The first line in~\cref{eq:sourcepostma_odd} again coincides with the standard VIA result~\cite{Postma:2019scv}, while the second line includes the residues coming from the previously neglected poles of the $f(k_0)$-function, found by using $f(i\omega_n + \delta z) = -1/\delta z + {\mathcal O}(\delta z^0)$. When all contributions included, we find that the integrals are again the same: $I^\CPslash_{\gamma,1}= I_{\gamma_,1}^{\rm P}=I^\CP_{\gamma,1}$. We show again the equivalence of these quantities at the level of the momentum distributions in the right panel of figure~\cref{fig:vev-insertion-sources}.

%%%%%%%%%%%%%%%%%%%%%%%%%%%%%%%%%%%%%%%%%%%%%%%%%%%%%%%%%%%%%%%%%%%%%%%%%%%%%%%%%%%%%%%%%%%%%%
\paragraph{The "UV-singularity"}
%%%%%%%%%%%%%%%%%%%%%%%%%%%%%%%%%%%%%%%%%%%%%%%%%%%%%%%%%%%%%%%%%%%%%%%%%%%%%%%%%%%%%%%%%%%%%%

The term proportional to $s$ in~\cref{eq:sourcepostma_odd} has raised some due concerns in the VIA-literature, because it is UV-divergent. This divergence is manifest in figure~\cref{fig:vev-insertion-sources} as a very slow convergence of the standard VIA-result for large $|{\bm k}|$ (red dashed line). The common {\em belief} has been that this divergence is removed by a renormalisation counter term\footnote{For example ref.~\cite{Postma:2019scv} (among others) makes this claim citing~\cite{Liu:2011jh}. However, ref.~\cite{Liu:2011jh} does not present a proof, but refers to a private communication with C. Lee, where the trail appears to end. This explanation should have been discarded as untenable from the outset, because the counter-term required to cancel the UV-divergence in $I_{\gamma,0}^\CPslash$ should be proportional to a $T$-dependent thermal damping rate.}, which has prompted people to adopt a further {\em recipe} of setting $s\rightarrow 0$ in~\cref{eq:sourcepostma_odd}. We show the VIA-source with $s\equiv 0$ by dotted magenta line in figure~\cref{fig:vev-insertion-sources}. It is now evident that this explanation is wrong: the correct one is that when performing integrations this particular order, the implicit restriction to quasiparticle pole contributions does not give a finite result.  Moreover, setting $s=0$ by hand just replaces the UV-divergence by an IR-divergence, because in the massless limit the $s$-term is needed to keep the second term in the first line of~\cref{eq:sourcepostma_odd} IR-finite. This divergence is again hidden by the thermal masses, but its presence is a yet another sign of the inherent inconsistency of the scheme. 

%%%%%%%%%%%%%%%%%%%%%%%%%%%%%%%%%%%%%%%%%%%%%%%%%%%%%%%%%%%%%%%%%%%%%%%%%%%%%%%%%%%%%%%%%%%%%%
%
\subsection{Mass mixing models}
\label{sec:inconsistency}
%
%%%%%%%%%%%%%%%%%%%%%%%%%%%%%%%%%%%%%%%%%%%%%%%%%%%%%%%%%%%%%%%%%%%%%%%%%%%%%%%%%%%%%%%%%%%%%%

The model building efforts with the VIA-method often involve more complicated self-energy structures, where mass-insertions enter as spatially varying (Dirac) mass correction onto a propagator with constant diagonal (Majorana) masses.  This has led to effective VIA-self-energy operators of the form~\cite{Riotto:1997vy,Carena:1996wj,Carena:1997gx,Lee:2004we,Chiang:2016vgf,Blum:2010by}
\begin{equation}
i\Sigma_{a,b}^{\l,\g} = -m_\L(x)m_\R^*(y) P_\L iS_{b,a}^{\l,\g}(x,w) P_\L  
                        -m_\R(x)m^*_\L(y) P_\R iS_{b,a}^{\l,\g}(x,w) P_\R.
\end{equation}
The labels $a, b$ refer to the left- and right handed Majorana fields. It is essential that now, unlike in~\cref{eq:sigma_m}, one has the same chiral projector on both sides of the propagator. This construction gives the following CP-violating source for the species $a$:
\begin{equation}
  S_{a}^\CPslash(x) = -2 \int {\rm d}^4w \Theta_{xw} {\rm Im}(m_{Rx}m_{Iw}^*)
             {\rm Im}\big[ P_\L iS_{b,xw}^\g P_\L iS^\l_{a,wx} - < \leftrightarrow >  \big] + L \leftrightarrow R .
\label{eq:Source-equationsA1}
\end{equation}
Using massive equilibrium correlation functions 
\begin{equation}
iS_a^{\l,\g}(k) = 2\pi (\dag{k} + M_a) \, {\rm sgn}(u\cdot k) f_a^{\l,\g}(u\cdot k+\mu_a) \delta (k^2-M_a^2).
\label{eq:eqslessmass}
\end{equation}
one immediately finds:
\begin{equation}
   S_a^\CPslash(x) = 4 M_aM_b |m_\R||m_\L|\partial_\mu\theta 
   \int {\rm d}^4r \theta(r_0) r^\mu \int_{k_1,k_2} c_\R(k_1,k_2) \sin((k_1-k_2)\cdot r),
\label{eq:Source-equationsAb2}
\end{equation}
where $\theta$ is the phase of $m_\R(x)m_\L^*(x)$. Proceeding as in section~\cref{sec:pinch} one can write~\cref{eq:Source-equationsAb2} as follows:
\begin{equation}
   S_a^\CPslash = -v_w\gamma_w M_aM_b |m_R||m_L|\theta' J^{ab}_\gamma,
\label{eq:Source-equations5}
\end{equation}
where the phase space integral is
\begin{align}
J^{ab}_{\gamma} &= 8\pi^2 \int \frac{{\rm d}^4k}{(2\pi)^4}\;[{\rm sgn}(k_0)\delta_\gamma(k^2-M_a^2)][{\rm sgn}(k_0)\delta_\gamma(k^2-M_b^2)] f'(k_0).
\label{eq:Igamma2a}
\end{align}
This integral is even more ambiguous than~\cref{eq:Igamma}, evaluating either to zero, when $M_a \neq M_b$, or to infinity, when $M_a = M_b$. The integral~\cref{eq:Igamma2a} can of course be regulated using different options discussed in the previous section, with a corresponding spectrum of different results, where all regulated quantities display an enhancement when $|M_a-M_b| \lsim \gamma$, originating from the pinch singularity in~\cref{eq:Igamma2a} in the degenerate limit and for $M_a=M_b$ all results become strictly infinite in the zero coupling limit. Sources with this apocryphal behaviour are often reported in the VEV-insertion literature~\cite{Riotto:1995hh,Riotto:1997vy,Carena:1996wj,Carena:1997gx,Lee:2004we,Ramsey-Musolf:2017tgh,Chiang:2016vgf,Blum:2010by}, where the phenomenon has been dubbed a {\em resonance enhancement}~\cite{Lee:2004we}.

%%%%%%%%%%%%%%%%%%%%%%%%%%%%%%%%%%%%%%%%%%%%%%%%%%%%%%%%%%%%%%%%%%%%%%%%%%%%%%%%%%%%%%%%%%%%%%
\paragraph{Discussion}
\label{VEV:discussion}
%%%%%%%%%%%%%%%%%%%%%%%%%%%%%%%%%%%%%%%%%%%%%%%%%%%%%%%%%%%%%%%%%%%%%%%%%%%%%%%%%%%%%%%%%%%%%%

We have seen that the CTP-derivation of the VEV-insertion mechanism is based on an ad hoc ansaz, where the effect of a singular mass-operator is modelled by a nonlocal self-energy term and a memory integral that contains a pinch singularity.  Attempts to regulate the singularity by a finite width and thermal mass effects were shown to give ambiguous results, even after some errors permeating the VIA-literature were corrected.

VEV-insertion mechanism was introduced differently in articles~\cite{Huet:1994jb,Huet:1995mm,Huet:1995sh} and in~\cite{Riotto:1997gu,Carena:1997gx}. Instead of current divergences these works attempt to model the CP-violating currents $j^\mu_\CPslash$ emanating from the transition front. The zero component of the current is then promoted to a source on the current equations by a phenomenological relation $S_\CPslash \equiv j^0_\CPslash/\tau$, where $\tau$ is some ``typical'' thermalisation scale in the process~\cite{Carena:1997gx}. In the later work~\cite{Riotto:1997gu,Carena:1997gx} the current was computed starting from the CTP-formalism, and is subordinate to the same criticism as the CTP-based derivation of the current divergence equations. The very early work~\cite{Huet:1994jb,Huet:1995mm,Huet:1995sh} on VIA approach was quite different. There one computed reflection currents off the wall in the semiclassical limit, based on a Dirac equation and taking account of decohering collisions using a somewhat nebulous concept of ``an effective mean free path of a particle between successive reflections". We shall not discuss these phenomenological approaches in more detail here.

While the the ``CTP-derivation" of the VIA-method was here found to be deeply in want of rigour, one might still ask if it still stumbled on something interesting: could a source parametrically of the form~\cref{eq:Source-equations5} arise in a more consistent approach? Luckily a formalism where this question can be studied already exists. The semiclassical method~\cite{Kainulainen:2001cn,Kainulainen:2002th} is also based on the CTP-formalism and it treats the singular mass term consistently in a controlled expansion in gradients; in section~\cref{sec:fickslaw} we will show that it is in perfect agreement with the exact current divergence equations~\cref{eq:ax-current}. However, the current formulation of the SC-method does not include thermal corrections for the quasiparticles. In what follows we first review the derivation of the SC method and then extend it to encompass all thermal corrections that were invoked to regularise the VIA-sources in this section. Armed with these results we will be able to unambiguously decide whether the VIA-sources exist. (The answer will be they do not).

%%%%%%%%%%%%%%%%%%%%%%%%%%%%%%%%%%%%%%%%%%%%%%%%%%%%%%%%%%%%%%%%%%%%%%%%%%%%%%%%%%%%%%%%%%%%%%
%%%%%%%%%%%%%%%%%%%%%%%%%%%%%%%%%%%%%%%%%%%%%%%%%%%%%%%%%%%%%%%%%%%%%%%%%%%%%%%%%%%%%%%%%%%%%%
%
\section{The Semiclassical Formalism}
\label{sec:SC}
%
%%%%%%%%%%%%%%%%%%%%%%%%%%%%%%%%%%%%%%%%%%%%%%%%%%%%%%%%%%%%%%%%%%%%%%%%%%%%%%%%%%%%%%%%%%%%%%
%%%%%%%%%%%%%%%%%%%%%%%%%%%%%%%%%%%%%%%%%%%%%%%%%%%%%%%%%%%%%%%%%%%%%%%%%%%%%%%%%%%%%%%%%%%%%%

Semiclassical method was derived from the CTP-formalism in~\cite{Kainulainen:2001cn,Kainulainen:2002th}. We will outline this derivation for completeness in section~\cref{sec:SC_wkp-qp}. We then generalise the free WKB-quasiparticle picture first to thermal WKB-quasiparticles and then to damped WKB-quasiparticles. We start by rewriting the KB-equations~\cref{eq:pole,eq:kb} in the mixed representation, by performing the Wigner transformation
\begin{equation}
    S(k,x) \equiv \int\mathrm{d}^4r\,{\mathrm e}^{{\mathrm i} k \cdot r}S\bigl(
        x + \sfrac{r}{2}, x - \sfrac{r}{2}
    \bigr).
\label{eq:wignertot}
\end{equation}
A particularly useful form for the mixed representation KB-equations, from the point of view of gradient expansions, was derived in~\cite{Jukkala:2019slc}:
\def\sSigma{{\scriptscriptstyle \Sigma}}
\begin{alignat}{3}
   \slashed {\hat K} & S^{p}
   &&{}-   {\mathrm e}^{-\ihalf\partial_x^\sSigma \cdot \partial_k}
           [\Sigma^{p}_{\rm out} ({\hat K},x) S^p]
   && =    1 \text{,} 
\label{PoleEqMix2} \\
   \slashed {\hat K} & S^{s}
   && {}- {\mathrm e}^{-\ihalf\partial_x^\sSigma \cdot \partial_k}
          [\Sigma^{\mathrm{r}}_{\rm out} ({\hat K},x) S^{s} ]
   && = 
          {\mathrm e}^{-\ihalf\partial_x^\sSigma \cdot \partial_k}
          [\Sigma^{s}_{\rm out} ({\hat K},x) S^{\mathrm{a}}] \text{,} 
\label{StatEqMix2}
\end{alignat}
where ${\hat K}^\mu \equiv k^\mu + \ihalf \partial^\mu_x$ and the functions $\Sigma^a_{\rm out}$ are defined as
\begin{equation}
    \Sigma^a_{\rm out}(k,x) \equiv \int{\rm d}^4z\,{\mathrm e}^{{\mathrm i} k\cdot(x-z)} \Sigma(x,z)
    = {\mathrm e}^{\ihalf\partial_x^\sSigma \cdot \partial_k^\sSigma} \Sigma^a(k,x), 
\label{eq:SigmaOut}
\end{equation}
where $\Sigma^a(k,x)$ are the usual Wigner functions. Equations~\cref{PoleEqMix2,StatEqMix2} can be used to develop a more complete formalism that includes the quantum coherence effects beyond the semiclassical picture, as has already been done in the spatially homogeneous situations~\cite{Herranen:2008hi,Herranen:2008hu,Herranen:2008di,Herranen:2009zi,Herranen:2009xi,Herranen:2010mh,Fidler:2011yq}. Here they serve as a complete reference and starting point for SC-formalism with interactions.

%%%%%%%%%%%%%%%%%%%%%%%%%%%%%%%%%%%%%%%%%%%%%%%%%%%%%%%%%%%%%%%%%%%%%%%%%%%%%%%%%%%%%%%%%%%%%%
%
\subsection{The WKB-quasiparticle picture}
\label{sec:SC_wkp-qp}
%
%%%%%%%%%%%%%%%%%%%%%%%%%%%%%%%%%%%%%%%%%%%%%%%%%%%%%%%%%%%%%%%%%%%%%%%%%%%%%%%%%%%%%%%%%%%%%%

One can derive the collisionless equation by Wigner transforming the free equation $\big(i\deldag_x - m(x) \big)S^\l(x,y) = 0$ directly, or from equations~\cref{PoleEqMix2,StatEqMix2} by turning off the interactions. Indeed, replacing $\Sigma^p \rightarrow m_\R + i\gamma^5m_\I$, setting other self-energy functions to zero and noting that $m_{\rm out}(\hat K,x) = m(x)$, equation~\cref{StatEqMix2} for the Wightmann function $S^\l$ reduces to:
\begin{equation}
\Big( \dag{k} + \sfrac{i}{2}\deldag_x - m_\R(x-\sfrac{i}{2}\partial_k) 
                          - i\gamma^5 m_\I(x-\sfrac{i}{2}\partial_k) \Big) S^\l(x,k) = 0.
\label{eq:SCexact}
\end{equation}
This equation is still exact for the mass operators. Gradient expansions are easily obtained by Taylor expanding to the desired order. Before doing this, we simplify the equation using the symmetries. Assuming a stationary solution, moving to the wall frame and using planar symmetry, the differential operator in~\cref{eq:SCexact} reduces to $\deldag_x = \gamma^3 \partial_z$ and the mass operators to $m_{\L,\R}(z+\sfrac{i}{2}\partial_{k_z})$. These operators commute with boosts in directions parallel to wall, so we can move into the frame where the parallel momentum $\evec{k}_\PA$ vanishes. This can be done with the boost $L_\PA$ that transforms $L_\PA (k_0\gamma^0 - \evec{k}_\PA\cdot \evec{\gamma})L_\PA^{-1} = \tilde k_0\gamma_0$,  with
$\tilde k_0 \equiv {\rm sgn}(k_0)(k_0^2-\evec{k}^2_\PA)^{1/2}$. Explicitly
\begin{equation}
L_\PA^{\pm 1} = \frac{1}{\sqrt{2}} \big( \sqrt{\gamma_\PA \! + \!1}  
                \mp \sqrt{\gamma_\PA\!-\!1} \;\evec{\alpha}\cdot \evec{\hat k}_\PA \big),
\label{L-Lambda}
\end{equation}
where $\evec{\alpha} = \gamma^0\evec{\gamma}$ and $\gamma_\PA = k_0/\tilde k_0$. In the doubly boosted frame the KB-equation for boosted correlation functions $\bar S^\l_{\PA s}$ can be written
\begin{equation}
\Big(\tilde k_0 - \gamma^0\gamma^3(k_z-\sfrac{i}{2}\partial_z) - \gamma^0 {\hat m}_\R
 - i\gamma^0 \gamma^5 {\hat m}_\I \Big) S_{\PA}^\l = 0,
\label{eq:SCdbf}
\end{equation}
where $\hat m_{\L,\R} = m_{\L,\R}(z+\sfrac{i}{2}\partial_{k_z})$. The operator acting on 
$S^\l_{\PA s}$ in~\cref{eq:SCdbf} commutes with the operator $S_z=\gamma^0\gamma^3\gamma^5$, implying that the spin along the $z$-direction is conserved in boosted frame. We can use this to break $S^\l_\PA$ into components obeying $P_{s}S^\l_\PA P_{s'} = S^\l_{\PA s}\delta_{ss'}$, where  $P_{\PA s} = \sfrac{1}{2}(1 + sS_z)$. It is now clear that $S^\l$ has only eight independent components and it can be parametrised as
\begin{equation}
iS^\l_\PA \;=\; \frac{1}{2}\sum_s\big( \tilde g^\l_{0s}\gamma^0 + s\gamma^3 \tilde g^\l_{3s} + \tilde g^\l_{1s} + i \gamma^5 \tilde g^\l_{2s} \big)P_{\PA s}.
\label{eq:SlessPAs}
\end{equation}
The wall frame functions can now be obtained by inverse boost:
\begin{equation}
iS^\l_{w} = L_\PA^{-1} iS^\l_\PA  L_\PA = \frac{1}{2}\sum_s\big( \dag{u}_\PA \tilde g_{0s}^< + s\gamma^3 \tilde g^\l_{3s} + \tilde g^\l_{1s} + i \gamma^5 \tilde g^\l_{2s} \big)P_{ws},
\label{es:Slesswall}
\end{equation}
where $u_\PA^\mu \equiv \gamma_\PA(1,\evec{v}_\PA) = \gamma_\PA(1,\evec{k}_\PA/k_0)$. 
Note that $S^\l_w=\sum_sS^\l_{ws}$ is still block diagonal in the $z$-spin, obeying $P_{ws}S^\l_wP_{ws'} = S^\l_{ws}\delta_{ss'}$, where the boosted spin operator is\footnote{A slightly different, but equivalent form was given in~\cite{Kainulainen:2002th}, using $\dag{u}_\PA\gamma^3\gamma^5 = \gamma_\PA[S_z - i(\evec{v}_\PA \times \evec{\alpha})_z]$.}:
\begin{equation}
P_{ws} = L_\PA^{-1} P_{\PA s} L_\PA = \sfrac{1}{2}\big( 1 + s \dag{u}_\PA\gamma^3\gamma^5 \big).
\label{eq:Pws}
\end{equation}
Finally, the $s$-spin projected wall frame correlation functions $S^\l_{ws}$ then satisfy the equation
\begin{equation}
\Big(k_0 - \evec{\alpha}\cdot\evec{k}_\PA - \alpha^3(k_z-\sfrac{i}{2}\partial_z) - \gamma^0 {\hat m}_\R - i\gamma^0 \gamma^5 {\hat m}_\I \Big) S_{ws}^\l = 0,
\label{eq:SCwf}
\end{equation}
where $\hat m_{\L,\R}$ are the same as in equation~\cref{eq:SCdbf}. Note that $\gamma^3 P_{ws} = s \dag{u}_{\PA} \gamma^5P_{ws}$, showing that one could set $\gamma^3\rightarrow s \dag{u}_{\PA} \gamma^5$ in the wall frame.

%%%%%%%%%%%%%%%%%%%%%%%%%%%%%%%%%%%%%%%%%%%%%%%%%%%%%%%%%%%%%%%%%%%%%%%%%%%%%%%%%%%%%%%%%%%%%%
%
\paragraph{Gradient expansion} 
\label{sec:grad_expansion}
%
%%%%%%%%%%%%%%%%%%%%%%%%%%%%%%%%%%%%%%%%%%%%%%%%%%%%%%%%%%%%%%%%%%%%%%%%%%%%%%%%%%%%%%%%%%%%%%

Equation~\cref{eq:SCwf} is a full quantum equation with a complete spinor structure. We want to reduce it into a semiclassical equation for a scalar particle density. To this end we introduce a decomposition:
\begin{equation}
\smash{
\bar S^\l_{w} \equiv iS^\l_{w}\gamma^0 \equiv \sum_s\sfrac{1}{4}g^\sl_{ab} (\rho^a \otimes \sigma^b)},
\label{eq:decomposition}
\end{equation}
where $\rho^a$ and $\sigma^b$ are Pauli matrices, which carry chiral and spin indices, respectively, in the Weyl basis. The components $g^\sl_{ab}$ are simply related to $\tilde g^\l_{sa}$ through equation~\cref{es:Slesswall}. Using the form~\cref{eq:decomposition}, taking trace of equation~\cref{eq:SCwf} and separating the real and imaginary parts one finds
(sum over $i$ is implied):
\begin{align}
k_0 g^\sl_{00} + k_i g^\sl_{3i} - \cos\diamond (m_R g^\sl_{10}) + \cos\diamond (m_I g^\sl_{20}) &= 0
\label{eq:wfconstr}
\\
\sfrac{1}{2}\partial_z g^\sl_{33} + \sin\diamond (m_R g^\sl_{10}) - \sin\diamond (m_I g^\sl_{20}) &= 0,
\label{eq:wfevoeq}
\end{align}
where the action of the Moyal operator is\footnote{This appears as only ``half'' of the full Moyal operator~\cite{Prokopec:2003pj}, because the mass operator is $k_z$-independent.} 
\begin{equation}
\diamond^n(mg) \equiv \frac{1}{2^n}(\partial_z^n m) (\partial_{k_z}^ng).
\label{eq:diamond}
\end{equation}
To proceed we need to solve the components $g^\sl_{3i}$, $g^\sl_{10}$ and $g^\sl_{20}$ in terms of $g^\sl_{00}$, using other projections of the original matrix equation. This can be done directly in the wall frame, or by going to the doubly boosted frame, solving $\tilde g^\l_{1s,2s,3s}$ in terms of $\tilde g^\l_{0s}$ from~\cref{eq:SCdbf,eq:SlessPAs} and translating results to the wall frame by use of~\cref{es:Slesswall}; in particular $g^\sl_{00}=\gamma_\PA\tilde g^\l_{s0}$, $g^\sl_{33}=s\tilde g^\l_{s3}$  and $g^\sl_{i0}=\tilde g^\l_{si}$, for $i=1,2$. The procedure is similar to one in refs.~\cite{Kainulainen:2001cn,Kainulainen:2002th} and we just quote the relevant results in the wall frame variables\footnote{The formulae~\cref{eq:wfconstr,eq:wfevoeq,eq:gsconnections} were given only to the second order in gradients in~\cite{Kainulainen:2002th}, but the extension to arbitrary order is obvious. Note that our definition of $\bar S^\l$ is different from~\cite{Kainulainen:2002th}, whereby our functions $g^s_{2,3}$ have opposite signs compared to those of~\cite{Kainulainen:2002th}. This results in a number of sign differences in the intermediate results.}:
\begin{subequations}
\label{eq:gsconnections}
\begin{align}
k_0 g^\sl_{10} &= \phantom{+} 
\sfrac{s}{2}\gamma_\PA\partial_z g_{20}^\sl + \cos \diamond (m_\R g_{00}^\sl) 
+ s\gamma_\PA \sin\diamond (m_\I g_{33}^\sl) 
\\*
k_0 g^\sl_{20} &= -
 \sfrac{s}{2}\gamma_\PA\partial_z g_{10}^\sl - \cos \diamond (m_\I\, g_{00}^\sl) 
 + s\gamma_\PA \sin\diamond (m_\R g_{33}^\sl)
\\*
k_0g_{33}^\sl &=
-k_zg_{00}^\sl - s\gamma_\PA \sin\diamond (m_\I g_{10}^\sl) - s\gamma_\PA \sin\diamond (m_\R g_{20}^\sl).
\label{eq:g33-constraint}
\end{align}
\end{subequations}
In addition we have the exact relation $g^\sl_{3i} = -(k_i/k_0)g^\sl_{00}$ for $i=1,2$. So far no approximations have been made. Next one expands equations~\cref{eq:wfconstr,eq:wfevoeq} in gradients and uses equations~\cref{eq:gsconnections} iteratively to express $g^\sl_{i\alpha}$ in terms of $g^\sl_{00}$ to a given order in gradients. Working to the lowest nontrivial order this procedure leads~\cite{Kainulainen:2002th} to the constraint equation
\begin{equation}
\Omega^2_sg^\sl_{00} \equiv \Big(k_0^2 - \omega_{0}^2  + s \gamma_\PA\frac{|m|^2\theta'}{k_0}\Big) g^\sl_{00} = 0,
\label{eq:wfconstrb}
\end{equation}
where $\theta$ is the phase of the mass term $m = |m|e^{i\theta}$, and the evolution equation:
\begin{equation}
{\cal D}_sg^\sl_{00} \equiv \frac{k_z}{k_0} \partial_z g^\sl_{00} + \Big( -\frac{|m|^{2\prime}}{2k_0} 
+ s \gamma_\PA\frac{(|m|^2\theta')'}{2k_0^2} \Big) \partial_{k_z} g^\sl_{00}  = 0.
\label{eq:wfevoeqb}
\end{equation}
where $\omega_0^2 = \evec{k}^2 + |m|^2$. The structure of these equations implies that $g^{s\l}_{ab}$ are generalised functions, which have to be defined and analysed with some care.

%%%%%%%%%%%%%%%%%%%%%%%%%%%%%%%%%%%%%%%%%%%%%%%%%%%%%%%%%%%%%%%%%%%%%%%%%%%%%%%%%%%%%%%%%%%%%%
%
\paragraph{Spectral solutions} 
\label{sec:specsol}
%
%%%%%%%%%%%%%%%%%%%%%%%%%%%%%%%%%%%%%%%%%%%%%%%%%%%%%%%%%%%%%%%%%%%%%%%%%%%%%%%%%%%%%%%%%%%%%%

Equation~\cref{eq:wfconstrb} has a spectral solution $g^\sl_{00} \sim f^\l_s\delta (\Omega^2_s)$, and since ${\cal D}_s \Omega^2_s =0$, the coefficient functions $f^\l_s$ satisfy a differential equation ${\cal D}_{s\pm} f^\l_{s\pm} = 0$, where ${\cal D}_{s\pm}$ is projected onto the energy shell given by the dispersion relation:
\begin{equation}
\Omega_s^2 = 0 
\; \Rightarrow \; \omega_{s\pm} \approx \omega_{0} \mp s \gamma_\PA\frac{|m|^2\theta'}{2\omega_{0}^2}.
\label{eq:scdr}
\end{equation}
The normalisation of the coefficient functions $f^\l_{s\pm}$ is set by the thermal limit, where the KMS-condition imposes $S^\l_{\rm th} = -2if^\l_{\rm th}(k_0){\cal A}$, with $f^<_{\rm th}(k_0) \equiv f_\FD(p_0)$, where $p_0 = \gamma_w(k_0+v_wk_z)$ and $f_\FD$ is the thermal Fermi-Dirac distribution $f_\FD(p_0) = (e^{\beta p_0}+1)^{-1}$. 

The spectral function is defined as ${\cal A}_s =(i/2)(S^r_s-S^a_s)$, where retarded and advanced functions $\bar S^{r,a} = iS^{r,a}\gamma^0$, projected to a given spins, $s$ satisfy
\begin{equation}
\Big(k_0 \pm i\eta - \evec{\alpha}\cdot\evec{k}_\PA - s\gamma^5(k_z-\sfrac{i}{2}\partial_z) - \gamma^0 {\hat m}_\R
 - i\gamma^0 \gamma^5 {\hat m}_\I \Big) \bar S_{ws}^{r,a} = i\gamma^0P_{ws}\gamma^0.
\label{eq:SCwfra}
\end{equation}
Here $\eta$ is an infinitesimal number and positive (negative) sign refers to retarded (advanced) function. Introducing a similar decomposition to~\cref{eq:decomposition} for $\bar S_{ws}^{r,a}$ and taking the trace, we find:
\begin{equation}
(k_0 \pm i\eta )g^{r,a}_{00,s} + k_i g^{r,a}_{3i,s} - \cos\diamond (m_R g^{r,a}_{10,s}) + \cos\diamond (m_I g^{r,a}_{20,s}) = 2i.
\label{eq:polecomponent00}
\end{equation}
Equation~\cref{eq:polecomponent00} is in fact the analytic continuation of the real part of the trace equation to complex frequency, which provides the retarded and advanced functions with correct boundary conditions. Other equations for $g^{r,a}_{ab,s}$ are identical to~\cref{eq:gsconnections}, except for the analytic continuation $k_0 \rightarrow k_0 \pm i\eta$. Computing again to the leading nontrivial order in gradients, we get
\begin{equation}
g_{00,s}^{r,a} = \frac{2ik_0}{\Omega^2_s(k_0 \pm i\eta)}.
\label{eq:specfungra}
\end{equation}
The corresponding 00-component of the spectral function ${\cal A}_s\gamma^0 \equiv \sfrac{1}{4}a^s_{ab} (\rho^a \otimes \sigma^b$) is
\begin{equation}
a_{00}^s = 2\pi |k_0| \delta(\Omega_s^2)
\,=\,\pi\sum_{\pm} Z_{s\pm} \delta(k_0 \mp \omega_{s\pm}),
\label{eq:specfuna0}
\end{equation}
\vskip-0.2cm\noindent where the wave function renormalisation factor is\footnote{The spectral function ${\cal A}$ must obey the spectral sum rule $\sfrac{1}{\pi}\int {\rm d}k_0{\cal A} = \gamma^0$. Indeed, it is easy to see that $\sfrac{1}{\pi}\int {\rm d}k_0 {\rm Tr}[{\cal A}\gamma^0] = \sum_{s\pm}Z_{s\pm} =4$. Other components $a^s_{ab}$ are related to $a^s_{00}$ as in the case of $g^\sl_{ab}$ and one can readily show that the $k_0$-integrals over them vanish to the order we are working.}
\begin{equation}
Z_{s\pm} \approx 1 \pm s\frac{|m|^2\theta'}{2 \tilde \omega_0^3}.
\label{eq:wkbZs}
\end{equation}
The correctly normalised 00-component of the Wightmann function then is:
\begin{equation}
g_{00}^\sl = 4\pi\sum_{\pm} f^\l_{s\pm} \theta(\pm k_0) |k_0|\delta(\Omega^2_s) 
\, \approx \, 2\pi\sum_{\pm} Z_{s\pm} f^\l_{s\pm} \delta(k_0 \mp \omega_{s\pm}),
\label{eq:specfunf}
\end{equation}
\vskip-0.2truecm \noindent
where $f^\l_{s\pm} \rightarrow f_\FD(\pm\gamma_w(\omega_{s\pm}+v_wk_z))$ in thermal limit. Note that the full thermal correlation function $S^\l_{\rm th}= -2if^\l_{\rm th}(k_0){\cal A}$ contains gradient corrections beyond the the zeroth order thermal propagators~\cref{eq:eqsless} and \cref{eq:eqslessmass}, through non-trivial relations between the different components of the spectral function ${\cal A}$. 

%%%%%%%%%%%%%%%%%%%%%%%%%%%%%%%%%%%%%%%%%%%%%%%%%%%%%%%%%%%%%%%%%%%%%%%%%%%%%%%%%%%%%%%%%%%%%%
%
\paragraph{Collision terms} 
\label{sec:collisiontermsone}
%
%%%%%%%%%%%%%%%%%%%%%%%%%%%%%%%%%%%%%%%%%%%%%%%%%%%%%%%%%%%%%%%%%%%%%%%%%%%%%%%%%%%%%%%%%%%%%%

The spectral solution~\cref{eq:specfunf} allows a quasiparticle description of the fermions interacting with the wall. In particular it allows one to compute also the collision integrals explicitly. Collision terms can in fact be included into equations~\cref{eq:wfconstr,eq:wfevoeq,eq:gsconnections} from the outset, without spoiling the WKB-picture: treating the leading collision terms as being of order $\sim \mu \sim \partial_z^2$, one can show that the dispersion relation~\cref{eq:wfconstrb} still holds, as does the equation~\cref{eq:wfevoeqb} apart from emergence of the collision term with the expected from on the {\em r.h.s.} (for more details see appendix~\cref{sec:appendixA}):
\begin{equation}
{\cal D}_s g_{00}^\sl = {\mathrm e}^{\partial_z^\Sigma\partial_{k_z}}{\rm Tr}\big[ (\hat \Sigma^\g S^\l_w - \hat\Sigma^\l S^\g_w)P_{ws} \big] \equiv C_{\mathbb 1}^\sl[f].
\label{eq:Coll-term1}
\end{equation}
where $\hat \Sigma \equiv \Sigma_{\rm out}(z,k-\sfrac{i}{2}\partial_z)$. The precise form of the self-energy functions $\Sigma^{\l,\g}$ depends on the problem. To the lowest order in gradients the collision term becomes: 
\begin{equation}
C_{\mathbb 1}^\sl[f] \approx {\rm Tr}\big[ (\Sigma^\g S^\l_w - \Sigma^\l S^\g_w)P_{ws} \big] .
\label{eq:Coll-term2}
\end{equation}
Given the connection formulae between the $g_{ab}^\sl$ and $g_{00}^\sl$, the collision integral can always be reduced to a functional of $g_{00}^\sl$ only. Then, the spectral form~\cref{eq:specfunf} allows performing all frequency integrals, which further reduces the collision term~\cref{eq:Coll-term2} into a functional of the generalised Boltzmann distributions $f^\l_{s\pm}$. This procedure is a straightforward generalisation of the usual Boltzmann theory to the case of WKB-quasistates.

%%%%%%%%%%%%%%%%%%%%%%%%%%%%%%%%%%%%%%%%%%%%%%%%%%%%%%%%%%%%%%%%%%%%%%%%%%%%%%%%%%%%%%%%%%%%%%
%
\paragraph{Helicity eigenbasis} 
%
%%%%%%%%%%%%%%%%%%%%%%%%%%%%%%%%%%%%%%%%%%%%%%%%%%%%%%%%%%%%%%%%%%%%%%%%%%%%%%%%%%%%%%%%%%%%%%

So far we have labeled our states by their spin in the $z$-direction in the doubly boosted frame. It is also possible to work in the helicity basis, which is more directly related to chirality. Helicity spinors are the eigenspinors of the operator $\hat{\evec{h}}\equiv \hat{\evec{k}}\cdot\evec{\alpha}\gamma^5$. In the doubly boosted frame the helicity $h_\PA$ and spin $s$ are then simply related:
\begin{equation}
s = h_\PA {\rm sgn}(k_z).
\end{equation}
Going from $s$ to $h_\PA$ eigenstates is just a matter of relabelling in equations~\cref{eq:wfconstrb,eq:scdr,eq:wfevoeqb}. A little more thought is needed to extend the formulae for the {\em wall frame} helicity states as this needs a statistical interpretation. Indeed, in the (semi)classical picture it is consistent to compute the force acting on a helicity state $h$ as the sum of forces acting on the projections of the $h$-state onto $s$-states. This corresponds to setting~\cite{Cline:2017qpe}
\begin{equation}
s \; \rightarrow\;  s_h \; \equiv \; 
\langle k,h | \gamma_\PA \big(S_z - i(\evec{v}_\PA \times \evec{\alpha})_z\big)|k,h\rangle 
\; =\;  h \gamma_\PA \frac{k_z}{|\evec{k}|} 
\; \equiv \; h s_{\rm k}.
\label{eq:helicity_correspondence}
\end{equation}
Because $s_h$ is already multiplying a gradient correction term, it was sufficient to compute the projection using the lowest order adiabatic helicity eigenstates. From the quantum point of view the helicity states behave {\em on average}, in the sense of a statistical ensemble, as if they were subject to a force and a dispersion relation where $s\rightarrow hs_{\rm k}$. 
The $s$-eigenstates are very close to the $h$ eigenstates and the relation becomes exact in the massless limit. In what follows we shall label the states by $h$ and include the factor $s_{\rm k}$ explicitly. However, going from $h$ basis to the $s$-basis is a simple matter of resetting $s_{\rm k}={\rm sgn}(k_z)$.

%%%%%%%%%%%%%%%%%%%%%%%%%%%%%%%%%%%%%%%%%%%%%%%%%%%%%%%%%%%%%%%%%%%%%%%%%%%%%%%%%%%%%%%%%%%%%%
%
\subsection{Semiclassical Boltzmann equation} 
\label{sec:SCBE}
%
%%%%%%%%%%%%%%%%%%%%%%%%%%%%%%%%%%%%%%%%%%%%%%%%%%%%%%%%%%%%%%%%%%%%%%%%%%%%%%%%%%%%%%%%%%%%%%

We are now ready to put everything together. Integrating~\cref{eq:wfevoeqb} over the frequency with the spectral form~\cref{eq:specfunf} and using the helicity basis, one gets the semiclassical Boltzmann equation for the WKB-quasiparticle distribution functions $f_{h\pm}$:
\begin{equation}
v_{h\pm}\partial_z f_{h\pm} + F_{h\pm} \partial_{k_z} f_{h\pm} = {\cal C}_{h\pm}[f],
\label{eq:bolzmann_equation}
\end{equation}
where $v_{h\pm}$ is a velocity factor in the $z$-direction and $F_{h\pm}$ is the related semiclassical force term~\cite{Cline:1997vk,Cline:2000nw,Kainulainen:2002th,Cline:2017qpe,Cline:2020jre}:
\begin{equation}
  v_{h\pm} = \frac{k_z}{\omega_{h\pm}} \qquad {\rm and} \qquad
  F_{h\pm} = -\frac{|m|^{2\prime}}{2\omega_{h\pm}} \pm h s_k \gamma_\PA {(|m|^2\theta')'\over 2 \omega_0^2}.
\label{eq:vgF_2b}
\end{equation}
with
$\omega_{h\pm} \approx \omega_{0} \mp h s_{\rm k} \gamma_\PA|m|^2\theta'/(2\omega_{0}^2)$
and finally  
\begin{equation}
{\cal C}_{h\pm}[f] \equiv \pm \frac{1}{Z_{h\pm}}\int \frac{{\rm d}k_0}{2\pi}\rho_{h\pm}(k_0)C^{h}_{\mathbb 1}[f].
\label{eq:collterm}
\end{equation}
where $\rho_{h\pm}(k_0)$ is a weight function which singles out a given frequency and helicity solution  (such as a narrow top-hat distribution with $\rho_{h\pm}(\omega_{h\pm}) = 1$). This notation formalises the usual on-shell projection in the spectral limit.

Contrary to the common identification in the SC-literature, $v_{h\pm}$ is {\em not} the group velocity. Instead, it is easy to show that $v_{gh\pm} \equiv \partial_{k_z} \omega_{h\pm} = Z_{h\pm}v_{h\pm}$ and similarly, the actual semiclassical force is $-\partial_{z} \omega_{h\pm} = Z_{h\pm}F_{h\pm}$, consistent with canonical equations. However, it is perfectly consistent (and practical) to divide both sides of the projected equation~\cref{eq:wfevoeqb} by a common factor $Z_{h\pm}$, which leads to the standard SC-equations~\cref{eq:vgF_2b}. The explicit $Z_{h\pm}^{-1}$~factor in the collision term~\cref{eq:collterm} is usually omitted, but this is consistent to the order we are working: since $C^{h}_{00}[f]$ vanishes in thermal equilibrium, it must be proportional to the perturbation generated by the source and corrections from the $Z_{h\pm}^{-1}$-factor are thus of higher order in gradients. Curiously this factor also gets cancelled by another $Z_{h\pm}$-factor contained in $C^{h\pm}_{00}[f]$, as we shall see later in sections~\cref{sec:SC_collisional_damping} and~\cref{sec:SC_collisional}.

Equation~\cref{eq:bolzmann_equation} is not very useful as such. One has to first
separate the non-trivial equilibrium part of the distribution from the out-of-equilibrium perturbation. It is easy to see that $v_{s\pm}\partial_z \omega_{s\pm} + F_{s\pm} \partial_{k_z}\omega_{s\pm} = 0$, which then implies $v_{s\pm}\partial_z f_\FD(\pm \omega_{s\pm}) + F_{s\pm} \partial_{k_z}f_\FD(\pm\omega_{s\pm})= 0$ for the equilibrium distribution with $v_w=0$. The point of this excercise is to emphasise that a lot of the work done by the semiclassical force goes into setting up the non-trivial local equilibrium including the vacuum state. The CP-violating changes in the equilibrium state do not lead to any physical effect however, such as biasing of physical rates including the sphaleron rate. However, in the  case of a {\em moving} wall, the equilibrium distribution no longer satisfies the Liouville equation. Indeed, let us define
\begin{equation}
f_{h\pm} = f_\FD(\gamma_w(\omega_{h\pm}+v_wk_z)) + \Delta f_{h\pm}
\equiv f_\FD^{h\pm} + \Delta f_{h\pm}.
\label{eq:fplupert}
\end{equation}
Inserting~\cref{eq:fplupert} into equation~\cref{eq:bolzmann_equation} one gets the equation for the perturbation $\Delta f_{h\pm}$:
\begin{equation}
v_{h\pm}\partial_z \Delta f_{h\pm} + F_{h\pm} \partial_{k_z} \Delta f_{h\pm} = - 
v_w\gamma_w F_{h\pm} (f_\FD^{h\pm})^\prime + {\cal C}_{h\pm}[f].
\label{eq:bolzmann_equationdel}
\end{equation}
where prime acting on $f$ denotes $\partial f/\partial (\gamma_w\omega)$. Thus, the perturbation $\Delta f_{h\pm}$ around the local equilibrium is sourced by a velocity suppressed force term related to the derivative of the equilibrium distribution.

%%%%%%%%%%%%%%%%%%%%%%%%%%%%%%%%%%%%%%%%%%%%%%%%%%%%%%%%%%%%%%%%%%%%%%%%%%%%%%%%%%%%%%%%%%%%%%
%
\paragraph{CP-violating perturbation}
\label{sec:CPpert}
%
%%%%%%%%%%%%%%%%%%%%%%%%%%%%%%%%%%%%%%%%%%%%%%%%%%%%%%%%%%%%%%%%%%%%%%%%%%%%%%%%%%%%%%%%%%%%%%

The perturbation $\Delta f_{h\pm}$ contains a CP-conserving and a CP-violating part. The former is sourced by the force $F_0=-\m2p/(2\omega_0)$, which is first order in gradients, while the CP-odd force arises only at second order in gradients. The larger CP-even perturbation is mainly responsible for the friction that determines the speed and the shape of the phase transition wall (for non-runaway walls), while the CP-violating perturbation will eventually bias sphalerons to create a baryon asymmetry. We separate the two by defining:
\begin{equation}
\Delta f_{h\pm} = \Delta f \pm \Delta f_h.
\end{equation}
Only the CP-odd perturbation $\Delta f_h$ depends on helicity. One can derive equations for $\Delta f$ and $\Delta f_h$ by taking the sum  and the difference of the equation~\cref{eq:bolzmann_equationdel}. These equations mix in general, but this occurs only at the third order or higher in gradients~\cite{Fromme:2006wx,Cline:2020jre}, so we can treat the equations independently. Expanding consistently to second order in gradients, one finds
\begin{equation}
\frac{k_z}{\omega_0} \partial_z \Delta f_h - \frac{\m2p}{2\omega_0} \partial_{k_z} \Delta f_h = {\cal S}_h + {\cal C}_h[f],
\label{eq:bolzmann_equationdel2}
\end{equation}
where the collision term is ${\cal C}_h = ({\cal C}_{h+}-{\cal C}_{h-})/2$ and the CP-violating source is:
\begin{equation}
{\cal S}_h 
=-v_w \gamma_w h s_{\rm k} \,\gamma_\PA \left[ \frac{(|m|^2\theta')'}{2 \omega_0^2}f'_{0w}
   - \frac{\m2p |m|^2\theta^\prime}{4 \omega_0^4}
	\left(f'_{0w} - \gamma_w \omega_0 f''_{0w}\right) \right] ,
\label{eq:CPodd_source}
\end{equation}
where $f_{0w} \equiv f_\FD(\gamma_w(\omega_0 + v_wk_z))$. The CP-even perturbation $\Delta f$ satisfies a similar equation with the replacement ${\cal S}_h \rightarrow \sfrac{1}{2}v_w\gamma_w(|m^2|'/\omega_0)f'_{0w}$. 

%%%%%%%%%%%%%%%%%%%%%%%%%%%%%%%%%%%%%%%%%%%%%%%%%%%%%%%%%%%%%%%%%%%%%%%%%%%%%%%%%%%%%%%%%%%%%%
%
\subsection{Moment expansion}
\label{sec:moments}
%
%%%%%%%%%%%%%%%%%%%%%%%%%%%%%%%%%%%%%%%%%%%%%%%%%%%%%%%%%%%%%%%%%%%%%%%%%%%%%%%%%%%%%%%%%%%%%%

One usually solves the semiclassical equation~\cref{eq:bolzmann_equationdel} in moment expansion, by singling out the integrated perturbation as a pseudo chemical potential\footnote{Note that $\Delta f$ is a perturbation around the actual equilibrium distribution $f_\FD^{h\pm}$, which is different from the zeroth order quantity $f_{0w}$. This difference is the source for the $f''$-term in~\cref{eq:CPodd_source}. However, in the~\cref{eq:division} $f_\FD^{h\pm}\approx f_{0w}$ to the order we are working.}: 
\begin{equation}
\Delta f_{h} \equiv -\mu_h f_{0w}^{\prime} + \delta f_{h},
\label{eq:division}
\end{equation}
where $\int {\rm d}^3k \,\delta f_h \equiv 0$ and when acting on $f_{0w}$, prime denotes $\partial/\partial(\gamma_w\omega)$. Integrating~\cref{eq:bolzmann_equationdel} over the spatial momenta weighted by $(k_z/\omega_0)^n$, one obtains a set of equations for $\mu_h$ and the higher moments of the perturbation $\delta f_h$. This procedure was revisited recently in~\cite{Cline:2020jre} and we collect just the main results here. We set
\begin{equation}
\int \frac{{\rm d}^3k}{(2\pi )^3} \delta f_{h\pm} \equiv 0 \qquad {\rm and} \qquad
u_h \equiv \frac{1}{N_1}\int \frac{{\rm d}^3k}{(2\pi )^3}  \frac{k_z}{\omega_0}\delta f_{h\pm}.
\end{equation}
with $N_1 \equiv -2\pi^2\gamma_wT^2/3$. The first equation defines the chemical potential and the second the first velocity moment of the perturbation. Truncating to the two lowest moments, the SC-equations read
\begin{equation}
\left( \!\! \begin{array}{cc}-D_1 &  1\\
            -D_2 &  -v_w \end{array}\!\right)
                    \left( \begin{array}{cc} \mu_h'\\
	                 u_h' \end{array}\! \right)
+ \m2p \left( \! \begin{array}{cc}v_w\gamma_w Q_1&   0 \\
	                               v_w\gamma_w Q_2 &  \bar R \end{array}\!\right)
                    \left( \begin{array}{cc} \mu_h\\
	                 u_h \end{array}\! \right)
=                  \left(  \begin{array}{cc} {\cal S}_{h1}\\
	                 {\cal S}_{h2} \end{array}\! \right)
+                  \left(   \begin{array}{cc} \delta {\cal C}_{h1}\\
	                 \delta {\cal C}_{h2} \end{array}\! \right),
\label{eq:moment_equations}
\end{equation}
where the semiclassical source functions are
\begin{equation}
{\cal S}_{h\ell} = -v_w \gamma_w h \Big[ (|m|^2\theta^\prime)^\prime Q^{8o}_\ell - \m2p |m|^2\theta^\prime 
Q^{9o}_\ell \Big] .
\label{eq:sterms}
\end{equation}
The kinematic integral functions $D_\ell$, $Q_\ell$, $\bar R$ and $Q^{8o}_\ell$ and $Q^{9o}_\ell$ are defined in~\cite{Cline:2020jre}. Finally, computing to the lowest order in gradients (see section~\cref{sec:SC_collisional}), the collision integrals are given by~\cite{Cline:2000nw,Cline:2020jre},
\begin{align}
\delta  {\cal C}_{h1} &= K_0 \sum_i\Gamma_i \sum_j s_{ij}\frac{\mu_j}{T}\,,\nn\\
\delta  {\cal C}_{h2} &= - \Gamma_\TOT\,u - v_w \delta {\cal C}_{h1}\,.
\label{eq:collision_terms}
\end{align}
where $s_{ij} = 1$ ($-1$) for a species in the initial (final) state in the inelastic channel with the rate $\Gamma_i$, and $\Gamma_\TOT = \sum_i\Gamma_i$ is the total interaction rate, including elastic channels, and $K_0$ is another kinematic function defined in~\cite{Cline:2020jre}. These equations are valid also for the CP-even perturbations, when one replaces the source functions by $S^e_\ell = v_w \gamma_w \m2p Q^e_\ell$.

%%%%%%%%%%%%%%%%%%%%%%%%%%%%%%%%%%%%%%%%%%%%%%%%%%%%%%%%%%%%%%%%%%%%%%%%%%%%%%%%%%%%%%%%%%%%%%%
%
\section{Correct treatment of current divergence equations}
\label{VEV:correct}
%
%%%%%%%%%%%%%%%%%%%%%%%%%%%%%%%%%%%%%%%%%%%%%%%%%%%%%%%%%%%%%%%%%%%%%%%%%%%%%%%%%%%%%%%%%%%%%%

In section~\cref{sec:currents} we pointed out that the current divergence equations~\cref{eq:vec-current} and~\cref{eq:ax-current} are highly truncated equations that contain essential unconstrained degrees of freedom in the the singular mass term contributions. However, having set up the SC-formalism, we can now evaluate these terms consistently. We again introduce the Wigner-transform variables $X = \sfrac{1}{2}(x+y)$ and $r=x-y$ and expand the mass terms in the operators in the first lines of equations~\cref{eq:vec-current} and~\cref{eq:ax-current} as $m(x) = {\mathrm e}^{ \frac{r}{2}\partial_\mu} m(X)$ and $m(y) = {\mathrm e}^{-\frac{r}{2}\partial_\mu} m(X)$, where $\partial_\mu = \partial_{X^\mu}$. At the end of the calculation we can take $X \rightarrow x$. Following this procedure one finds the exact relation
\begin{align}
 \lim_{y\rightarrow x} {\rm Tr} \Big[ (m(x)- m(y))S^<(x,y)\Big] 
  &= -2\int_k  \sin\diamond {\rm Tr}\big[m(x)iS^<(k,x)]\big],
\nonumber \\
  &= \sum_s \int_k  \big[ - 2\sin\diamond (m_R g^\sl_{10}) + 2\sin\diamond (m_I g^\sl_{20}) \big]
\label{eq:current_eq-limit}
\end{align}
where we used $m(x)=m_\R(x)+im_\I(x)\gamma^5$ and introduced the the diamond operator defined in~\cref{eq:diamond}. Similarly, the mass correction in the axial current equation becomes:
\begin{equation}
 \lim_{y\rightarrow x} {\rm Tr} \Big[ \gamma^5(m(x)+m(y)) S^<(x,y) \Big] 
=  \sum_s\int_k \big[ 2 \cos\diamond (m_R g^\sl_{20}) + 2\cos\diamond (m_I g^\sl_{10}) \big].
\label{eq:current_scA_exa1}
\end{equation}
Using~\cref{eq:currents_def} for currents and noting that in the planar symmetric case $\partial_\mu j^\mu = -\partial_z g^s_{33}$ and $\partial_\mu j_5^\mu = -s\partial_z g^s_{00}$, we can write the divergence equations as
\begin{align}
  \sum_s \int_k \big[ \partial_z g^\sl_{33} + 2\sin\diamond (m_R g^\sl_{10}) 
                                      - 2\sin\diamond (m_I g^\sl_{20}) + C_{\mathbb 1}^\sl \big] &= 0
\label{eq:vector_current_eq-limit2} \\    
  \sum_s \int_k \big[ s\partial_z g^\sl_{00} + 2\gamma_\PA \cos\diamond (m_R g^\sl_{10}) 
                                      + 2\gamma_\PA \sin\diamond (m_I g^\sl_{20}) + C_{\gamma^5}^\sl\big] & = 0,\label{eq:axial_current_eq-limit2}
\end{align}
where $C_{\mathbb 1}^\sl$ is the collision term defined in~\cref{eq:Coll-term1} and $C_{\gamma^5}^\sl$ corresponds to~\cref{eq:Coll-term1} with the operator in trace multiplied by $\gamma^5$. The vector current divergence equation~\cref{eq:vector_current_eq-limit2} is obviously just the integral over the SC-equation equation of motion~\cref{eq:wfevoeq}, augmented with the collision term as shown in~\cref{eq:Coll-term1}.

Equations~\cref{eq:vector_current_eq-limit2} and \cref{eq:axial_current_eq-limit2} contain the full contribution from singular mass term to all orders in gradients, but they also contain four independent scalar functions for each momenta and spin. We have already shown that the differential operator in~\cref{eq:vector_current_eq-limit2} can be reduced to $-{\cal D}_sg^\sl_{00}$, defined in equation~\cref{eq:wfevoeqb}, in a controlled expansion in gradients. Similarly, using the constraint equations~\cref{eq:gsconnections}, and working up to second order in gradients, one can show that the differential operator in~\cref{eq:axial_current_eq-limit2} reduces to $(k_z/k_0){\cal D}_sg^\sl_{00}$. The current equations can eventually be written as
\begin{align}
  \sum_s \int_k \big[{\cal D}_sg^\sl_{00} - C_0^\sl \big] = 0 
\qquad {\rm and} \qquad
  \sum_s \int_k \frac{k_z}{k_0}\big[{\cal D}_sg^\sl_{00} - C_0^\sl \big] = 0.
\label{eq:comparison}
\end{align}
This result was already pointed out in~\cite{Kainulainen:2002th} in the collisionless limit, but it works out consistently also when the collision integral is included. We give nontrivial details of the reduction of the axial vector current with the collision term in appendix~\cref{sec:appendixA}. 

We have shown that the current divergence equations are consistent with the semiclassical equation of motion\footnote{The reduction of the axial current divergence holds at the level of unintegrated functions, and hence it in fact merely proves that the equations appearing as integrands in equations~\cref{eq:vector_current_eq-limit2} and~\cref{eq:axial_current_eq-limit2} are equivalent.}, and exactly reproduce its two lowest moment equations given in~\cref{eq:moment_equations}. So the only source terms in current equations are the semiclassical sources~\cref{eq:sterms} with $\ell = 1$ for vector and $\ell = 2$ in the axial current equation. We stress that {\em both} current divergences contain non-vanishing semiclassical sources. Yet the vector current is conserved, so using Fick's law to turn the vector current divergence equation into an evolution equation following the VIA-method, would yield {\em no} source term for vector current. This contradiction suggests that there is a subtlety in the use of Fick's law that has been overlooked in the VIA literature.

%%%%%%%%%%%%%%%%%%%%%%%%%%%%%%%%%%%%%%%%%%%%%%%%%%%%%%%%%%%%%%%%%%%%%%%%%%%%%%%%%%%%%%%%%%%%%%
%
\subsection{Diffusion equations and the Fick's law}
\label{sec:fickslaw}
%
%%%%%%%%%%%%%%%%%%%%%%%%%%%%%%%%%%%%%%%%%%%%%%%%%%%%%%%%%%%%%%%%%%%%%%%%%%%%%%%%%%%%%%%%%%%%%%

Fick's law is a phenomenological relation, which connects the diffusive flux to the rate of change of the concentration: $\evec{j} = -D\nabla n$. In the VIA-formalism one associates the current and the number density appearing in this formula with the components of  the total 4-currents\footnote{In VIA-method one considers chirality instead of helicity, which obscures the treatment further, but the main idea is the same.} $j_h^\mu = (n_{h};\evec{j}_{h})$. However, as pointed out above, when current is conserved $\partial_\mu j_h^\mu = 0$, employing Fick's law naively gives a diffusion equation with no source: $\dot n_h + D\nabla^2 n_{h} = 0$. The problem is that one is not correctly identifying the diffusion current and the associated out-of-equilibrium concentration. In reality the vector current $\evec{j}_{h\pm}$ consists of three distinct pieces: the diffusion current, advective current and a drag term due to the semiclassical force. In order to see this more clearly we rewrite the current divergence using the ansaz~\cref{eq:division}:
\begin{equation}
\partial_z j_{h\pm}^z = \partial_z \int \!\frac{{\rm d}^3k}{(2\pi)^3} v_{h\pm}\big( f_\FD^{h\pm} - \mu_h f_{0w}^\prime + \delta f_{h\pm}  \big).
\label{eq:currentdivergencediv}
\end{equation}
The first term in equation~\cref{eq:currentdivergencediv} can be rewritten as an integral over the force term, and it returns the SC source:
\begin{equation}
\partial_z j^z_{h\pm, \rm force} \equiv  \int \!\frac{{\rm d}^3k}{(2\pi)^3} v_w \gamma_w F_{h\pm} f_{0w}^\prime = -{\cal S}^{n}_{h1\pm}.
\end{equation}
The second term produces, up to negligible corrections of order $(\mu |m|^2\theta')'$, the advection current:
\begin{equation}
\partial_z j^z_{h\pm,\rm adv} \approx  -v_w \partial_z \delta n_{h\pm},
\end{equation}
Finally, the third term in~\cref{eq:currentdivergencediv} is the true diffusion current, related to the non-equilibrium concentration $\delta n_{h\pm}$, to which the Fick's law can be consistently applied:
\begin{equation}
\partial_z j^z_{h\pm,\rm diff} \equiv \partial_z \int \!\frac{{\rm d}^3k}{(2\pi)^3} v_{h\pm} \delta f_{h\pm} \stackrel{\scriptscriptstyle \rm FL}{\equiv} -D\partial_z^2\delta n_{h\pm}.
\label{eq:fick1}
\end{equation}
Note that the diffusion current coincides up to a normalisation factor with the first velocity moment in the moment expansion: $j_{h\pm,\rm diff}^z = N_1u_{h\pm}$. Taking the difference of the particle and antiparticle equations, we get a diffusion equation for the CP-violating perturbation:
\begin{equation}
- v_w \delta n'_h - D\delta n^{\prime\prime}_h = {\cal S}^{n}_{h1} + {\cal C}^{n}_{h1}.
\label{eq:diffeq}
\end{equation}
where $'=\partial_z$. Here ${\cal S}^n_{h1} = N_1 {\cal S}_{h1}$ and $\delta n_h = -N_1D_0\mu_h$, and we finally added the collision term to the current divergence equation, given by $\delta {\cal C}^n_{h1} = -N_1\delta {\cal C}_{h1}$. This shows that the current conservation equation is fully consistent with the SC-equations, with a non-trivial diffusion current and a non-vanishing source in the diffusion equation. Note however, that our phenomenological use of the Fick's law left the diffusion constant $D$ still unspecified. 

%%%%%%%%%%%%%%%%%%%%%%%%%%%%%%%%%%%%%%%%%%%%%%%%%%%%%%%%%%%%%%%%%%%%%%%%%%%%%%%%%%%%%%%%%%%%%%
%
\paragraph{Improved Fick's law}
%
%%%%%%%%%%%%%%%%%%%%%%%%%%%%%%%%%%%%%%%%%%%%%%%%%%%%%%%%%%%%%%%%%%%%%%%%%%%%%%%%%%%%%%%%%%%%%%

Equation~\cref{eq:diffeq} is actually a poor approximation to the underlying SC Boltzmann equation, in particular for small wall velocities, where $S^{n}_{h1}$ is strongly suppressed (this is due to antisymmetry of $F_s\pm$ in reflection $k_z\rightarrow -k_z$ for $v_w = 0$, whereby $S^{n}_{h1} \propto v_w^2$). The problem is in the Fick's law itself and a better diffusion approximation can be derived from the moment equations~\cref{eq:moment_equations}. First, neglecting the $\m2p$-dependent terms one can write the first moment equation as 
\begin{equation}
j^{z\prime}_{h,\rm diff} = v_w \delta n_h' + {\cal S}^n_{h1} + \delta {\cal C}^n_{h1}.
\label{eq:difffistmoment}
\end{equation}
In the same approximation the second moment equation can be shown to give a corrected Fick's law\footnote{One should not confuse the kinematic $D_\ell$-functions in the moment equations with the diffusion coefficient $D_{\rm eff}$. Note also the role of the the axial current (first moment) equation is not to provide an evolution equation, given the Fick's law, but indeed to provide the definition for the (improved) Fick's law itself.}:
\begin{equation}
j^z_{h,\rm diff} = -\frac{D_2 + v_w D_1}{D_0\Gamma_\TOT}\delta n_h' + \frac{{\cal S}^n_{h2} + v_w {\cal S}^n_{h1}}{\Gamma_\TOT}.
\label{eq:jfick}
\end{equation}
In addition to giving explicitly the diffusion coefficient appearing in the Fick's law: $D_{\rm eff} =(D_2 + v_w D_1)/D_0\Gamma_\TOT$, this equation has a source $S^n_{h2}$ which is less strongly suppressed by the wall velocity. Differentiating and inserting~\cref{eq:jfick} back to~\cref{eq:difffistmoment} one gets an improved diffusion approximation to the semiclassical Boltzmann equation:
\begin{equation}
-D_{\rm eff} \delta n_h'' - v_w \delta n_h' = {\cal S}^n_{h,\rm eff} + \delta {\cal C}^n_{h1},
\label{eq:difffistmoment2}
\end{equation}
where
\begin{equation}
{\cal S}^n_{h,\rm eff} = \frac{{\cal S}^n_{h1}}{D_0} -
\frac{{\cal S}^{n\prime}_{h2} + v_w {\cal S}^{n\prime}_{h1}}{D_0\Gamma_\TOT}.
\label{eq:difffistmoment3}
\end{equation}
Despite the additional suppression by an extra derivative, the ${\cal S}_{h2}^{n\prime}$-source is by far dominant for non-relativistic wall velocities and it used to be the only one accounted for by the SC-method, before the more complete calculation was introduced in ref.~\cite{Cline:2020jre}. We wish to stress that solving the moment equations~\cref{eq:moment_equations} directly is much more accurate than using the diffusion approximation, in particular for large wall velocities~\cite{Cline:2020jre}. We went through the exercises in this section just to show the intricacies of the use of Fick's law, and its consistency with the SC-equations.

%%%%%%%%%%%%%%%%%%%%%%%%%%%%%%%%%%%%%%%%%%%%%%%%%%%%%%%%%%%%%%%%%%%%%%%%%%%%%%%%%%%%%%%%%%%%%%
%
\section{Thermal corrections so the SC-method}
\label{sec:SC_thermal}
%
%%%%%%%%%%%%%%%%%%%%%%%%%%%%%%%%%%%%%%%%%%%%%%%%%%%%%%%%%%%%%%%%%%%%%%%%%%%%%%%%%%%%%%%%%%%%%%

The next two chapters are the most important part of this work. We will include thermal corrections to the semiclassical formalism in a series of steps: we will start with dispersive corrections, which generalise the WKB-states to thermal WKB-quasiparticles including new collective hole solutions at soft momenta $p\sim g_sT$. We then include the collisional damping and finally the finite width effects. After these generalisations the SC-formalism will contain all thermal corrections that were invoked in the calculation of the VIA-source terms. Here the corrections are included in a consistent expansion in gradients and coupling constants. Before detailed calculations, we will study the generic structure of the thermally corrected equations.

We will assume that the self-energy operators are slowly varying as a function of $z$, so that they can be considered to the lowest order in spatial gradients. However, we we wish to keep the momentum gradients of self-energy operators. This means that in equations~\cref{PoleEqMix2,StatEqMix2} we can everywhere replace
\begin{equation}
{\mathrm e}^{-\ihalf\partial_x^\sSigma \cdot \partial_k}[\Sigma_{\rm out} ({\hat K},x) S] \rightarrow \Sigma (k+\sfrac{i}{2}\partial_x) S(k,x) \equiv \hat\Sigma S.
\label{eq:replacement_rule}
\end{equation}
This is an excellent approximation in the EWBG problem, where the main source of spatial dependence in the self-energy functions is the (almost constant) temperature. Moreover, to the order we are working, even the truncated gradient expansion~\cref{eq:replacement_rule} will only be needed for the Hermitean self-energy function $\hat\Sigma_\H$. On the other hand, we will keep the full gradient expansion {\em w.r.t.}~the mass operators, which are here included in $\hat\Sigma_\H$. All other self-energy functions can be eventually computed to the lowest order in gradients. In this case the Wigner space KB-equations for pole functions~\cref{PoleEqMix2} reduce to
\begin{align}
  (\hat S_0^{-1} - \hat\Sigma_\H) S_\H &= 1 - \SigA{\cal A} 
\label{PoleEqMix2a}
\\
  (\hat S_0^{-1} - \hat\Sigma_\H) {\cal A} &= \SigA S_\H,
\label{PoleEqMix2b}
\end{align}
where $\hat S_0^{-1} = \dag{k} + \sfrac{i}{2}\deldag_x - \hat m_\R - i\gamma^5 \hat m_\I$ is the free operator appearing in equation~\cref{eq:SCexact} and we used the decompositions $S^{r,a}=S_\H \mp i{\cal A}$ and $\Sigma^{r,a}=\Sigma_\H \mp i\SigA$.  The equation~\cref{StatEqMix2} for the Wightmann function likewise becomes
\begin{equation}
(\hat S_0^{-1} - \hat\Sigma_\H) S^\l = \Sigma^\l S_\H  - i\SigA S^\l + i \Sigma^\l {\cal A}.
\label{eq:StatEqMix2b}
\end{equation}
In all these equations the Hermitean self-energy $\hat\Sigma_\H$ induces thermal dispersion relations for  quasiparticles. The $S_\H$ term in equation~\cref{eq:StatEqMix2b} and the $\SigA$-terms in the pole equations~\cref{PoleEqMix2b} account for the damping, which gives finite width to pole functions and to thermal parts of the Wightmann functions. Finally, the last two terms in~\cref{eq:StatEqMix2b} describe hard collisions between the quasiparticles. In this section we will further assume a thermal self energy function, which obeys the KMS-condition $\Sigma^\g_{\rm th} = e^{\beta p_0}\Sigma^\l_{\rm th}$. Generalisation to non-thermal self-energies will be considered in section~\cref{sec:SC_collisional}.

Let us now split the full solution $S^<$ into a thermal part and a perturbation\footnote{The division~\cref{eq:divisionofSless} is obviously connected to the division~\cref{eq:fplupert}; here it is simply made from the outset.}:
\begin{equation}
S^\l = S^\l_{\rm th} + \delta S^\l \equiv -2i f^\l_{\rm th} {\cal A} + \delta S^\l.
\label{eq:divisionofSless}
\end{equation}
where $f^\l_{\rm th}(k_0) = f_\FD(p_0)$, with $p_0 = \gamma_w(k_0+v_wk_z)$ and ${\mathcal A}$ is the solution to the SC-pole equations. This thermal ansaz makes the collision terms vanish in~\cref{eq:StatEqMix2b}, as can be seen from the KMS-property of the thermal self energy function $\Sigma^\l_{\rm th} = -2if_{\rm th}^\l\Sigma^{\rm th}_{\cal A}$. Using this form and the decomposition~\cref{eq:divisionofSless}, as well as the pole equation for the spectral function~\cref{PoleEqMix2b}, one can write~\cref{eq:StatEqMix2b}, separately for different spin states as
follows:
\begin{equation}
(\hat S_0^{-1} - \hat\Sigma_{\H,{s}}^{\rm th}) \delta S^\l_{s} = 
{\mathcal S}^s_M -i\Sigma^{\rm th}_{{\cal A},s} \delta S^\l_{s},
\label{eq:StatEqMix2d}
\end{equation}
where the matrix valued {\em source term} is defined as
\begin{align}
{\mathcal S}^s_M
   &\equiv 2i\big(\hat S_r^{-1} (f^\l_{\rm th} {\mathcal A}_s) - f^\l_{\rm th}  \hat S_r^{-1} {\mathcal A}_s\big)
\nonumber \\
   &     = -2i\big( \delta \hat m (f^\l_{\rm th} {\mathcal A}_s) + f^\l_{\rm th}  \delta \hat m {\mathcal A}_s \big)
   \; \approx \; (\partial_{k_z}f^\l_{\rm th}) (m_\R^\prime + i\gamma^5 m_\I^\prime) {\mathcal A}_s,
\label{eq:sourceform}
\end{align}
where $\hat S_r^{-1} \equiv \hat S_0^{-1} - \hat\Sigma_{\rm H}$ and $\delta \hat m = \hat m- m$. To get the second line we used the fact that $f^\l_{\rm th}$ does not depend on $z$ and then at last expanded $\delta \hat m$ to first order in gradients. We will see that~\cref{eq:sourceform} generalises the source appearing in~\cref{eq:bolzmann_equationdel} to include thermal spectral corrections and finite width effects.
Note that perturbations $\delta S_s$ are only indirectly affected by finite width, or the {\em coherence damping}, through the source term, but they are directly subject to {\em collisional damping} in the dynamical evolution. The two damping effects are of course fundamentally related (see equations~\cref{eq:colltermD} and~\cref{eq:general_dampingrate} below) and their apparently distinct roles  here arise but from our chosen strategy to solve the full equations.

Equation~\cref{eq:StatEqMix2d} is our master equation for the thermal WKB-quasiparticles for the remainder of this section. We stress that the only approximations made in~\cref{eq:StatEqMix2d,eq:sourceform} are the assumption that the self-energy obeys the KMS-relation and that it is treated to the lowest order in spatial gradients. In particular we stress again that coherence damping affects the perturbations $\delta S^\l_{s}$ only through the dependence of the source on the width of the spectral function ${\cal A}_{s}$.

%%%%%%%%%%%%%%%%%%%%%%%%%%%%%%%%%%%%%%%%%%%%%%%%%%%%%%%%%%%%%%%%%%%%%%%%%%%%%%%%%%%%%%%%%%%%%%
%
\subsection{Thermal quasiparticles}
\label{sec:SC_disp}
% 
%%%%%%%%%%%%%%%%%%%%%%%%%%%%%%%%%%%%%%%%%%%%%%%%%%%%%%%%%%%%%%%%%%%%%%%%%%%%%%%%%%%%%%%%%%%%%%

\begin{figure}[t!]
\begin{center}
\includegraphics[scale=0.44]{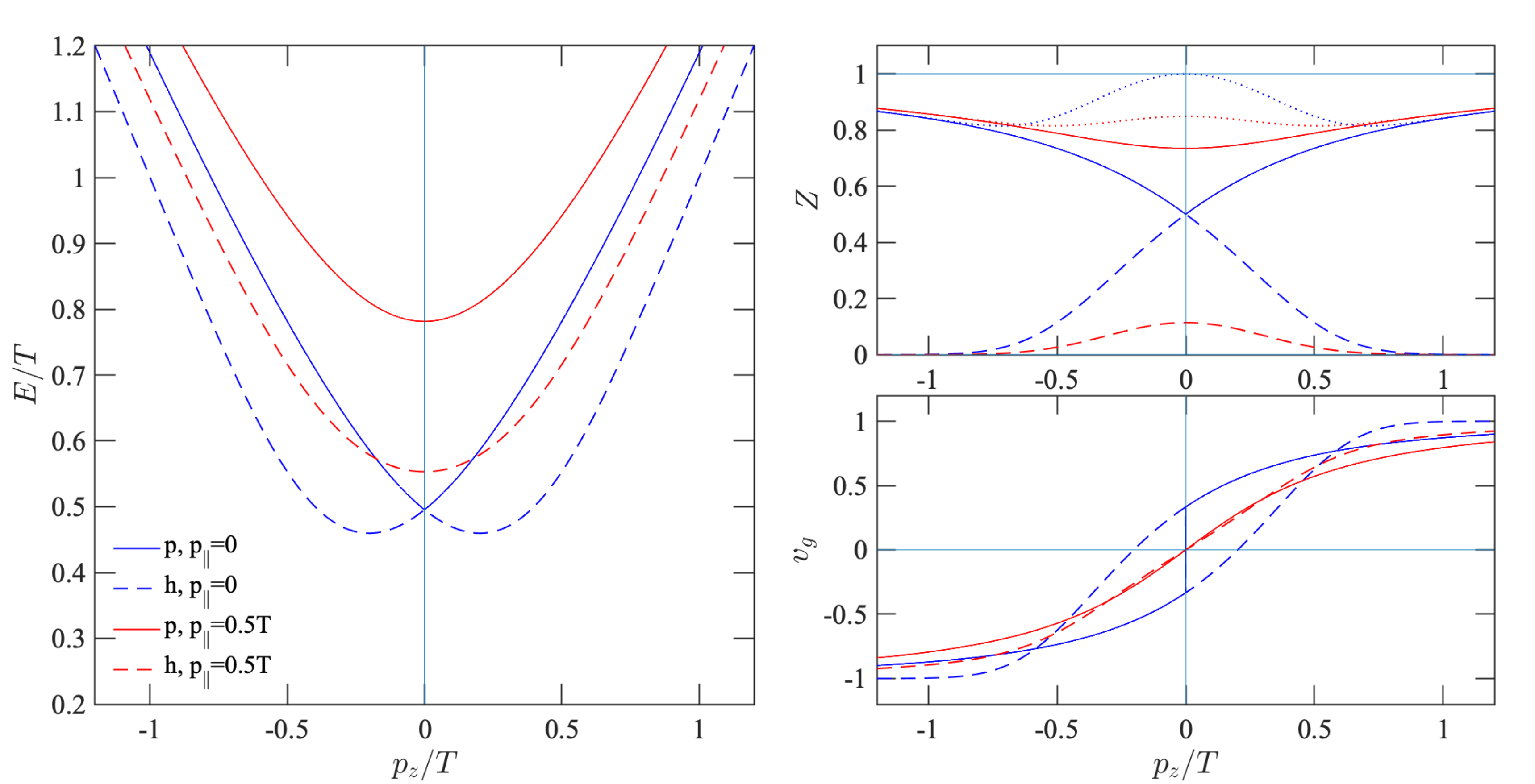}
\end{center}
\vskip-0.5truecm
\caption{Shown are the dispersion relations (left panel), wave function renormalisation factors (upper right panel) and group velocities (lower right panel) as a function of the wall frame momentum $p_z$ for $v_w=0$, $m=0$. Solid curves show the particle and dashed curves the hole solutions. Blue curves correspond to $p_\PA = 0$ and red ones to $p_\PA = 0.5T$. Dotted curves in the $Z$-plot display the sums over particle and hole wfr-factors.}
\label{fig:standardrel}
\end{figure}

We first review the usual quasiparticle picture in spatially and temporally constant system and derive the thermal quasiparticle dispersion relations and propagators in a frame moving with a velocity $v_w$ with respect to the plasma. We then extend the treatment to thermal WKB-quasiparticles in~\cref{sec:thSC_disp}. We consider only vector-like gauge interactions, for which the self energy has the generic form:
\begin{equation}
\Sigma_\H^{\rm th} = - a\dag{p} - b \dag{u} + c,
\label{eq:sigmaHgeneral}
\end{equation}
where $u^\mu$ is the plasma 4-velocity and the functions $a$, $b$ and $c$ were computed in~\cite{Weldon:1982bn}. We use the hard thermal loop (HTL) approximation for simplicity, so that $c(\op,\bp) \approx 0$ and
\begin{align}
a(\op,\bp) &= \phantom{-}\frac{{\cal M}_T^2}{\bp^2}\left(1-\frac{y}{2} \log\Big(\frac{y+1}{y-1}\Big)\right)
\nn\\
b(\op,\bp) &= \frac{{\cal M}_T^2}{\bp}\left(-y+\frac{1}{2}(y^2-1) \log\Big(\frac{y+1}{y-1}\Big)\right),
\end{align}
where $y\equiv\op/|\evec{p}|$ and $\op$ and $\evec{p}$ refer to the frequency and the 3-momentum in the plasma frame, and the thermal mass operator ${\cal M}_T^2 = \sfrac{1}{6}g_s^2T^2$ in QCD. To avoid confusion, we use the notation $p^\mu$ when referring to plasma frame and $k^\mu$ when referring to wall frame momentum. Given correction~\cref{eq:sigmaHgeneral}, the inverse propagator, computed to the lowest order in gradients, becomes:
\begin{equation}
S^{-1} = rnp_0\gamma^0 - r\evec{\gamma} \cdot \evec{p} - m_c,
\label{eq:invereseS0}
\end{equation}
where $m_c \equiv m+c$, $r \equiv 1+a$ and $rnp_0 \equiv rp_0 + b$. The propagator $S$ has two branches of poles given by:
\begin{equation}
  nr\op  = \pm \sqrt{|r{\bm p}|^2 + |m_c|^2} \quad \Rightarrow \quad \op = \opl(\bp,T).
\label{eq:QBEwith-dispersion}
\end{equation}
The positive sign corresponds to particle  and the negative sign to the hole solutions~\cite{Weldon:1989bg,Weldon:1989ys}. Both energies are positive, since $nr$ is positive in the particle branch and negative in the hole branch. Near these poles, the retarded and advanced propagators corresponding to~\cref{eq:invereseS0} can be written as
\begin{equation}
  iS^p(p_0,|\evec{p}|)\gamma^0 \,\approx\, \sum_\pm\frac{ Z_\pm P_{s_{p_0}}^\pm}{p_0 -s_{p_0}\opl + i\epsilon_p}, \;\; {\rm with} \;\; P_{p_0}^\pm = \frac{1}{2}\Big(\mathbb{1} + s_{p_0}\frac{H^\pm_{p_0}}{\opl}\Big),
\label{eq:thermalProp}
\end{equation}
where $s_{p_0} = {\rm sgn}(p_0)$ and $\epsilon_p = \pm\epsilon$ with $+$ for the retarded and $-$ for the advanced propagator and $P_{p_0}^\pm$ is the projection operator into the quasiparticle and -antiparticle states, written in terms of the effective Dirac Hamiltonian in the thermal background:
\begin{equation}
H^\pm_{p_0} \equiv \frac{1}{rn}\big( r\evec{\alpha}\cdot \evec{p} + m_c^\dagger\gamma^0 \big)\Big|_{p_0=\opl}.
\end{equation}
Finally, the thermal wave function renormalisation factors are given by
\begin{equation}
  Z_\pm^{-1} = \frac{\partial}{\partial p_0}\Big(rnp_0 \mp \sqrt{|r\evec{p}|^2 + |m_c|^2}\Big)\Big|_{|p_0| =\opl}.
\label{eq:Zpf}
\end{equation}
This approximation for propagators neglects the discontinuity, which represents inherently collective phenomena in the full spectral function. This results in the well known inequality $Z_+ + Z_- < 1$ for $|\evec {p}| \sim g_sT$, and it can be seen in upper left panel of Fig.~\cref{fig:standardrel}, which shows the wave function renormalisation factors in a specific example.
%

%==============================================================================================
%
%
%==============================================================================================

One usually displays $\opl$ as a function of $|\evec{p}|$, but for us it is more relevant to know the dispersion relations as a function of $p_z$. We show $\opl(p_z,p_\PA)$ in the left panel of figure~\cref{fig:standardrel} for different $p_\PA$ when $m=0$. Solid curves correspond to particles and dashed ones to holes. Blue curves have $p_\PA = 0$, so this case, restricted to $p_z\ge 0$ the corresponds to the usual case found in~\cite{Weldon:1989bg,Weldon:1989ys}. Curiously, the particle solutions on one side smoothly connect to the hole solutions on the other side across $p_z=0$. The same phenomenon is observed in the right panels of figure~\cref{fig:standardrel}, which shows the wave function renormalisation factors (upper panel) and the group velocities (lower panel): 
\begin{equation}
v_{g\pm} \equiv \partial_{p_z}\opl.
\label{eq:thermal_group_velocity}
\end{equation}
For $p_\PA =0$ the particle and hole group velocities remain nonzero for $p_z = 0$, while both $Z$-factors for go to 1/2. In this region of momenta the plasma is thus heavily influenced by collective effects.  
For nonzero $p_\PA$ or $|m|$ a gap is formed however, which leads to a smooth joining of the particle and hole solutions across $p_z=0$. For typical $p_\PA$ the gap is large and the $Z$-factor for particles tends to unity and that for holes tend to zero everywhere. Also, the peculiar behaviour of the group velocities is restricted to $p_\PA \ll T$ and $|m| \ll T$, as is clearly illustrated by red curves, corresponding to $p_\PA = 0.5T$ in Fig.~\cref{fig:standardrel}. 

%%%%%%%%%%%%%%%%%%%%%%%%%%%%%%%%%%%%%%%%%%%%%%%%%%%%%%%%%%%%%%%%%%%%%%%%%%%%%%%%%%%%%%%%%%%%%%
%
\paragraph{Thermal wall frame}
\label{sec:wall_frame}
%
%%%%%%%%%%%%%%%%%%%%%%%%%%%%%%%%%%%%%%%%%%%%%%%%%%%%%%%%%%%%%%%%%%%%%%%%%%%%%%%%%%%%%%%%%%%%%%

Including thermal corrections in wall frame requires some care, because the plasma breaks the Lorentz invariance. We start therefore by including thermal corrections in the wall frame to the lowest order in gradients. The scalar functions $a$, $b$ can be expressed in wall frame variables (denoted by $k^\mu$) using the Lorentz relations 
\begin{align}
p_0 &= \gamma_w(k_0 + v_wk_z)
\nn\\ 
p_z &= \gamma_w(k_z + v_wk_0) \qquad {\rm and}\qquad  p_\PA = k_\PA \,.
\label{eq:lorenz1a}
\end{align}
The get the wall frame dispersion relations it is easiest to first invert numerically the relation $k_z = \gamma_w(p_z - v_w\opl(p_z,k_\PA;T))$ to get $p_{z\pm}=p_{z\pm}(k_z,k_\PA;T)$ and then compute the wall frame energy function as $\omega_{\pm}(k_z,k_\PA;T) = \gamma_w(\opl - v_w p_{z\pm})$. 
We can also use the Lorenz-covariance to write the inverse propagator~\cref{eq:invereseS0} directly in the wall frame, where the plasma 4-velocity is $u^\mu = \gamma_w(1,-v_w\hat{\evec{u}}_z)$. The result is
\begin{equation}
S^{-1} = q_0\gamma^0 - \evec{\gamma} \cdot \evec{q} - m_c,
\label{eq:invereseS0wf}
\end{equation}
\vskip -0.2cm \noindent where $m_c = m + c$ and
\begin{equation}
q_0 \equiv r k_0 + \gamma_w b, \quad q_z \equiv rk_z - v_w\gamma_w b
\quad {\rm and} \quad \evec{q}_\PA = r\evec{k}_\PA.
\label{eq:q1}
\end{equation}
Note that $q_0$ and $q_z$ are conjugate to the plasma frame variables $nrp_0$ and $rp_z$: $nr p_0 = \gamma_w(q_0 + v_wq_z)$ and $rp_z = \gamma_w(q_z + v_wq_0)$.  It is now obvious that the wall frame dispersion relations could also be solved from equations $q_0  = \pm \sqrt{\evec{q}^2 + |m_c|^2}$. 

The thermally corrected retarded and advanced propagators can be written near their wall frame poles as
\begin{equation}
  iS_w^p(k_0,|\evec{k}|)\gamma^0 \,\approx \, \sum_\pm\frac{ Z_{w\pm} P_{w{k_0}}^\pm}{k_0 - s_{k_0}\omega_\pm + i\epsilon_p}, \;\; {\rm with} \;\;  P_{wk_0}^\pm = \frac{1}{2}\Big(\mathbb{1} + s_{k_0}\frac{H^\pm_{wk_0}}{\omega_\pm}\Big),
\label{eq:thermalPropwf}
\end{equation}
where again $s_{k_0} = {\rm sgn}(k_0)$ and the effective wall frame Dirac Hamiltonian is
\begin{equation}
  H^\pm_{wk_0} \equiv \frac{1}{rn_w}\big( \evec{\alpha}\cdot \evec{q} + m_c^\dagger \gamma^0 \big)\Big|_{|k_0|=\omega_\pm},
\label{eq:wall-frame-hamiltonian}
\end{equation}
where $n_w \equiv q_0/k_0 = 1 + \gamma_w b/(rk_0)$ and
\begin{equation}
  Z_{w\pm}^{-1} = \frac{\partial}{\partial k_0}\Big(q_0 \mp \sqrt{|\evec{q}|^2 + |m_c|^2}\Big)\Big|_{|k_0| =\omega_\pm}.
\label{eq:Zpfwf}
\end{equation}
Using~\cref{eq:thermalPropwf} one can construct the spectral function ${\cal A} = \sfrac{i}{2}(S^r-S^a)$ and finally the thermal quasiparticle Wightman-functions:
\begin{align}
  iS_{w,\rm th}^{\l,\g}  &\approx \sum_{k_0,\pm} 
  2\pi Z_{w\pm} P^\pm_{wk_0}\gamma^0 f^{\l,\g}_{\rm th} \delta(|k_0| - \omega_\pm)
\nn \\
  &= 
  2\pi {\rm sgn}(q_0)(\dag{q} + m_c^\dagger)f^{\l,\g}_{\rm th} \delta(q^2 - |m_c|^2),
\label{eq:thermalWightmannwf}
\end{align}
where $f^\l_{\rm th} = f_\FD(\gamma_w(k_0 + v_wk_z))$ is the boosted thermal distribution and $f^\g_{\rm th} = 1-f^\l_{\rm th}$. The second line is a covariant form for the thermal propagator in arbitrary Lorentz frame.

%%%%%%%%%%%%%%%%%%%%%%%%%%%%%%%%%%%%%%%%%%%%%%%%%%%%%%%%%%%%%%%%%%%%%%%%%%%%%%%%%%%%%%%%%%%%%%%
%
\subsection{Thermal WKB-quasiparticles}
\label{sec:thSC_disp}
% 
%%%%%%%%%%%%%%%%%%%%%%%%%%%%%%%%%%%%%%%%%%%%%%%%%%%%%%%%%%%%%%%%%%%%%%%%%%%%%%%%%%%%%%%%%%%%%%%

We now extend the formalism to include thermal corrections to the WKB-quasiparticles. Our treatment includes all thermal effects invoked to regulate the VIA-method, which allows us to make a definite statement of the nonexistence of the VIA-sources. We continue to work in the HTL-approximation, with the self-energy function of the generic form~\cref{eq:sigmaHgeneral} (with $c=0$). In order to get consistent picture for the thermal WKB-states including the non-trivial wave-function renormalisation factors for the WKB-hole states, we need to expand the Hermitean self-energy operator up to first order in gradients:
\begin{equation}
\hat\Sigma_\H^{\rm th}(k_0;{\bm k}_\PA,k_z -\sfrac{i}{2}\partial_z) \approx - a\dag{k} - b \dag{u} + \frac{i}{2}
\Big[ a\gamma^3 + (\partial_{k_z} a) \dag{k} + (\partial_{k_z} b) \dag{u} \Big]\partial_z.
\label{eq:sigmaHgeneral_wkb}
\end{equation}
The gradient term in~\cref{eq:sigmaHgeneral_wkb} is the lowest nontrivial correction in a resummation discussed earlier in the context of quantum transport theory for homogeneous and isotropic systems in~\cite{Herranen:2010mh,Fidler:2011yq,Herranen:2011zg}. In the fully quantum treatment this resummation must be carried out to all orders in gradients, and it is an integral part of the cQPA quantum transport formalism for leptogenesis~\cite{Jukkala:2021cys}, particle production~\cite{Jukkala:2019slc} and for reheating after inflation~\cite{Kainulainen:2021zbf}. 

Including the corrections from the self-energy operator~\cref{eq:sigmaHgeneral_wkb} we can write the collisionless wall frame equation as follows:
\begin{equation}
\Big( q_0 - \evec{\alpha} \cdot \evec{q}_\PA   - \alpha^3(q_z- \sfrac{i}{2} {\mathcal R}\,\partial_z) - \frac{i}{2}{\mathcal G} \partial_z - \hat{m}_\R\gamma^0 - i\gamma^0\gamma^5 \hat{m}_\I \Big) S_w^\l(x,k) = 0,
\label{eq:SCexactDC}
\end{equation}
where $q_0$, $q_z$ and $\evec{q}_\PA$ were defined in~\cref{eq:q1} and the gradient corrections from~\cref{eq:sigmaHgeneral_wkb} are collected in functions ${\mathcal R}$ and ${\mathcal G}$:
\begin{align}
{\mathcal R}   &= (\partial_{k_z} q_z) = r + k_z(\partial_{k_z}a) - v_w\gamma_w(\partial_{k_z}b) 
\nonumber\\
{\mathcal G} &= (\partial_{k_z} q_0) - {\bm \alpha}\cdot{\bm k}_\PA (\partial_{k_z}a).
\label{eq:grad_operators}
\end{align}
Collision terms can be included in~\cref{eq:SCexactDC} later, exactly analogously to the vacuum case. Apart from the coefficient ${\mathcal R}$ and the operator ${\mathcal G}\partial_z$ equation~\cref{eq:SCexactDC} has the same form as~\cref{eq:SCwf} and we can proceed analogously to section~\cref{sec:SC_wkp-qp}. In fact, the treatment becomes exactly the same, when we treat the operator ${\mathcal G}\partial_z$ as a perturbation. This is sensible, because we are expanding the effect of this operator to a finite order in gradients\footnote{One could include also the effect of the ${\mathcal G}\partial_z$ term in the definition of the boost by an iterative procedure, where $q_0\rightarrow \hat q_0 \rightarrow q_{0,\rm eff}$ and similarly for ${\bm q}_\PA$. Here the first arrow indicates the promotion of $q_0$ to an operator and the second arrow the fact that in the iterative procedure this operator would become a $c$-number function with an increasing complexity of gradient corrections. Thus the ${\mathcal G}\partial_z$ operator does not spoil the diagonalisability of the correlation function in $z$-spin in the boosted frame; it merely changes the identity of that frame. But this correction is easier to compute by the perturbative approach implemented here.}. Then, neglecting at first step the ${\mathcal G}\partial_z$-term, we can again define a boost operator, similar to~\cref{L-Lambda}, but replacing $k_\mu\rightarrow q_\mu$ everywhere:
\begin{equation}
 \quad  L_{q\PA}^{\pm 1} = 
   \frac{1}{\sqrt{2}} \big( \sqrt{\gamma_{q\PA} \! + \!1}  
                \mp \sqrt{\gamma_{q\PA}\!-\!1} \;\evec{\alpha}\cdot \evec{\hat q}_\PA \big),
\label{L-Lambdaq}
\end{equation}
where $\gamma_{q\PA} = q_0/\tilde q_0$ and $\tilde q_0 = {\rm sgn}(q_0)(q_0^2-q_\PA^2)^{1/2}$. This boost removes the $\evec{q}\PA\cdot \evec{\gamma}$ term from equation~\cref{eq:SCexactDC} and allows us to find the parametrisation for the wall-frame correlation function in terms of eight functions $g^\sl_{ab}$:
\begin{equation}
S^\l_{w} = \frac{1}{2}\sum_s\big( \sfrac{1}{\gamma_{q\PA}}\dag{u}_{q\PA} g_{00}^\sl + \gamma^3 g^\sl_{33} + g^\sl_{10} + i \gamma^5 g^\sl_{20} \big)P_{qws} ,
\label{es:Slesswallq}
\end{equation}
where $u_{q\PA}^\mu \equiv \gamma_{q\PA}(1,\evec{v}_{q\PA})=\gamma_{q\PA}(1,\evec{q}\PA/q_0)$ and $P_{qws} = \sfrac{1}{2}\big( 1 + s\dag{u}_{q\PA}\gamma^3\gamma^5 \big)$. These functions now obey equations similar to~\cref{eq:gsconnections}, where $k_\mu\rightarrow q_\mu$ and one includes the corrections from the ${\mathcal R}$ and the ${\mathcal G}\partial_z$ operators defined in~\cref{eq:grad_operators}. Implementing the former is a simple question of multiplication of some derivative terms with ${\mathcal R}$ and the latter are easiest to compute treating ${\mathcal G}\partial_z$ formally as an interaction term and using the technique explained in the appendix~\cref{sec:appendixA}. The final result is quite simple and we again display only the relevant equations. The basis equations for the dispersion relation~\cref{eq:wfconstr} and the evolution equation~\cref{eq:wfevoeq} now become:
\begin{align}
q_0 g^\sl_{00} + q_i g^\sl_{3i} - \cos\diamond (m_R g^\sl_{10}) + \cos\diamond (m_I g^\sl_{20}) &= 0
\label{eq:th_wfconstr}
\\
\sfrac{1}{2}\partial_z g^\sl_{33} + \sin\diamond (m_R g^\sl_{10}) - \sin\diamond (m_I g^\sl_{20}) &= -\sfrac{1}{2}\gamma_{q\PA}^{-1}(\partial_{k_z}\tilde q_0)\partial_z g^\sl_{00},
\label{eq:th_wfevoeq}
\end{align}
and the constraint equations~\cref{eq:gsconnections} generalise to:
\begin{subequations}
\label{eq:th_gsconnections}
\begin{align}
q_0 g^\sl_{10} &= \phantom{+} 
\sfrac{s}{2}\gamma_{q\PA} {\mathcal R}\,\partial_z g_{20}^\sl + \cos \diamond (m_\R g_{00}^\sl) 
+ s\gamma_{q\PA} \sin\diamond (m_\I g_{33}^\sl) 
\\*
q_0 g^\sl_{20} &= -
 \sfrac{s}{2}\gamma_{q\PA} {\mathcal R}\, \partial_z g_{10}^\sl - \cos \diamond (m_\I\, g_{00}^\sl) 
 + s\gamma_{q\PA} \sin\diamond (m_\R g_{33}^\sl)
\\*
q_0g_{33}^\sl &=
-q_zg_{00}^\sl - s\gamma_{q\PA} \sin\diamond (m_\I g_{10}^\sl) - s\gamma_{q\PA} \sin\diamond (m_\R g_{20}^\sl).
\label{eq:th_g33-constraint}
\end{align}
\end{subequations}

These equations are again solved by simple iterative procedure expanding consistently in spatial gradients. First, equation~\cref{eq:th_wfconstr} reduces to the dispersion relation:
\begin{equation}
\Omega_{qs}^2 \,\equiv\, q_0^2 - q_z^2 - q_\PA^2 - |m|^2  + s {\mathcal R}\,\gamma_{q\PA} \frac{|m|^2\theta'}{q_0} = 0.
\label{eq:wfdrq}
\end{equation}
If we neglect the $\theta'$-term, this equation reproduces the usual thermal quasiparticle equation~\cref{eq:QBEwith-dispersion}, and if we drop thermal corrections instead, it reduces to the vacuum WKB-dispersion relation~\cref{eq:wfconstrb}. Near the quasiparticle pole we can write~\cref{eq:wfdrq} as follows:
\begin{equation}
2q_{0\pm} s_\CP Z_{w\pm}^{-1}(\omega_{s\pm} - \omega_{0\pm}) + s {\mathcal R}\frac{|m|^2\theta'}{\tilde q_{0\pm}} \approx 0,
\label{eq:wfdrqb}
\end{equation}
where $\omega_{0\pm}$ the leading order thermal quasiparticle energy in the wall frame derived in section~\cref{sec:SC_disp} and $q_{0\pm} = r_\pm\omega_{0\pm} + \gamma_w b_\pm$ and $\tilde q_{0\pm} \equiv {\rm sgn}(q_{0\pm})({q_{0\pm}^2 - (r_\pm k_\PA)^2})^{1/2}$, and finally $s_\CP = 1 (-1)$ for particles (antiparticles). The dispersion relation for the thermal WKB quasiparticles then is, to the accuracy we are working, given by:
\begin{equation}
\omega_{s\pm} \approx \omega_{0\pm} - s s_\CP Z_{w\pm}{\mathcal R}\frac{|m|^2\theta'}{2(q_0\tilde q_0)_\pm}.
\label{eq:wfdrq_sol}
\end{equation}
The wave function renormalisation factor $Z_{w\pm}$ is essential in the WKB-correction. Without it the WKB-hole dispersion relation would go to the unphysical spacelike region for large momenta. Here and below we economise the notation in quantities like $\omega_{s\pm}$ by letting index $s$ to represent the product $ss_\CP$. It is slightly more involved, but still straightforward to show that the evolution equation~\cref{eq:th_wfevoeq} for $g^\sl_{00}$ reduces to:
\begin{align}
\Big[q_z {\mathcal R} - \tilde q_0(\partial_{k_z}\tilde q_0)\Big(1-s{\mathcal R}\frac{|m|^2\theta'}{2\tilde q_0^3}\Big) - s(\partial_{k_z}{\mathcal R})\frac{|m|^2\theta'}{2\tilde q_0} \Big] \partial_z \Big(\frac{g^\sl_{00}}{q_0}\Big) &
\nonumber\\
+ \Big[ -\frac{|m|^{2\prime}}{2} 
+ s{\mathcal R}\gamma_{q\PA}\frac{(|m|^2\theta')'}{2q_0} \Big] \partial_{k_z} \Big(\frac{g^\sl_{00}}{q_0}\Big)  &= 0.
\label{eq:wfevoeqc}
\end{align}
This equation may look a little cumbersome, but one should keep in mind that all terms appearing in the square bracket terms are known functions. Moreover, equation can actually be written in a very simple form
using the operator $\Omega_{qs}^2$ defined in~\cref{eq:wfdrq}:

\begin{equation}
- \frac{1}{2}\Big( [\partial_{k_z}\Omega_{qs}^2] \partial_z - [\partial_z \Omega_{qs}^2] \partial_{k_z}\Big) \big(\frac{g^\sl_{00}}{q_0}\big) = 0.
\label{eq:wfevoeqcb}
\end{equation}
It should be appreciated that without the gradient corrections in~\cref{eq:sigmaHgeneral_wkb}, equation~\cref{eq:wfevoeqc} would not be consistent with the dispersion relation~\cref{eq:wfdrq_sol} including the thermal wave-function renormalisation factor. This consistency is ensured by the correction terms $\sim (\partial_{k_z}\tilde q_0)$ and $\sim (\partial_{k_z}{\mathcal R})$ in the square bracket in the first line of the equation~\cref{eq:wfevoeqc}. Despite the notation, all $k_z$-derivatives here, in the definition of ${\mathcal R}$ and the quantities appearing in equation~\cref{eq:wfevoeqc}, must be understood as total derivatives. Deriving the SC Boltzmann equations now proceeds similarly to sections~\cref{sec:SCBE}-\cref{sec:CPpert}.

%%%%%%%%%%%%%%%%%%%%%%%%%%%%%%%%%%%%%%%%%%%%%%%%%%%%%%%%%%%%%%%%%%%%%%%%%%%%%%%%%%%%%%%%%%%%%%%%%
\paragraph{SC Boltzmann equation for thermal WKB-quasiparticles}
%%%%%%%%%%%%%%%%%%%%%%%%%%%%%%%%%%%%%%%%%%%%%%%%%%%%%%%%%%%%%%%%%%%%%%%%%%%%%%%%%%%%%%%%%%%%%%%%%

Equation~\cref{eq:wfdrq} clearly has a spectral solution $g^\sl_{00} \sim f^\l_sq_0\delta (\Omega^2_{qs})$, and since ${\cal D}_{qs} \Omega^2_{qs} = 0$ to the order we are working, the coefficient functions satisfy a differential equation ${\cal D}_{sq\pm} f^\l_{s\pm} = 0$, where the operator ${\cal D}_{s\pm}$ is projected on-shell~\cref{eq:wfdrq_sol}. As before, the normalisation of the functions $f^\l_{s\pm}$ is set by the thermal limit, which implies:
\begin{equation}
g_{00}^\sl = 4\pi\sum_{\pm,s_\CP} f^\l_{s\pm} \theta(s_\CP k_0) |q_0|\delta(\Omega^2_{qs}) 
\, \approx \, 2\pi\sum_{\pm,s_\CP} Z_{ws\pm}  f^\l_{s\pm} \delta(k_0 - s_\CP \omega_{s\pm}),
\label{eq:specfunf2}
\end{equation}
where the wall frame wave function renormalisation factor is
\begin{equation}
Z_{ws\pm}^{-1} = \Big|\frac{1}{2q_0}\frac{\partial \Omega_{qs}^2}{\partial k_0}\Big|_{k_0=s_{\rm CP} \omega_{s\pm}} \approx Z_{w\pm}^{-1} + \Big(\frac{\partial}{\partial k_0}\frac{\mathcal R}{\tilde q_0}\Big)_\pm s \frac{|m|^2\theta'}{2 q_{0\pm}},
\label{eq:Zgrand}
\end{equation}
where $Z_{w\pm}$ is given by~\cref{eq:Zpfwf}. Here, as well as in equations~\cref{eq:Zpf,eq:Zpfwf} 
$\partial_{k_0}$ is really a partial derivative by definition. This expression again obviously reduces to~\cref{eq:wkbZs} for the free WKB-states when thermal corrections are dropped, and to~\cref{eq:Zpfwf} for thermal quasistates when gradient corrections are dropped. Using~\cref{eq:specfunf2} and integrating over frequencies the equation~\cref{eq:wfevoeqc} becomes
\begin{equation}
- Z_{ws\pm}\frac{1}{2|q_{0\pm}|}\Big( [\partial_{k_z}\Omega_{qs}^2]_\pm \partial_z - [\partial_z \Omega_{qs}^2]_\pm \partial_{k_z}\Big) f_{s\pm} = 0.
\label{eq:seed-equation}
\end{equation}
Moreover, using the fact that near the quasiparticle poles $[\partial_{x}\Omega_{qs}^2]_\pm \approx 2q_{0\pm}Z_{ws\pm}^{-1}\partial_{x}\omega_{s\pm}$ we can write~\cref{eq:seed-equation} equivalently as 
\begin{equation}
\Big( (\partial_{k_z}\omega_{s\pm}) \partial_z - (\partial_z \omega_{s\pm}) \partial_{k_z} \Big) f_{s\pm} = 0.
\end{equation}
From this form it is evident that the evolution equation admits a thermal background solution with a zero wall velocity: $f_{s\pm} = f_{\rm th}(\omega_{s\pm})$, where energy is given precisely by the WKB-dispersion relation~\cref{eq:wfdrq_sol}. We can thus again define a perturbative solution around the equilibrium as
\begin{equation}
f_{s\pm} = f_{\rm th}(\omega_{s\pm}) + \Delta f_{s\pm},
\end{equation}
where we are still economising the notation, letting $s$ refer to combination $ss_\CP$.
Inserting the collision term into~\cref{eq:seed-equation}, dividing by $Z_{ws\pm}$ and then 
taking the difference between the positive and negative frequency sectors, the the semiclassical equation for the CP-violating perturbations can be written as:
\begin{equation}
\Big[ {\mathcal R}_\pm\frac{q_{z\pm}}{|q_{0\pm}|} + \frac{(\partial_{k_z} \tilde q_{0\pm}^2)}{2|q_{0\pm}|}\Big] \partial_z \Delta f_{s\pm} - \frac{\m2p}{2|q_{0\pm}|} \partial_{k_z} \Delta f_{s\pm} = {\cal S}_{qs\pm} + {\cal C}_{s\pm}[f].
\label{eq:bolzmann_equationdewKB}
\end{equation}
Here the source that includes thermal dispersive corrections at one-loop level is
\begin{equation}
{\cal S}_{qs\pm} 
= \mp v_w \gamma_w s {\mathcal R}_\pm\left[ \frac{(|m|^2\theta')'}{2 q_0\tilde q_0}f'_{0w}
    - \frac{|m|^2\m2p\theta^\prime}{4 q_0^3\tilde q_0}
	\left(f'_{0w} - \gamma_w\omega_0 f''_{0w}\right) \right]_\pm,
\label{eq:CPodd_source}
\end{equation}
with $f_{0w\pm} \equiv f_\FD(\gamma_w(\omega_{0\pm} + v_wk_z))$ and the CP-odd collision terms for particles and holes are:
\begin{equation}
{\cal C}_{s\pm}[f] \equiv \pm \sum_{s_\CP} \frac{1}{2Z_{ws\pm}}\int \frac{{\rm d}k_0}{2\pi}\rho_{s\pm}(k_0)C^{s}_{\mathbb 1}[f].
\label{eq:colltermB}
\end{equation}
where the weight function $\rho_{s\pm}(k_0)$ formalises the on-shell projection as in~\cref{eq:collterm}.
When deriving~\cref{eq:bolzmann_equationdewKB} we again treated the CP-even perturbations as being of first order in the gradient expansion, which allowed us to separate the CP-even and CP-odd sectors and drop many terms as higher order gradient corrections. Note that the sources for particles and holes have opposite signs and that the hole perturbations $\Delta f_{s-}$ are not suppressed in comparison with the particle perturbations $\Delta f_{s+}$ at the level of the semiclassical equation.

%=============================================================================================
%
\begin{figure}[t!]
\begin{center}
\includegraphics[scale=0.43]{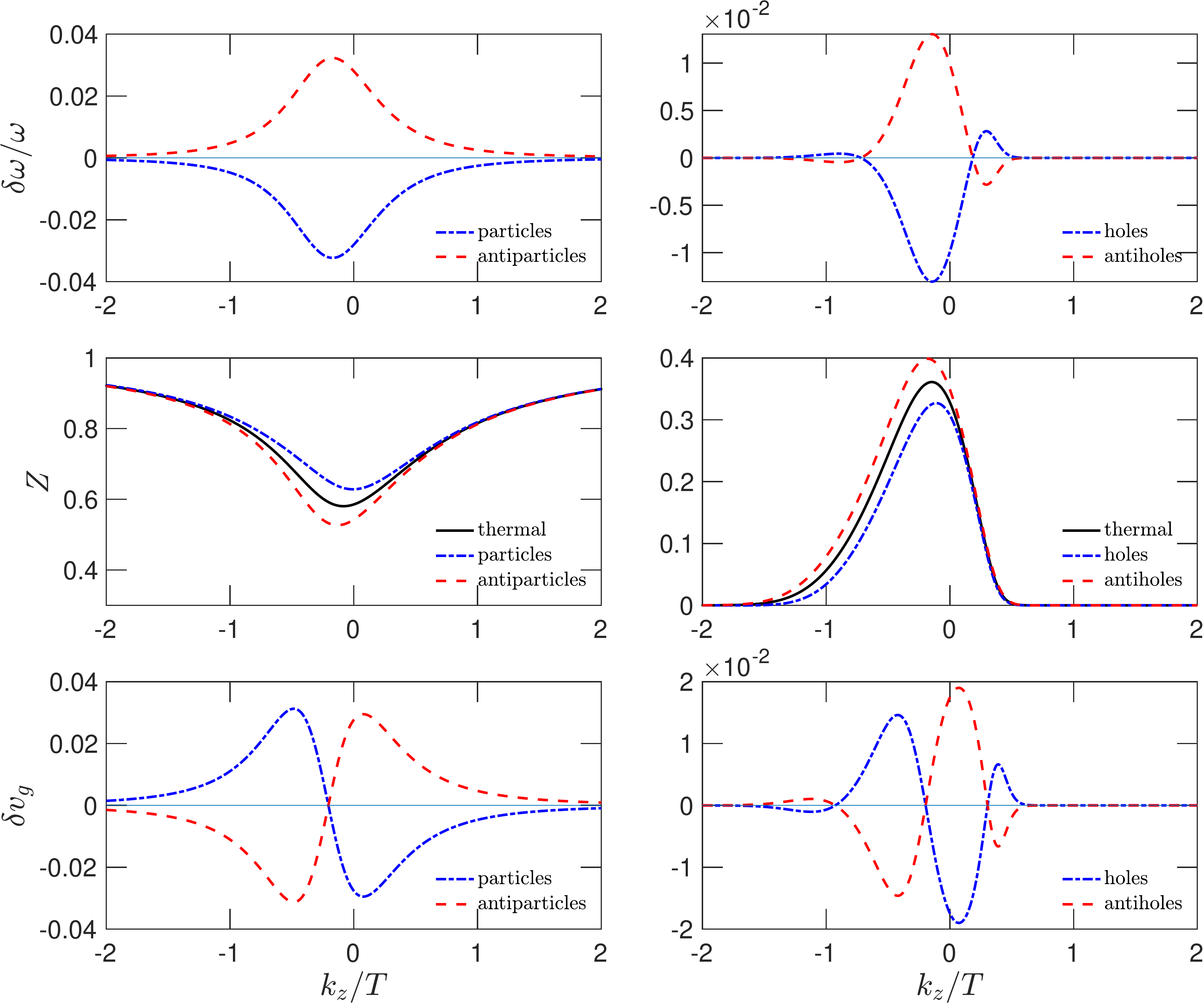} \;
\end{center}
\vskip-0.5truecm
\caption{Upper panels show the relative change in thermal WKB-quasiparticle (left) and thermal WKB-quasihole (right) dispersion relations in comparison to usual thermal quasiparticles. In all panels WKB-states correspond to blue dash-dotted lines and WKB-antistates to red dashed lines. Middle panels show the wave function renormalisation factors including also usual thermal quasistates (solid black line). Lowest panels show the difference between the usual and the thermal WKB-quasiparticle group velocities. In all cases we assumed a constant spin $s=1$, wall velocity $v_w=0.5$, the mass parameters $|m|=0.3T$ and $|m|^2\theta'=0.1T^3$ and finally vanishing parallel momentum $k_\PA = 0$.} 
\label{fig:cqpastrb}
\end{figure}

%%%%%%%%%%%%%%%%%%%%%%%%%%%%%%%%%%%%%%%%%%%%%%%%%%%%%%%%%%%%%%%%%%%%%%%%%%%%%%%%%%%%%%%%%%%%%%%%%
\paragraph{Thermal WKB helicity states.}
%%%%%%%%%%%%%%%%%%%%%%%%%%%%%%%%%%%%%%%%%%%%%%%%%%%%%%%%%%%%%%%%%%%%%%%%%%%%%%%%%%%%%%%%%%%%%%%%%

As we explained when deriving equation~\cref{eq:helicity_correspondence}, the interpretation in terms of helicity must be made in the statistical sense. In the thermal wall-frame the quasiparticle helicity eigenspinors are defined by the 4-momentum $q^\mu$, {\em i.e.}~as eigenstates of the operator $\hat{\evec{h}}_q \equiv \hat{\evec{q}}\cdot\evec{\alpha}\gamma^5$ (this is the operator that commutes with the wall-frame Hamiltonian~\cref{eq:wall-frame-hamiltonian}). The semiclassical force acting on these states can be found as before, boosting the $S_z$-operator back to the wall frame by $L^{-1}_{q\PA}$ and computing the expectation value of the boosted operator in the helicity basis. This corresponds to setting
\begin{equation}
s \; \rightarrow\; 
\langle q,h | \gamma_\PA \big(S_z - i(\evec{v}_\PA \times \evec{\alpha})_z\big)|q,h\rangle 
\; =\;  h \gamma_\PA \frac{q_z}{|\evec{q}|} 
\; \equiv \; h s_{\rm q}
\label{eq:sq}
\end{equation}
everywhere in the above formulae. This replacement rule holds true for other physical variables as well, such as the dispersion relations and the group velocities. In each case the actual derivation of the quantity must be made using the spin $s$ and only in the end one uses the statistical interpretation leading to rule~\cref{eq:sq}.

%%%%%%%%%%%%%%%%%%%%%%%%%%%%%%%%%%%%%%%%%%%%%%%%%%%%%%%%%%%%%%%%%%%%%%%%%%%%%%%%%%%%%%%%%%%%%%
\paragraph{Physical perturbations}
%%%%%%%%%%%%%%%%%%%%%%%%%%%%%%%%%%%%%%%%%%%%%%%%%%%%%%%%%%%%%%%%%%%%%%%%%%%%%%%%%%%%%%%%%%%%%%

While the hole perturbations $\Delta f_{s-}$ are not suppressed at the level of equations (because we divide the $Z_{ws\pm}$-factor out), their effect on physical quantities, such as the collision rates, is suppressed by the wave function renormalisation factor, as is evident from~\cref{eq:specfunf2}. If we use the same division as in~\cref{eq:division}: $\Delta f_{h\pm} \equiv -\mu_{h\pm} f_{0w}^\prime + \delta f_{h\pm}$, we see that in particular the CP-violating chemical potentials entering the physical reaction rates will contain the wave-function normalisation factors: $\mu_{h} \rightarrow Z_{wh\pm} \mu_{h\pm} \approx Z_{w\pm} \mu_{h\pm}$. We emphasise that while the gradient corrections in the $Z_{wh\pm}$'s can be consistently dropped in collision integrals, the thermal part must indeed be kept: only the vanishing of the hole wave-function renormalisation factor $Z_{w-}$  guarantees that holes do not contribute to physical processes at large momentum region.

In the upper panels of figure~\cref{fig:cqpastrb} we show the relative change in the thermal WKB-quasiparticle dispersion relations~\cref{eq:wfdrq_sol} compared to the usual thermal wall frame quasistates: $\delta \omega_{s\pm}/\omega \equiv \omega_{s\pm}/\omega_{0\pm}-1$, for $s=1$, $v_w=0.5$ and $k_\PA = 0$. The values for the mass parameters are given in the figure caption. Naturally in all cases the change in antiparticles is the negative of the change in the particle sector. In the middle panels we display the wave function renormalisation factors for the WKB-quasistates and for the usual thermal quasiparticles (solid black lines). Finally, in the lowest panels we show the difference in the group velocities between the usual and the thermal WKB-quasistates. This difference quantitatively displays the physical effect that gives rise to the charge separation in the semiclassical mechanism. The value of parallel momentum $k_\PA = 0$ was chosen to maximise the thermal effects visually. However, for a typical parallel momentum $k_\PA \gsim T$ the perturbations in the hole sector essentially vanish, so that in practical calculations the hole sector can be neglected to a good approximation.

%%%%%%%%%%%%%%%%%%%%%%%%%%%%%%%%%%%%%%%%%%%%%%%%%%%%%%%%%%%%%%%%%%%%%%%%%%%%%%%%%%%%%%%%%%%%%%
%
\subsection{Collisional damping by a thermal operator}
\label{sec:SC_collisional_damping}
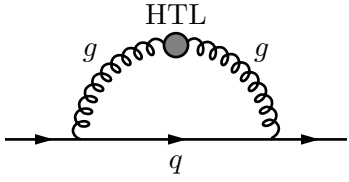
\begin{figure}

\def\mytip{
    \path [fill, scale=1, xshift=4pt] (0,0) -- ++(-7pt,2pt) -- ++(0,-4pt) -- cycle;}
\tikzset{
    myblob/.style={draw=black, circle, fill=gray, minimum size=#1, inner sep=0pt},
    myarrow/.style={decoration={markings, mark = at position #1 with {\mytip};},
        postaction={decorate}}}

\centering
\begin{tikzpicture} [line width=1.0pt]
  \def\dist{2.5cm}

  \coordinate (x);
  \coordinate [right = \dist of x]     (y);
  \coordinate [right = \dist/2 of x]   (z1);
  \coordinate [above = \dist/2 of z1]  (z);
  \coordinate [left  = \dist*0.4 of x] (in);
  \coordinate [right = \dist*0.4 of y] (out);
  % Fermion lines
  \draw [myarrow=0.5] (in) -- (x);
  \draw [myarrow=0.5] (y)  -- (out);
  \draw [myarrow=0.5] (x)  -- (y) node [midway, below=2pt] {$q$};
  % The coil 
  \draw [decorate,decoration={coil,aspect=0.7,amplitude=3pt,
                              segment length=5pt,pre length=0pt,post length=0pt}]
      (x) arc (180:90:\dist/2) node [midway, above left]  {$g$};
  \draw [decorate,decoration={coil,aspect=0.7,amplitude=3pt,
                              segment length=5pt,pre length=3.5pt,post length=0pt}]
      (z) arc (90:0:\dist/2)   node [midway, above right] {$g$};
  % Vertices 
  \node [myblob = 9pt]   at (z) {};
  \node [above  = 4pt]   at (z) {$\mathrm{HTL}$};
\end{tikzpicture}

\caption{The self-energy function that gives dominant damping for quarks.}
\label{fig:qcdloop}
\end{figure}

%
%%%%%%%%%%%%%%%%%%%%%%%%%%%%%%%%%%%%%%%%%%%%%%%%%%%%%%%%%%%%%%%%%%%%%%%%%%%%%%%%%%%%%%%%%%%%%%
In this section we consider the collisional damping of perturbations by a thermal operator. We keep also the dispersive self-energy corrections but continue to ignore the finite width effects. In this case the Kadanoff-Baym equation~\cref{eq:StatEqMix2b} becomes
\begin{equation}
\Big(q_0 - \evec{\alpha}\cdot\evec{q}_\PA - \alpha^3(q_z-\sfrac{i}{2}r\partial_z) - \gamma^0 {\hat m}_\R - i\gamma^0 \gamma^5 {\hat m}_\I \Big) S_w^< 
= -i\gamma^0\Sigma^{\rm th}_{{\cal A}} \big( S_w^< - S^<_{w,{\rm th}}).
\label{StatEqMix2e}
\end{equation}
The right hand side of~\cref{StatEqMix2e} vanishes identically in thermal equilibrium, so it is clear that a thermal self energy function cannot give rise to collisional sources for  WKB-quasiparticles. In our prototype case the leading correction to $\Sigma^{\rm th}_{{\cal A}}$ comes from the diagram shown in figure~\cref{fig:qcdloop}. For light quarks it can be written as 
\begin{equation}
\Sigma_{\cal A}^{\rm th} \;=\; - a_I\dag{k} - b_I \dag{u} \;\equiv\; -\dag{q}_I,
\label{eq:thermalAI}
\end{equation}
\vskip -0.1cm \noindent where
\begin{equation}
q_{I0} = a_Ik_0 + \gamma_wb_I, \quad
\evec{q}_{I\PA} = a_I \evec{k}_\PA
\quad {\rm and} \quad q_{Iz} = a_Ik_z -v_w\gamma_wb_I.
\label{eq:q2}
\end{equation}
We did not introduce the re-summed self-energy function as in~\cref{eq:sigmaHgeneral_wkb}, because here the derivatives are acting on a perturbation that is already of second order in gradients. Moreover, the resulting corrections would enter in equations~\cref{eq:wfconstr,eq:g33-constraint,eq:polecomponent00}, where they would be two orders down in gradient expansion from the order we are working on. Functions $a_I$ and $b_I$ were computed for example in~\cite{Thoma:1990fm}, but their specific forms are not relevant for us. Given~\cref{eq:thermalAI}, the collision integral~\cref{eq:colltermB} becomes
\begin{equation}
{\cal C}_{s\pm}[f] = \mp\sum_{s_\CP}\frac{1}{2Z_{ws\pm}}\int \frac{{\rm d}k_0}{2\pi}\rho_{s\pm}(k_0)
{\rm Tr}\big[ \Sigma_{{\cal A}}^{\rm th}\delta S^<_{w} P_{ws}\big],
\label{eq:colltermC}
\end{equation}
where $s$ again tracks the combination $ss_\CP$ and $\pm$ refers to particles and holes.
Using the form~\cref{eq:thermalAI} and the explicit expression~\cref{es:Slesswallq} for $S^\l_w$, one finds
\begin{equation}
{\rm Tr}\big[ \Sigma_{{\cal A}}^{\rm th}\delta S^<_{w}P_{ws} \big] 
\approx -\frac{q\cdot q_I}{q_0} (g^\sl_{00}-g^\sl_{00,{\rm th}}),
\label{eq:colltermC1}
\end{equation}
up to third order in gradients. When we perform the $k_0$-integration in~\cref{eq:colltermC} using expression~\cref{eq:specfunf2} for $g_{00}^\sl$, the overall $Z_{s\pm}^{-1}$ factor in~\cref{eq:colltermC} is cancelled as suggested in section~\cref{sec:SCBE} and we get:
\begin{equation}
{\cal C}_{s\pm}[f] \approx 
\mp\big(v_q \cdot q_I\big)_\pm \Delta f_{0s\pm}^\l ,
\label{eq:colltermD}
\end{equation}
where $v_q\equiv (1,\evec{q}/q_0)$ and we take $v_{q\pm}$ and $q_{I\pm}$ to correspond to the solution of~\cref{eq:wfdrq} with no gradient corrections. Indeed, all neglected terms are of higher in gradients, because $\Delta f_{0s}^\l$ are only sourced by terms of order ${\cal O}(v_w(|m|^2\theta')')$. The collision integral can then be computed by standard techniques to the zeroth order in the gradient expansion. In principle, collision integrals can create new sources due to CP-biased rate of decay of perturbations, but these effects are always of higher order in gradients. Note that despite the simple notation, equation~\cref{eq:colltermD} is a fully general collision integral for thermal WKB-quasiparticles due to a thermal self energy function.

%=============================================================================================
%
%=============================================================================================

%%%%%%%%%%%%%%%%%%%%%%%%%%%%%%%%%%%%%%%%%%%%%%%%%%%%%%%%%%%%%%%%%%%%%%%%%%%%%%%%%%%%%%%%%%%%%%
%
\subsection{Coherence damping: finite width effects}
\label{sec:SC_coherence_damping}
%
%%%%%%%%%%%%%%%%%%%%%%%%%%%%%%%%%%%%%%%%%%%%%%%%%%%%%%%%%%%%%%%%%%%%%%%%%%%%%%%%%%%%%%%%%%%%%%

As we explained in~\cref{sec:SC_thermal}, interactions affect the inhomogeneous solutions very differently from the perturbations $\delta S^\l$. Our treatment here introduces the coherence damping effects to  thermal WKB-quasistates. In practice, we need to generalise our treatment in section~\cref{sec:thSC_disp} for a {\em complex} momentum vector $Q_\mu$, defined as in~\cref{eq:q1}, but this time with $r = 1 + \hat a_R \pm ia_I$ and $b = \hat b_R \pm ib_I$, where +(-) refer to retarderd (advanced) pole function and $\hat a_R$ and $\hat b_R$ are operators that include the gradient corrections discussed in section~\cref{sec:thSC_disp},  That is, $\hat Q_\mu = \hat q_\mu + iq_{I\mu}$ in our earlier notation defined in~\cref{eq:q1} and~\cref{eq:q2}. Using the constraints~\cref{eq:th_gsconnections}, generalised for the complex $Q^\mu$, we can solve the similarly generalised constraint equation~\cref{eq:th_wfconstr} to find:
\begin{equation}
\hat Q_0 g^{r,a}_{00} + \sum_i\hat Q_i g^{r,a}_{3i} - \cos\diamond (m_R g^{r,a}_{10}) + \cos\diamond (m_I g^{r,a}_{20}) = 2i.
\label{eq:gzeroforpeqs2}
\end{equation}
Working to first order in gradients this gives:
\begin{equation}
2Q_{0\pm} \approx \Big(Q_{0\pm}^2 - \evec{Q}_\pm^2 - |m|^2 + s{\mathcal R}\frac{|m|^2\theta'}{\tilde Q_{0\pm}}\Big) g^{r,a}_{00,s} 
\equiv \Omega^2_{Q\pm}g^{r,a}_{00,s}.
\label{eq:thermalPropwf1}
\end{equation}
For simplicity we set $q_I^2 = 0$ and assume that $\evec{q}\cdot\evec{q}_I\ll q_0^2$. Then in the neighbourhood of $\Omega_{Q\pm}^2 = 0$
\begin{equation}
(Q_0^{-1}\Omega_Q^2)_{s\pm\pm^\prime} \approx \; 2\Big(Z^{-1}_{ws\pm}(k_0-s_\CP\omega_{s\pm}) \pm' i(v_q\cdot q_I)_{s\pm}\Big),
\end{equation}
where $+'$ $(-')$ refers to retarded (advanced) solution, $\pm$ refers to particles and holes and again the label $s$ tracks the product $ss_\CP$. Moreover, $\omega_{s\pm}$ is as in~\cref{eq:wfdrq} and $Z_{s\pm}$ is given by~\cref{eq:Zgrand}. We then find that near the quasi-particle poles:
\begin{equation}
g^{r,a}_{00,s\pm} \approx  
  \frac{1}{2}\, \frac{Z_{ws\pm}}{k_0 - s_\CP\omega_{s\pm} \pm' i\gamma_{s\pm}}  \,,
\label{eq:thermalPropwf2}
\end{equation}
where
\begin{equation}
\gamma_{s\pm}(k_0,k_\PA,k_z) \equiv Z_{ws\pm} \big(v_q \cdot q_I\big)_{s\pm}
\label{eq:general_dampingrate}
\end{equation}
is the thermal WKB-quasiparticle damping rate at finite temperature. We again defined $v_q \equiv (1,\evec{q}/q_0)$. To the lowest order in gradients $\gamma_{s\pm}$ coincides with the collision rate in equation~\cref{eq:colltermD} (up to the multiplicative scale factor $Z_{ws\pm}$ which was explicitly divided out there) as it should. If we further set $v_w=0$, ignore gradient corrections and take the limit $k_\PA,k_z\rightarrow 0$, where $Z_{s\pm} \rightarrow 1/2$, and $k_0\rightarrow s_\CP{\cal M}_T$ (see section~\cref{sec:SC_disp}), the rate $\gamma_{s\pm}$ reduces to the well known gauge invariant quark damping rate first computed in ref.~\cite{Braaten:1992gd}:
\begin{equation}
\gamma_{s\pm}(s_\CP{\cal M}_T,0,0) = \frac{1}{2} q_{I0} = \frac{1}{8}{\rm Tr} 
       \big[\gamma^0\Sigma_{\cal A}^{\rm th}\big] \approx 0.151g_s^2 T \;\equiv \; \gamma.
\label{eq:braaten}
\end{equation}
One often makes the approximation $\gamma_{s\pm}\rightarrow \gamma$ throughout the kinematic region, as we indeed did implicitly in section~\cref{sec:simple}, with the VIA-method. However, nothing prevents one from keeping the fully $k$-dependent damping rate here, even with the gradient corrections:
\begin{equation}
a_{00s\pm} =  \frac{1}{2}Z_{ws\pm}{\rm Re}\Big(\frac{-i}{k_0-s_\CP\omega_{s\pm} - i\gamma_{s\pm}}\Big).
\label{eq:azerozero}
\end{equation}
The $Z_{ws\pm}$-dependence of the damping term~\cref{eq:general_dampingrate} is very important the hole states. Indeed, for holes $\omega_{s-}\rightarrow |\evec{k}|$ exponentially fast for large $|\evec{k}|$~\cite{Weldon:1989ys}. However, $Z_-$ vanishes exponentially at the same time, so that the hole spectral function approaches exponentially quickly a delta-distribution. This ensures that the hole spectral function does not "leak" below the light-cone.

%%%%%%%%%%%%%%%%%%%%%%%%%%%%%%%%%%%%%%%%%%%%%%%%%%%%%%%%%%%%%%%%%%%%%%%%%%%%%%%%%%%%%%%%%%%%%%
%
\subsection{Thermal SC-source including finite width}
\label{sec:SC_finalSource}
%
%%%%%%%%%%%%%%%%%%%%%%%%%%%%%%%%%%%%%%%%%%%%%%%%%%%%%%%%%%%%%%%%%%%%%%%%%%%%%%%%%%%%%%%%%%%%%%

We are now ready to compute the SC source for the thermal WKB states in the Boltzmann equation including finite width on the spectral function. The matrix valued source term ${\mathcal S}_M$ in~\cref{eq:sourceform} again has to be run through the now familiar reduction process to get the scalar valued source in the equation for the perturbation $\delta g_{00}^\sl$. In this process, it is sufficient to use the simple first-order expanded form for ${\mathcal S}_M$, which makes calculation quite simple. The final result for the source that appears in the non-integrated (over $k_0$) for the perturbation $\delta g_{00}^s$ is
\begin{equation}
{\cal S}^\gamma_{00s}(k_0) = v_w\gamma_w (2a_{00s}) 
\Big(\frac{|m|^{2\prime}}{2q_0} 
- s{\mathcal R}\gamma_{q\PA} \frac{(|m|^2\theta')'}{2q_0^2} \Big) (f_\FD)^\prime,
\label{eq:finalsorsa3}
\end{equation}
where we used $\partial_{k_z}f_{\rm th}^\l = v_w\gamma_w f_{\rm FD}^\prime$ and prime again refers to $\partial_{\gamma_w\omega}$. If we take the limit $\gamma \rightarrow 0$ then $a_{00s}/Z_{ws}$ becomes a delta function at the quasiparticle shells. Then integrating~\cref{eq:finalsorsa3} over $k_0$, dividing with the wave-function renormalization factors and finally taking the difference of positive and negative frequency sectors would give our old source term~\cref{eq:CPodd_source}, here written for spin $s$ rather than for helicity. For a finite $\gamma$ the integral can be performed by complex contour integration, which picks the poles of the spectral function~\cref{eq:azerozero}. The result is simple:
\begin{equation}
  {\cal S}^\gamma_{qs\pm} = {\rm Re}\big[ {\cal S}_{qs\pm}( \omega_{s\pm} + i\gamma_{s\pm} )\big],
\label{eq:full_thermal_source}
\end{equation}
where ${\cal S}^\gamma_{qs\pm}$ is the source function defined in~\cref{eq:CPodd_source}.
This is the {\em only} effect that coherence damping has on the semiclassical equations. It only modifies only the energy-dependence in the source function, whose parametric dependence on gradients remains unchanged. Also, the relevance of $\gamma$ here depends on how it numerically compares to the total energy instead of the gradient corrections as in the VIA-case. In the limit where $\gamma < T$ one then finds
\begin{equation}
  {\cal S}^\gamma_{qs\pm} \approx {\cal S}_{qs\pm} + \frac{1}{2}[\partial_{k_0}^2{\cal S}_{qs\pm}]_{|k_0|=\omega_{s\pm}} (z_\pm\gamma)^2 + \cdots.
\label{eq:expanded_sorsa}
\end{equation}
That is, the damping correction to the SC-source term is suppressed by a factor $\sim (\gamma/T)^2$.
It does not give rise to a new source with the parametric dependence predicted by the VIA-mechanism.

%%%%%%%%%%%%%%%%%%%%%%%%%%%%%%%%%%%%%%%%%%%%%%%%%%%%%%%%%%%%%%%%%%%%%%%%%%%%%%%%%%%%%%%%%%%%%%
\paragraph{On-shell projection by complex integration}
%%%%%%%%%%%%%%%%%%%%%%%%%%%%%%%%%%%%%%%%%%%%%%%%%%%%%%%%%%%%%%%%%%%%%%%%%%%%%%%%%%%%%%%%%%%%%%

\begin{figure}[t!]
\begin{center}
\includegraphics[scale=0.15]{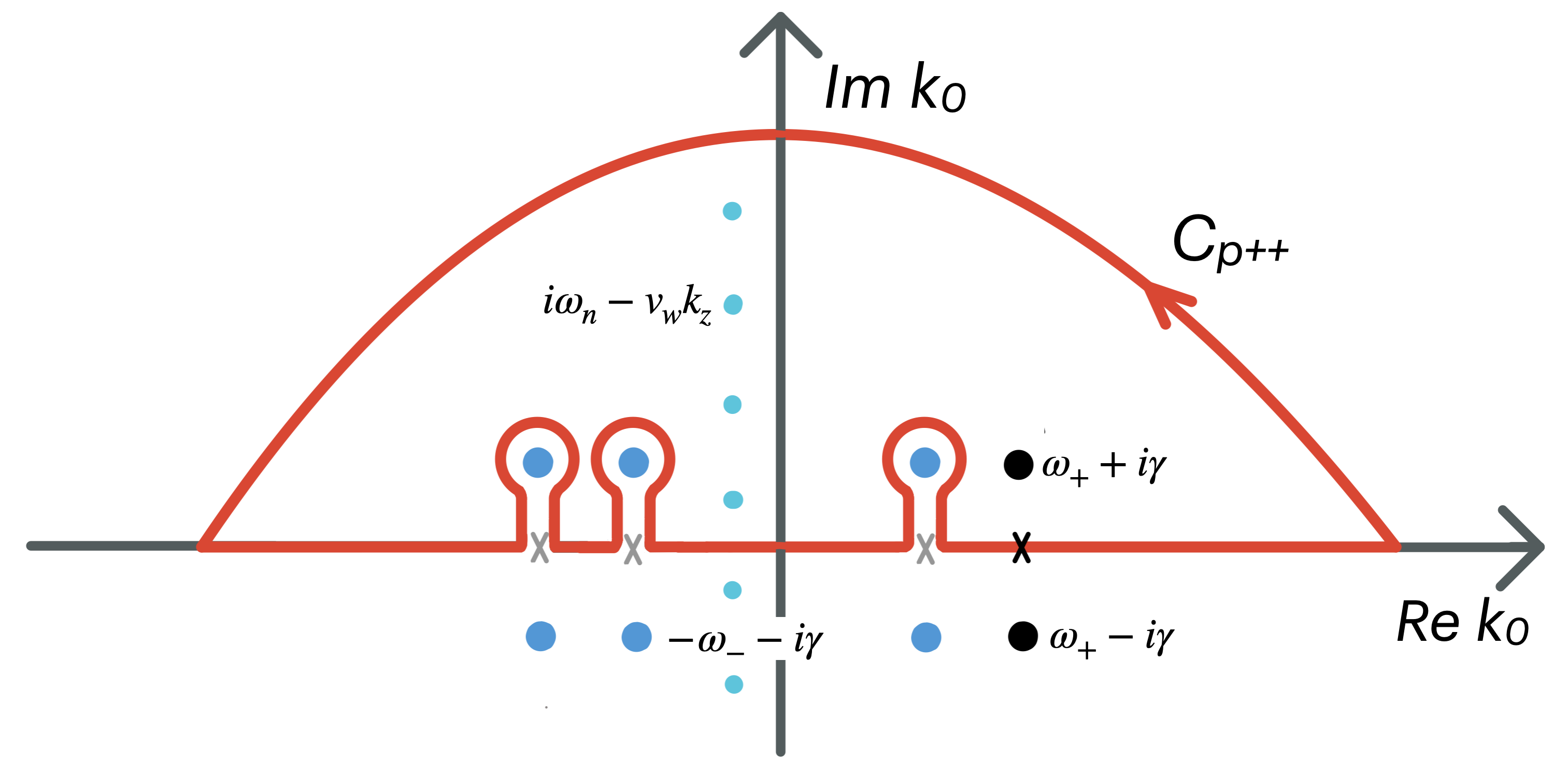}
\end{center}
\vskip-0.5truecm
\caption{A schematic complex contour for the on-shell projection when the source term has a finite width. The contour $C_{p++}$ picks the contribution to the particle source from the spectral function pole denoted by $\omega_+ +i\gamma$ and the spectral shell from the distribution $\delta g_{00}^s$ indicated by the black cross. Hole solutions correspond to $\omega_-$ and the fine structure due to spin is not displayed. The light dots at $k_0 = i\omega_n-v_wk_z$ display the poles of the thermal distribution function, which do not contribute to the integral.}
\label{fig:contour2}
\end{figure}

Let us take a closer look on the contour integration used to obtain~\cref{eq:full_thermal_source}. Until now we have introduced simple weight functions $\rho(k_0)$ to formally induce the on-shell projection under the frequency integration. Given the finite width of the spectral function defining the source term, we need to refine this procedure by introducing contour integration:
\begin{equation}
\int \frac{{\rm d}k_0}{2\pi} \rho_\alpha(k_0) \rightarrow \int_{\mathcal C_\alpha} \frac{{\rm d}k_0}{2\pi}, 
\end{equation}
where the contour ${\mathcal C_\alpha}$ should pick up only the contributions for a given quasistate. We show in figure~\cref{fig:contour2} and example of a path that picks the mass-shell and the source contributions for a particle branch. With this construction the  result~\cref{eq:full_thermal_source} follows immediately. The only non-trivial observation here is that the special points of the thermal distribution function at $k_{0n} = (2n+1)i\pi T-v_wk_z$ do not contribute to integral. Indeed, by the KMS-condition $\Sigma_{\rm th}^\g = - \Sigma_{\rm th}^\l$ and hence $\Sigma_{\rm th}^{\mathcal A} = 0$, which implies that ${\mathcal A} = 0$ at these points. Then, using the general expression on the first line of~\cref{eq:sourceform}, one can show that the matrix valued source function vanishes as well ${\mathcal S}_M(k_{0n})=0$. Of course we are neglecting all nontrivial complex structure arising from thermal corrections beyond the quasiparticle poles, which would contain more information of collective phenomena. However, these effects are numerically even more suppressed than the hole-contributions and can be safely ignored here.

%=============================================================================================
%
%
%=============================================================================================

%%%%%%%%%%%%%%%%%%%%%%%%%%%%%%%%%%%%%%%%%%%%%%%%%%%%%%%%%%%%%%%%%%%%%%%%%%%%%%%%%%%%%%%%%%%%%%
\paragraph{Discussion}
%%%%%%%%%%%%%%%%%%%%%%%%%%%%%%%%%%%%%%%%%%%%%%%%%%%%%%%%%%%%%%%%%%%%%%%%%%%%%%%%%%%%%%%%%%%%%%

We have derived  semiclassical Boltzmann equations for thermal WKB-quasi-particles including thermal corrections up to one loop order in the HTL-approximation, in a controlled expansion in coupling constants and and working up to second order in gradients. We found no sources that would be parametrically of the VIA-form: $\sim |m|^2\theta'$ in the current divergence equations. This leaves no room for speculation that the VIA-method would somehow incorporate different physics from the semiclassical approach.

Equation~\cref{eq:bolzmann_equationdewKB} equipped with the source~\cref{eq:full_thermal_source} and a collision integral~\cref{eq:colltermB}, which can be computed to the lowest order in gradients, but keeping the thermal corrections, is the main result of this paper. In the next section we shall supplement this result by proving the vanishing of the collisional sources of the form reported in~\cite{Prokopec:2004ic}. This proves that this equation contains all gradient corrections to fermionic thermal WKB-states up to second order in gradients. 

Solving the~\cref{eq:bolzmann_equationdewKB} numerically would be somewhat more challenging than the corresponding semiclassical vacuum equation~\cref{eq:bolzmann_equationdel2}. In particular, implementing the moment expansion method of section~\cref{sec:moments} would suffer from the complicated momentum dependence of terms multiplying the $\Delta f_{s\pm}$-gradients in the flow term. Partly this  difficulty is only apparent: if one neglected the momentum dependence of the ${\mathcal R}$ and $(\partial_{k_z}\tilde q_0)$-terms in~\cref{eq:bolzmann_equationdewKB}, then the moment expansion could be easily implemented as an expansion in powers $(q_z/q_0)^\ell$, following a very similar procedure to that  we outlined in section~\cref{sec:moments}. Alternatively, one could solve equation~\cref{eq:bolzmann_equationdewKB} by writing $\Delta f_{h\pm} \equiv - \mu_{h\pm}f_{0w\pm}^\prime + \delta f_{h\pm}$ with $\int_{\bm k} \delta f_{h\pm} = 0$ and expanding $\delta f_{h\pm}$ in some suitable set of basis functions, and setting up a moment expansion to determine the coefficients in this expansion. This approach would be an extension of the fluid ansaz of~\cite{Moore:1995si} and it has been recently shown to work well in the vacuum case in~\cite{Dorsch:2021ubz}. Anyway, while solving~\cref{eq:bolzmann_equationdewKB} will still require more work, it is only a technical challenge; all conceptual issues in setting up the semiclassical Boltzmann equation for the CP-violating perturbation in thermal WKB-quasiparticles have been solved here.
 
%%%%%%%%%%%%%%%%%%%%%%%%%%%%%%%%%%%%%%%%%%%%%%%%%%%%%%%%%%%%%%%%%%%%%%%%%%%%%%%%%%%%%%%%%%%%%%
%%%%%%%%%%%%%%%%%%%%%%%%%%%%%%%%%%%%%%%%%%%%%%%%%%%%%%%%%%%%%%%%%%%%%%%%%%%%%%%%%%%%%%%%%%%%%%
%
\section{Non-thermal self-energy operators}
\label{sec:SC_collisional}
%
%%%%%%%%%%%%%%%%%%%%%%%%%%%%%%%%%%%%%%%%%%%%%%%%%%%%%%%%%%%%%%%%%%%%%%%%%%%%%%%%%%%%%%%%%%%%%%
%%%%%%%%%%%%%%%%%%%%%%%%%%%%%%%%%%%%%%%%%%%%%%%%%%%%%%%%%%%%%%%%%%%%%%%%%%%%%%%%%%%%%%%%%%%%%%

We now consider more general, non-thermal self energy operators, which necessarily arise due to contributions from perturbations $\delta S^{\l,\g}$ to self energies. The division of equations for the inhomogeneous background and for the transient is somewhat intricate, so we will study the issue in some detail. Using again the ansaz~\cref{eq:divisionofSless}, but expanding the self-energy functions around thermal operator as $\Sigma = \Sigma_{\rm th} + \delta \Sigma$, the equation~\cref{eq:StatEqMix2b} becomes:
\begin{align}
(\hat S_0^{-1} - \hat\Sigma^{\rm th}_\H) \delta S^\l &=
{\mathcal S}_M
- i\SigA\delta S^\l 
- i\delta \SigA S^\l_{\rm th} 
+ i \delta \Sigma^\l {\cal A}_{\rm th} + \delta\Sigma^\l S^{\rm th}_\H
\nn \\
 &=
{\mathcal S}_M
+ \sfrac{1}{2} \delta \big( \Sigma^\g S^\l - \Sigma^\l S^\g \big) 
+ \delta\Sigma^\l S^{\rm th}_\H.
\label{eq:StatEqMix3}
\end{align}
Except for the last term this equation is consistent with a semiclassical Boltzmann equation with a nontrivial source and the usual collision term. The $\delta\Sigma^\l S^{\rm th}_\H$-term is problematic however, because it does not allow a spectral reduction. Note however, that the leading term $\Sigma^\l_{\rm th} S^{\rm th}_\H$, associated with the finite width of the solution $S^\l_{\rm th}$, was removed from equations~\cref{eq:StatEqMix2d} and~\cref{eq:StatEqMix3} by the thermal pole equation. This suggests that $\delta\Sigma^\l S^{\rm th}_\H$-term is also associated with the finite width, and gets removed if we use a better approximation for the pole functions and the ensuing spectral function in the ansaz~\cref{eq:divisionofSless}: $S^\l = -2i f^\l_{\rm th} {\cal A} + \delta S^\l$.

It turns out that this is indeed the case. However, since the spectral function and the inhomogeneous part of $S^\l$ are related: ${\cal A} = S^r \SigA S^a$ and $S_{\rm inh}^\l = S^r \Sigma^\l S^a$, we can, consistently with the thermal ansaz, only resum to pole equations the part of the self-energy that satisfies the KMS-condition. Indeed, it is easy to see that in the approximation, where the pole equations satisfy
\begin{equation}
  (\hat S_0^{-1} - \hat\Sigma_\H \pm  i \tilde \Sigma_{\cal A}) S^{r,a} = 1,
\label{PoleEqMixNb}
\end{equation}
where $\tilde{\Sigma}_{\cal A} \equiv \sfrac{1}{2}(1 + e^{\beta p_0})\Sigma^\l$, the equation for 
$\delta S^\l$, defined by the ansaz~\cref{eq:divisionofSless}, becomes
\begin{equation}
(\hat S_0^{-1} - \hat\Sigma_\H) \delta S^\l =  
{\mathcal S}_M
- i\SigA\delta S^\l 
- i\delta \SigA S^\l
+ i \delta \Sigma^\l {\cal A} \,.
\label{eq:StatEqMix4}
\end{equation}
The problematic term has now disappeared proving its association with a finite width of the background solution. The price of this simplification is the implicit dependence of the spectral function on the perturbation $\delta S^\l$ through the resummed collision term $\tilde \Sigma_{\cal A}$. 

How important are these $\delta S$-induced corrections to self-energies? They can be formally expressed as shifts in thermal functions $a$, $b$, $a_I$ and $b_I$, and hence as an eventual shift in the complex momentum vector $Q=q+iq_I$, introduced above equation~\cref{eq:gzeroforpeqs2}. For complex parts these shifts are proportional to g $\delta Q \sim g_s^2\mu$ where $\mu \propto v_w(|m|^2\theta')'$ and can be neglected to the order we are working. Shifts $\delta q$ coming from $\delta \Sigma_H$ are different, but again controlled by an expansion in gradients, and we already have included these corrections up to next-to-leading order in our analysis of thermal WKB-states.  Similarly, a small shift is induced to the width: $\delta \gamma_{s\pm} \sim \mu\gamma_{s\pm}$, which again is negligible everywhere to the order we are working. So, in the end we can use the ansaz~\cref{eq:divisionofSless} and the equation~\cref{eq:StatEqMix4} with thermal self-energy functions $\Sigma^{\rm th}_H$, $\Sigma^{\rm th}_{\cal A}$ and a corresponding solution to pole-equations: ${\cal A}$. That is, to the order we are working, one can just drop the $\delta\Sigma^\l S^{\rm th}_\H$-term in equation~\cref{eq:StatEqMix3}. 

Let us clarify what we just proved. Ideally, one would like to solve equations~\cref{eq:pole,eq:kb} exactly finding exact inhomogeneous solutions for the pole functions and correspondingly $S^\l = S^r\!*\!\Sigma^\l\!*\!S^a$. However, this is just a fancy way of rewriting the initial problem as coupled integral equations and not useful in practice. Our solution is to introduce an approximate known inhomogeneous background solution that tracks the true solution for $S^<$ as well as possible and model the difference by a dynamical perturbation, whose source is defined by the chosen background solution. This procedure is not exact, but we have proved that it is consistent and quantitatively correct to the order we are working in the gradient expansion.

%%%%%%%%%%%%%%%%%%%%%%%%%%%%%%%%%%%%%%%%%%%%%%%%%%%%%%%%%%%%%%%%%%%%%%%%%%%%%%%%%%%%%%%%%%%%%%
%%%%%%%%%%%%%%%%%%%%%%%%%%%%%%%%%%%%%%%%%%%%%%%%%%%%%%%%%%%%%%%%%%%%%%%%%%%%%%%%%%%%%%%%%%%%%%
%
\subsection{Vanishing of collisional sources}
\label{sec:SC_collisional_vanish}
%
%%%%%%%%%%%%%%%%%%%%%%%%%%%%%%%%%%%%%%%%%%%%%%%%%%%%%%%%%%%%%%%%%%%%%%%%%%%%%%%%%%%%%%%%%%%%%%
%%%%%%%%%%%%%%%%%%%%%%%%%%%%%%%%%%%%%%%%%%%%%%%%%%%%%%%%%%%%%%%%%%%%%%%%%%%%%%%%%%%%%%%%%%%%%%

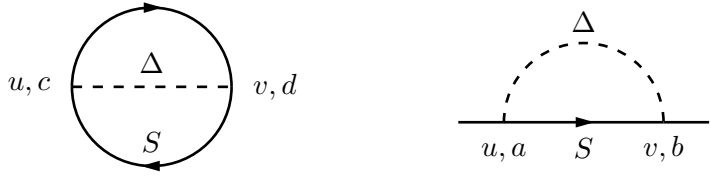
\begin{figure}[!t]

\def\mytip{
    \path [fill, scale=1, xshift=4pt] (0,0) -- ++(-7pt,2pt) -- ++(0,-4pt) -- cycle;}
\tikzset{
    myarrow/.style={decoration={markings, mark = at position #1 with {\mytip};},
        postaction={decorate}}}

 \centering
 \begin{subfigure}[c]{0.35\linewidth}
    \centering
    \begin{tikzpicture}[line width = 1.0pt]
        \coordinate (x);
        \coordinate [right = 2.1cm of x] (y);
        \draw [dash pattern=on 3.8pt off 4pt, dash phase=0pt]
               (x) node [left,   left  = 4pt] {$u,c$} 
            -- (y) node [midway, above = 1pt] {$\Delta$} 
                   node [right,  right = 4pt] {$v,d$};
        \draw [myarrow=0.5] 
               (x) arc (180:0:1.05cm); 
                   %node [midway, above = 1pt] {$S$};
        \draw [myarrow=0.5] 
               (y) arc (0:-180:1.05cm) 
               node [midway, above = 1pt] {$S$};
     \end{tikzpicture}
 \end{subfigure}
 ~    % This puts the subfigures sideways, otherwise on top of each others 
 \begin{subfigure}[c]{0.35\linewidth}
    \centering
    \begin{tikzpicture}[line width = 1.0pt, baseline=(bas)]
       \coordinate (x); 
       \coordinate [below = 3pt of x] (bas);
       \coordinate [right = 2.1cm  of x] (y);
       \coordinate [left  = 0.6cm  of x] (in);
       \coordinate [right = 0.6cm  of y] (out);
       \draw (in) -- (x);
       \draw (y) -- (out);
       \draw [myarrow=0.5]
             (x) node [left,   below = 4pt] {$u,a$}
          -- (y) node [midway, below = 2pt] {$S$}
                 node [left,   below = 1pt] {$v,b$};
       \draw [dash pattern=on 3.5pt off 4pt, dash phase=0pt] 
             (x) arc (180:0:1.05cm) node [midway, above=0.2pt] {$\Delta$};
    \end{tikzpicture}
 \end{subfigure}
 \caption{Leading contribution to the 2PI-action $\Gamma_2$ (left) and to fermion self
          energy (right) in the Yukawa model. $u$ and $b$ are space-time coordinates 
          and $a$ and $b$ labels on the complex time contour ${\cal C}$ (see   
          section~\cref{sec:CTP}.}
\label{fig:self-energy-2}

\end{figure}

It is clear from above that the collision term in~\cref{eq:StatEqMix4} is parametrically of order $\sim g_s^2 \mu$ and can be computed working to zeroth order in gradients, just as the thermal term in section~\cref{sec:SC_collisional_damping}. However, ref.~\cite{Prokopec:2004ic} reported a new source arising from the self-energy diagram shown in figure~\cref{fig:self-energy-2}, that is of lower order in gradients: $\propto v_w\gamma_w|m|^2\theta'y^2$, which is parametrically similar to the VIA source~\cref{eq:Source-equations4} and does not vanish in equilibrium. Although this term eventually does not source the divergence equation directly~\cite{Prokopec:2004ic}, its parametric form contradicts our general argument. To resolve the issue we now compute the collision term explicitly for the self-energy of figure~\cref{fig:self-energy-2}.

To keep the argument simple, and at the level of~\cite{Prokopec:2004ic}, we drop the thermal corrections. The interaction term ${\cal L} = -y\bar q_L \phi q_R + h.c.$, gives a two-loop contribution to the 2PI effective action:
\begin{equation}
\Gamma_2 = y^2\int_{\cal C} {\rm d}^4u{\rm d}^4v \sum_{cd}{\rm Tr}\big[S^{cd}P_RS^{dc}P_L\big]\Delta^{cd},
\end{equation}
where $P_{L,R} = \sfrac{1}{2}(1\mp\gamma^5)$ and indices $c,d = \pm$ refer to the position of the time-components of $u$ and $v$ on the complex time contour ${\cal C}$, such that $S^\l= S^{+-}$ and $S^\g=S^{-+}$. Self-energies can be computed from $\Gamma_2$ as variational derivatives: $\Sigma^{ab}(u,v) = -iab \delta \Gamma_2/\delta S^{ba}(v,u)$.
Going directly to the Wigner representation after the differentiation one finds
\begin{equation}
\Sigma^{\l,\g}_k = iy^2 \int_{k'}\big( P_L S_{k'}^{\l,\g} P_R \Delta^{\g,\l}_{k'-k} 
                                     + P_R S_{k'}^{\l,\g} P_L \Delta^{\l,\g}_{k-k'} \big),
\label{eq:sigmalgy}
\end{equation}
where $\int_{k'} \equiv \int \frac{{\rm d}^4k'}{(2\pi)^4}$ and $\Sigma^{\l,\g}_k \equiv \Sigma_w^{\l,\g}(k)$ with similar notation for propagators. Note that the projection operators $P_{L,R}$ eliminate the $g_{10}$- and $g_{20}$ terms from $S_{k'}^{\l,\g}$ and the same happens with the other propagator in~\cref{eq:Coll-term2} when one inserts~\cref{eq:sigmalgy} into the trace. Moreover, we assume that the scalar field is thermalised, so that $\Delta \rightarrow \Delta_{\rm th}$, which obeys $\Delta^{\l,\g}_{\rm th}(-k) =\Delta^{\g,\l}_{\rm th}(k)$. After these observations it is easy to see that the collision term~\cref{eq:Coll-term2} for the spin $s$ state becomes
\begin{equation}
C_{\mathbb 1}^\sl[f] =- \frac{iy^2}{2}\int_{k'}\big( 
   {\rm Tr}\big[S_{dk'}^{\l} S_{dk}^{\g} P_{sk} \big]\Delta^\l_{{\rm th},k-k'} 
 - {\rm Tr}\big[S_{dk'}^{\g} S_{dk}^{\l} P_{sk} \big]\Delta^\g_{{\rm th},k-k'} \big),
\end{equation}
where the index $d$ refers to the "diagonal" limit with $g^{\l,\g}_{10}\!=\!g^{\l,\g}_{20}\!\equiv \!0$. Given the explicit form for propagators~\cref{es:Slesswall} we can easily evaluate the traces: 
\begin{equation}
 {\rm Tr}\big[S_{dk'}^{\l} S_{dk}^{\g} P_{sk} \big] = \frac{1}{4} \big(1 + ss'u^k_\PA\cdot u^{k'}_\PA \big)\big[ \big(g^\sl_{33k} g^{s'\g}_{33k'} - \frac{ss'}{\gamma^k_\PA\gamma^{k'}_\PA}g^\sl_{00k} g^{s'\g}_{00k'}\big).
\label{eq:cfinaali}
\end{equation}
%

%=============================================================================================

%=============================================================================================

These results still agree with~\cite{Prokopec:2004ic}. Let us now be very precise of the definition of the thermal equilibrium values for various components $g^{s\l,\g}_{ab}$. Indeed, we have defined:
\begin{equation}
S^\l_{\rm th} \equiv -2i f^\l_{\rm th} {\cal A}       \quad {\rm and} \quad 
S^\g_{\rm th} \equiv -2i f^\g_{\rm th} {\cal A},
\label{eq:equilibriumvalues}
\end{equation}
where $f^<_{\rm th}(k_0)=f_\FD(p_0)$ and $f^\g_{\rm th}(k_0) = e^{\beta p_0}f^\l_{\rm th}(k_0)=1-f_\FD(p_0)$, where $p_0=\gamma_w(k_0 + v_wk_z)$. We then  have $\sfrac{i}{2}(S^\g_{\rm th} + S^\l_{\rm th}) = {\cal A}$ and $S^\g_{\rm th} = e^{\beta p_0}S^\l_{\rm th}$ as required by the KMS-condition. It is essential to understand that~\cref{eq:equilibriumvalues} are matrix equations that apply component by component, This implies for example that
\begin{equation}
g^{s\l}_{ab,{\rm th}}(k) g^{s'\g}_{ab,{\rm th}}(k')
= a^{s}_{ab}(k) a^{s'}_{ab}(k') f_\FD(p_0)(1-f_\FD(p_0')).
\end{equation}
Inserting this and similar expressions for other terms back to~\cref{eq:cfinaali}, we see that the equilibrium part of the collision term vanishes identically as it should. Note in particular that to first order in gradients equation~\cref{eq:equilibriumvalues} implies:
\begin{equation}
g_{33,{\rm th}}^{s\l,\g} \,=\, -2if^{\l,\g}_{0w} a^s_{33} 
\,=\,-2i f^{\l,\g}_{0w} \Big(\frac{k_z}{k_0} a^s_{00} - s \frac{|m|^2\theta'}{2k_0\tilde k_0}\partial_{k_z}a_{00}^s \Big).
\label{eq:gthrethree}
\end{equation}
The point where~\cite{Prokopec:2004ic} differs from our analysis is that instead of~\eqref{eq:gthrethree} they set (transferring signs to our notation)
\begin{equation}
g_{33,{\rm th}}^{s\l,\g} \rightarrow - \frac{k_z}{k_0}g^{s\l,\g}_{00,\rm th} - s \frac{|m|^2\theta'}{2k_0\tilde k_0}\partial_{k_z}g^{s\l,\g}_{00,\rm th} 
\label{eq:gthrethreewrong}
\end{equation}
and furthermore $\partial_{k_z} g^\g_{00,{\rm th}} \rightarrow \partial_{k_z}(e^{\beta p_0}g^\l_{00,{\rm th}})$. These choices are not consistent with the KMS condition. The definition~\cref{eq:gthrethreewrong} also assumes that $S_{\rm th}^\sl$ is a solution to SC-equations, which is not true: only the full $S^\sl$ and ${\cal A}_s$ solve the equations to the order we are working. It is precisely this failure that gives rise to the semiclassical source as we have shown above. The above rule for $\partial_{k_z} g^\g_{00,{\rm th}}$ then adds an extra term $\propto v_w\gamma_w|m|^2\theta'$ to the {\em r.h.s.}~of equation~\cref{eq:gthrethree} for $g^\sg_{33,{\rm th}}$, which breaks the KMS-condition explicitly and creates the fictitious collisional source reported in~\cite{Prokopec:2004ic}. 

To continue the explicit evaluation of~\cref{eq:cfinaali} we note that $\delta S^\g+\delta S^\l=0$, which implies $\delta g_{ab}^\sg = -\delta g_{ab}^\sl$. Then, using the explicit form~\cref{eq:specfunf} for $g^\l_{00}$ with the perturbations written in a form equivalent to~\cref{eq:division} and noting that in scalar particle decay channel kinematics requires ${\rm sgn}(k_0') = -{\rm sgn}(k_0)$, one readily finds:
\begin{equation}
 {\cal C}_{s\pm}[f] \approx  -\sum_{s'}\int_{{\bm k}'} |{\cal M}_{ss'}|^2 
 f_{0w}^k f_{0w}^{k'}(1+f_{\phi0}^{k'+k}) 
 \left[ \frac{\mu_{s}}{T} - \frac{\mu_{s'}}{T} 
 - \frac{\delta f_{sk}}{(f^k_{0w})'} 
 + \frac{\delta f_{s'k'}}{(f^{k'}_{0w})'} \right]_\pm ,
\end{equation}
where
\begin{equation}
|{\cal M}_{ss'}|^2= \frac{y^2}{8}
\big(1 + ss'u^k_\PA\cdot u^{k'}_\PA \big)\big[ \Big( \frac{k_zk_z'}{\omega_{sk}\omega_{sk'}} -\frac{ss'}{\gamma^k_\PA\gamma^{k'}_\PA}\Big).
\label{eq:melement}
\end{equation}
Neglecting the back-reaction term proportional to the integral over $\delta f_{s'k'}$, one can write the collision integral for the CP-odd part of the perturbations as
\begin{equation}
 {\cal C}_{s}[f] \approx  -\Gamma_{\rm flip}f_{0w}^k(\mu_{s}-\mu_{-s}) - \delta f_{sk} \Gamma_{\rm \scriptscriptstyle TOT}.
\label{eq:samplecterm}
\end{equation}
If one approximates $\Gamma_{\rm flip}$ and $\Gamma_{\rm \scriptscriptstyle TOT}$ by constants, one can show that the two lowest moments of~\cref{eq:samplecterm} gives rise to collision terms of the form shown in~\cref{eq:collision_terms} for moment equations. Of course it is not necessary to use the slightly cumbersome $s$-base in practical calculations; as stated several times before, one can easily move to helicity basis just by a simple redefinition of the force term.

%%%%%%%%%%%%%%%%%%%%%%%%%%%%%%%%%%%%%%%%%%%%%%%%%%%%%%%%%%%%%%%%%%%%%%%%%%%%%%%%%%%%%%%%%%%%%%
%%%%%%%%%%%%%%%%%%%%%%%%%%%%%%%%%%%%%%%%%%%%%%%%%%%%%%%%%%%%%%%%%%%%%%%%%%%%%%%%%%%%%%%%%%%%%%
%
\section{Conclusions}
\label{sec:conclusion}
%
%%%%%%%%%%%%%%%%%%%%%%%%%%%%%%%%%%%%%%%%%%%%%%%%%%%%%%%%%%%%%%%%%%%%%%%%%%%%%%%%%%%%%%%%%%%%%%
%%%%%%%%%%%%%%%%%%%%%%%%%%%%%%%%%%%%%%%%%%%%%%%%%%%%%%%%%%%%%%%%%%%%%%%%%%%%%%%%%%%%%%%%%%%%%%

We have derived the CP-violating transport equations for the electroweak baryogenesis in the limit of slowly varying background fields, including all thermal corrections to one loop order. Historically the transport problem has been studied using two competing methods, which have been shown~\cite{Cline:2020jre} to give very different answers to same physical questions: the VEV insertion approximation (VIA) and the semiclassical (SC) method. 

In the first part of this paper we carefully reviewed the VIA formalism. We showed that the method is based on an inconsistent representation of the singular mass operator by a nonlocal self-energy term and a memory integral containing a pinch singularity. It was shown that regulating the singularity by a finite width is  an inherently ambiguous process, where VIA literature has made several implicit and ad-hoc assumptions  that caused further inconsistencies including spurious ultraviolet and infrared divergences.

In the second part of the paper we carefully reviewed the semiclassical method and extended it to incorporate thermal corrections to dispersion relations and the finite width effects at one loop level in the HTL approximation. We derived of the dispersion relation and the semiclassical evolution equation for the thermal WKB-quasistates in a series of steps extending the usual concept of thermal quasiparticles to  moving frames and spatially varying backgrounds. Our equations encompass both particle and hole branches and expresses all CP-violating effects in a single source term that incorporates all dispersive and finite width corrections.  This is an important development on its own, and the most significant result of this paper. However, it also proves that the source predicted by the VIA-method does not arise in a consistent expansion in gradients and coupling constants. The VIA-source then is not merely ambiguous, but it simply does not exist.

We also made some clarifications to the earlier SC-literature, including a careful on-shell projection of the distributive equations and a precise identification of the true semiclassical force and collision terms, accounting for the wave function renormalisation factors that were earlier unappreciated. These conceptual issues have no practical effect on the final equations however. We also showed that there are no additional sources coming from collision terms in the semiclassical formalism up to second order in gradients. Such sources, with the same parametric form as the VIA-sources (albeit differently placed in moment equation hierarchy) were reported in~\cite{Prokopec:2004ic}. We showed both on general grounds and in the specific example studied in~\cite{Prokopec:2004ic} that these sources vanish in a correct treatment. This proves that collision integrals can be always computed working to the zeroth order in the gradient expansion in the SC formalism.

We also pointed out that the Fick's law is not consistently implemented in the VIA literature: it should not be applied to the total current, but to the diffusion current related to the non-equilibrium perturbation around the kinetic equilibrium (since in the kinetic equilibrium there is no flow). The total current contains two other parts: the advection current and a drag caused by the semiclassical force. When the correct diffusion current is identified, the Fick's law, applied in the current conservation equation  does give rise to a diffusion equation that is consistent with the SC-formalism. However, the SC-equations are more general, and they can be used to {\em derive} the Fick's law, including an explicit equation for the diffusion coefficient. 

In this paper we considered only fermions and only a single-flavour system. Extension to multi-flavour systems should be straightforward using the methods developed in~\cite{Jukkala:2021cys}. Generic source terms for a fermionic system with several flavours and dynamical flavour mixing have been already derived in~\cite{Kainulainen:2001cn,Kainulainen:2002th} in the flavour diagonal basis and the Liouville terms arising from the rotation to the flavour diagonal basis were identified in~\cite{Konstandin:2005cd}. We did not consider bosonic systems either, since is well known~\cite{Cline:1997vk,Konstandin:2013caa} that the CP-violating effects for bosons arise at higher order in gradients in the semiclassical formalism. However, the generic arguments presented here concerning the validity of the VIA-method apply also to the VIA-derivation of the bosonic source terms.

%%%%%%%%%%%%%%%%%%%%%%%%%%%%%%%%%%%%%%%%%%%%%%%%%%%%%%%%%%%%%%%%%%%%%%%%%%%%%%%%%%%%%%%%%%%%%%%
\section*{Acknowledgements} 
%
%%%%%%%%%%%%%%%%%%%%%%%%%%%%%%%%%%%%%%%%%%%%%%%%%%%%%%%%%%%%%%%%%%%%%%%%%%%%%%%%%%%%%%%%%%%%%%

I thank Larry McLerran for giving this excellent advice long time ago: {\em if something smells like a fish, it means there's a fish somewhere}. I also thank Jim Cline for spurring me to find out where this particular fish was {\em inserted}. I thank Jim also for a long and extensive collaboration related to the SC-method. I also thank Werner Porod, Mikko Laine, Olli Koskivaara, Pyry M. Rahkila and in particular Henri Jukkala for discussions and valuable comments on the manuscript. Large part of this work was done during a recent sabbatical leave at the CERN theory department, which I warmly thank for hospitality. This work was supported by the Academy of Finland grant 318319.

%%%%%%%%%%%%%%%%%%%%%%%%%%%%%%%%%%%%%%%%%%%%%%%%%%%%%%%%%%%%%%%%%%%%%%%%%%%%%%%%%%%%%%%%%%%%%%
%%%%%%%%%%%%%%%%%%%%%%%%%%%%%%%%%%%%%%%%%%%%%%%%%%%%%%%%%%%%%%%%%%%%%%%%%%%%%%%%%%%%%%%%%%%%%%

\section*{Appendices}
\addcontentsline{toc}{section}{\protect\numberline{}Appendices}
\appendix

%%%%%%%%%%%%%%%%%%%%%%%%%%%%%%%%%%%%%%%%%%%%%%%%%%%%%%%%%%%%%%%%%%%%%%%%%%%%%%%%%%%%%%%%%%%%%%
%
\section{Derivation of SC-equations with arbitrary collision term} 
\label{sec:appendixA} 
%
%%%%%%%%%%%%%%%%%%%%%%%%%%%%%%%%%%%%%%%%%%%%%%%%%%%%%%%%%%%%%%%%%%%%%%%%%%%%%%%%%%%%%%%%%%%%%%

In this appendix we show that the semiclassical equations of motion can be derived, and hold as such,  keeping the collisional damping terms in the reduction process. The fully general (we do not include thermal corrections here) starting point to derive the component equations can be written as:
\begin{equation}
{\rm Tr} \big[{\mathcal O}\big(\tilde k_0 \dag{u}_\PA - \gamma^3(k_z-\sfrac{i}{2}\partial_z) 
                              - {\hat m}_\R - i\gamma^5 {\hat m}_\I \big) 
iS_{ws}^\l \big] = iC_{\mathcal O}^\sl,
\label{eq:main_app}
\end{equation}
where the four operators that are needed can be chosen for example as ${\mathcal O} = {\mathbb 1}, \gamma^3, \gamma^5, \dag{u}_\PA$, and the corresponding collision integrals are:
\begin{equation}
C_{\mathcal O}^\sl \equiv {\rm Tr}[\sfrac{1}{2}(\Sigma^> S_{ws}^<-\Sigma^< S_{ws}^>)P_{ws}{\mathcal O}].
\end{equation}
Four out of the six relevant equations acquire contributions from collision integrals. First, the two (eventually equivalent) evolution equations associated with the vector and axial vector current divergences, as indicated in section~\cref{VEV:correct} become:
\begin{align}
\partial_z g^\sl_{33} + 2\sin\diamond (m_R g^\sl_{10}) - 2\sin\diamond (m_I g^\sl_{20})  &= -C^\sl_{\mathbb 1},
\label{eq:vector_current_eq_app} 
\\
s\partial_z g^\sl_{00} + 2\gamma_\PA \cos\diamond (m_R g^\sl_{10}) 
                       + 2\gamma_\PA \sin\diamond (m_I g^\sl_{20}) &= - C_{\gamma^5}^\sl,
\label{eq:axial_current_eq_app}
\end{align}
Similarly the two constraint equations giving $g^\sl_{10}$ and $g^\sl_{20}$ get updated:
\begin{subequations}
\label{eq:cont_app}
\begin{align}
k_0 g^\sl_{10} &= \phantom{+} 
\sfrac{s}{2}\gamma_\PA\partial_z g_{20}^\sl + \cos \diamond (m_\R g_{00}^\sl)  
+ s\gamma_\PA \sin\diamond (m_\I g_{33}^\sl) + \frac{i}{2}\gamma_\PA C_{\dag{u}_\PA}^\sl,
\\*
k_0 g^\sl_{20} &= -
 \sfrac{s}{2}\gamma_\PA\partial_z g_{10}^\sl - \cos \diamond (m_\I\, g_{00}^\sl) 
 + s\gamma_\PA \sin\diamond (m_\R g_{33}^\sl) - \frac{1}{2}s\gamma_\PA C_{\gamma^3}^\sl,
\label{eq:gsconnections_app}
\end{align}
\end{subequations}
On contrary, there are no corrections to the constraint equation~\cref{eq:wfconstr} for $g_{33}^\sl$ or to equation~\cref{eq:gsconnections} that serves as a starting point for deriving the dispersion relation. 

The trick is to treat collision terms formally as second order corrections in the gradient expansion. This follows from the fact that collision integrals vanish in the equilibrium and the (CP-odd) perturbations are eventually sourced by terms that are of at least of second order in gradients. It is easy to see that the derivation of the dispersion relation~\cref{eq:wfconstrb} is unchanged and moreover, all corrections from collision terms drop in the reduction of~\cref{eq:vector_current_eq_app}, so that the evolution equation~\cref{eq:Coll-term1} with the simple collision term and the free equation flow term~\cref{eq:wfevoeqb} emerge unchanged. Reduction of the second evolution equation~\cref{eq:axial_current_eq_app} to the form shown in~\cref{eq:comparison} is somewhat more involved and we give some intermediate steps here to the benefit of an interested reader. It is useful to note that one can still solve $g_{33}^\sl$ to the first order accuracy in gradients as follows:
\begin{equation}
g_{33}^\sl \approx - \frac{k_z}{k_0}g_{00}^\sl - s\gamma_\PA \frac{|m|^2\theta'}{2k_0^2}\partial_{k_z}g_{00}^\sl.
\label{eq:g3_app}
\end{equation}
Iterating the constraint equations~\cref{eq:g33-constraint} and~\cref{eq:cont_app} in~\cref{eq:axial_current_eq_app} one can easily rewrite~\cref{eq:axial_current_eq_app} as follows:
\begin{equation}
\frac{1}{k_0^2}(\Omega_s^2 + k_z^2)\partial_zg_{00}^\sl + F^s_{k_0}\big(\frac{g^\sl_{00}}{k_0} - \partial_{k_z}g_{33}^\sl\big) - s\gamma_\PA \frac{|m|^2\theta'}{2k_0^2}\partial_{k_z}\partial_zg_{33}^\sl = C^\sl_{\rm eff},
\label{eq:app_middle}
\end{equation}
where $\Omega_s^2$ was defined in~\cref{eq:wfconstrb}, $F^s_{k_0}$ is the force term appearing in~\cref{eq:wfevoeqb}, when written as ${\mathcal D}_s g_{00}^\sl = \frac{k_z}{k_0}\partial_zg_{00}^\sl + F_{k_0}^s\partial_{k_z}g_{00}^\sl$ and the effective collision integral is given by:
\begin{equation}
C^\sl_{\rm eff} = - \frac{s}{\gamma_\PA} C^\sl_{\gamma^5} + \frac{m_R}{k_0}C^\sl_{\gamma^3}
- s\frac{m_I}{k_0}C^\sl_{\dag{u}_\PA}.
\label{eq:Ceff_app}
\end{equation}
Observing that $(\Omega_s^2 + k_z^2)\partial_zg_{00}^\sl = \partial_z[(\Omega_s^2 + k_z^2)g_{00}^\sl] - (\partial_z\Omega_s^2)g_{00}^\sl = -k_z^2 g_{00}^\sl - 2k_0F^s_{k_0}g_{00}^\sl$, using~\cref{eq:g3_app} for $g^\sl_{33}$ in the second term and again~\cref{eq:vector_current_eq_app} to the first order accuracy in gradients in the third term, one can write the left hand side of~\cref{eq:app_middle} in the desired form: 
\begin{equation}
\frac{k_z^2}{k_0^2}\partial_zg_{00}^\sl + \frac{k_z}{k_0} F_{k_0}^s \partial_{k_z}g_{00}^\sl 
= \frac{k_z}{k_0}{\mathcal D}_s g_{00}^\sl.
\end{equation}
It remains to reduce the effective collision integral term $C_{\rm eff}^\sl$. To this end we note first that $iS_w^\l$ can be written as
\begin{equation}
iS_{ws}^\l = iS_w^\l P_{ws} = 
 \frac{1}{4}\Big(P^s_0 \frac{1}{\gamma_\PA}g^\sl_{00} 
               + P^s_3 g^\sl_{33} 
               + P^s_1 g^\sl_{10} 
               + P^s_2 g^\sl_{20} \Big),
\end{equation}
where the four orthogonal (under the trace operation) projection operators 
\begin{equation}
P^s_0 = \dag{u}_\PA + s\gamma^3\gamma^5, \quad
P^s_3 = \gamma^3 + s\dag{u}_\PA \gamma^5 \quad
P^s_1 = {\mathbb 1} + s\dag{u}_\PA \gamma^3\gamma^5, \quad
P^s_2 = \gamma^5 + s\dag{u}_\PA \gamma^3,
\end{equation}
obey the right multiplication rules given in table~\cref{tab:app}:
\begin{table}[t]
\begin{center}
\begin{tabular}{|c|c|c|c|c|}
 Operator $\backslash$ Right multiplier  &  $ 1   $  & $\gamma^3$  &  $\gamma^5$  &  $\dag{u}_\PA$  \\
\hline
$\dag{u}_\PA + s\gamma^3\gamma^5  $  &  $ P^s_0$  &  $  sP^s_2$  &  $ sP^s_3 $  &  $   P^s_1   $  \\
$\gamma^3 + s\dag{u}_\PA \gamma^5 $  &  $ P^s_3$  &  $  -P^s_1$  &  $ sP^s_0 $  &  $ -sP^s_2   $  \\
${\mathbb 1} + s\dag{u}_\PA \gamma^3\gamma^5$  &  $ P^s_1$  &  $   P^s_3$  &  $  P^s_2 $  &  $   P^s_0   $  \\
$\gamma^5 + s\dag{u}_\PA \gamma^3 $  &  $ P^s_2$  &  $ -sP^s_0$  &  $  P^s_1 $  &  $ -sP^s_3   $  \\
\end{tabular}
\caption{Shown are the spin-structures appearing in $iS_{ws}^\l$ and their right-multiplication rules by the operators indicated.}
\end{center}
\label{tab:app}
\end{table}
For an arbitrary self-energy, we may define the traces ${\rm Tr}[\Sigma^{sa}P^s_i] = c^{sa}_i$, where $a = <,>$. Using table~\cref{tab:app} it is then easy to show, working to lowest order in gradients that 
\begin{align}
C^\sl_1 &= \big(\sfrac{1}{\gamma_\PA} c_0^\sl - \sfrac{k_z}{k_0} c_3^\sl
              + \sfrac{m_R}{k_0} c_1^\sl  - i \sfrac{m_I}{k_0} c_2^\sl\big)g^\sl_{00} 
              \; + < \leftrightarrow >
\nonumber \\
C^\sl_{\gamma^3} &= \big(\sfrac{s}{\gamma_\PA} c_2^\sl + \sfrac{k_z}{k_0} c_1^\sl
              + \sfrac{m_R}{k_0} c_3^\sl  + is \sfrac{m_I}{k_0} c_0^\sl\big)g^\sl_{00} 
              \; + < \leftrightarrow >
\nonumber \\
C^\sl_{\gamma^5} &= \big(\sfrac{s}{\gamma_\PA} c_3^\sl - s\sfrac{k_z}{k_0} c_0^\sl
              + \sfrac{m_R}{k_0} c_2^\sl - i \sfrac{m_I}{k_0} c_1^\sl\big)g^\sl_{00} 
              \; + < \leftrightarrow >
\nonumber \\
C^\sl_{\dag{u}_\PA} &= \big(\sfrac{1}{\gamma_\PA} c_1^\sl + s\sfrac{k_z}{k_0} c_2^\sl
              + \sfrac{m_R}{k_0} c_0^\sl  + is \sfrac{m_I}{k_0} c_3^\sl\big)g^\sl_{00} 
              \; + < \leftrightarrow >
\end{align}
Using these relations in~\cref{eq:Ceff_app} is it is now easy to show that indeed $C^\sl_{\rm eff} = \frac{k_z}{k_0}C_{\mathbb 1}^\sl$ for an arbitrary self-energy function and hence for arbitrary interactions. Combining these results proves that equation~\cref{eq:axial_current_eq_app} is just equation~\cref{eq:vector_current_eq_app} multiplied by $k_z/k_0$ and hence redundant in the semiclassical analysis.

%%%%%%%%%%%%%%%%%%%%%%%%%%%%%%%%%%%%%%%%%%%%%%%%%%%%%%%%%%%%%%%%%%%%%%%%%%%%%%%%%%%%%%%%%%%%%%
%
\bibliography{VEV.bib}
%
%%%%%%%%%%%%%%%%%%%%%%%%%%%%%%%%%%%%%%%%%%%%%%%%%%%%%%%%%%%%%%%%%%%%%%%%%%%%%%%%%%%%%%%%%%%%%%

%%%%%%%%%%%%%%%%%%%%%%%%%%%%%%%%%%%%%%%%%%%%%%%%%%%%%%%%%%%%%%%%%%%%%%%%%%%%%%%%%%%%%%%%%%%%%%%
\end{document}